\documentclass[5p]{elsarticle}
\usepackage{lineno}
\usepackage{float}
\usepackage{graphicx}
\usepackage{hyperref}
\usepackage{booktabs}
\usepackage[tight]{subfigure}
\usepackage{caption}

\usepackage{afterpage}
\usepackage{graphicx}

\usepackage{amsmath,amssymb,amsfonts}
\usepackage{siunitx}
\usepackage{algorithmic}
\usepackage{textcomp}
\usepackage{xcolor}
\usepackage{tkz-euclide}
\usepackage{framed}
\usepackage{tikz}
\usepackage{adjustbox}
\usepackage{url}
\usepackage{libertine}
\usepackage{pdflscape}
\usepackage{enumitem}

\usetikzlibrary{positioning, arrows, shapes,chains}
\usepackage{color}
\usepackage{transparent}
\graphicspath{img/}
\usepackage{rotating}
\usepackage{multirow}
\usepackage{rotating}
\usepackage{cleveref}

\Crefmultiformat{section}{Sections~#2#1#3}{ and~#2#1#3}{, #2#1#3}{ and~#2#1#3}
\newenvironment{absolutelynopagebreak}
  {\par\nobreak\vfil\penalty0\vfilneg
   \vtop\bgroup}
  {\par\xdef\tpd{\the\prevdepth}\egroup
   \prevdepth=\tpd}

\modulolinenumbers[5]

\journal{Journal of Systems and Software -- Special Issue on Managing Technical Debt in Software-intensive Products and Services}

\newcommand{\ActionCycles}{five }
\newcommand{\StudyMonths}{16 }

\newcommand{\MeetingCnt}{35 }
\newcommand{\MeetingHrs}{25 }
\newcommand{\DoubleObservation}{five }
\newcommand{\ItemsObserved}{415 }

\newcommand{\TDSAGATMeetingCnt}{12 }
\newcommand{\TDSAGATMeetingAnswersCnt}{90 }
\newcommand{\TDSAGATMeetingAnswersAvg}{7.5 }

\newcommand{\BacklogitemsWhenTagged}{109 } %Count nach Konsolidierung-Meeting im Azure DevOps Tool
\newcommand{\TDitemsTagged}{65 } %Count nach Konsolidierung-Meeting im Azure DevOps Tool
\newcommand{\TDitemsTaggedExculdingDeleted}{40 } %Count nach Konsolidierung-Meeting im Azure DevOps Tool
\newcommand{\TDitemsDeleted}{50 } %Löschaktion 11-2023 (errechnet)
\newcommand{\TDitemsFinal}{102 } %aus Export 06-2024
\newcommand{\TDitemsOpenedFeb}{61 } %Refactoring Initiative OpenDate in 02-2024
\newcommand{\MeanEstablishmentEffort}{5.1 }
\newcommand{\MeanMaintenanceEffort}{4.67 }

\begin{document}

\begin{frontmatter}



\title{Establishing Technical Debt Management -- \\ A Five-Step Workshop Approach and an Action Research Study}


\author[mainaddress]{Marion Wiese\corref{mycorrespondingauthor}}
\cortext[mycorrespondingauthor]{Corresponding author}
\ead{marion.wiese@uni-hamburg.de}

\address[mainaddress]{University of Hamburg, Department of Informatics, Hamburg, Germany}
% \affiliation[inst1]{organization={Universität Hamburg, Department of Informatics},%Department and Organization
%             %addressline={Address One}, 
%             city={Hamburg},
%             %postcode={00000}, 
%             %state={State One},
%             country={Germany}}

\author[mainaddress]{Kamila Serwa}
\ead{kamila.serwa@uni-hamburg.de}

\author[mainaddress]{Anastasia Besier}
\ead{maria.besier@uni-hamburg.de}

\author[secondaddress]{Ariane S. Marion-Jetten}
\ead{ariane.marion-jetten@uni-hamburg.de}

\address[secondaddress]{University of Hamburg, Department of Health Science, Hamburg, Germany}

\author[mainaddress]{Eva Bittner}
\ead{eva.bittner@uni-hamburg.de}

%\address[scdaddress]{Warsaw University of Technology, Institute of Control and Computation Engineering, ul. Nowowiejska 15/19, 00-665 Warsaw, Poland}
%\address[gujsecondaryaddress]{https://www.guj.de/}

% \affiliation[inst2]{organization={Warsaw University of Technology, Institute of Control and Computation Engineering},%Department and Organization
%             %addressline={Address Two}, 
%             city={Warsaw},
%             %postcode={22222}, 
%             %state={State Two},
%             country={Poland}}

\begin{abstract}
% Context: The importance of the research questions addressed by the review. 
% 1. A general statement introducing the broad research area of the particular topic being investigated.
\textit{Context.} Technical debt (TD) items are constructs in a software system providing short-term benefits but hindering future changes. TD management (TDM) is frequently researched but rarely adopted in practice.
% Goal: 2. An explanation of the specific problem (difficulty, obstacle, challenge) to be solved.
% 3. A review of existing or standard solutions to this problem and their limitations. 
\textit{Goal.} 
This study aimed to establish a TDM process in an IT company based on a predefined workshop concept. We analyzed which research approaches practitioners adopted for each TD activity and the TDM's long-term effect on TD awareness.
% Methods:	Data Sources, Study selection, Quality Assessment and Data extraction. 
\textit{Method.} We used action research (\ActionCycles action cycles in \StudyMonths months) with an IT team that creates IT solutions for signal processing. To examine TD awareness, we (1) analyzed questionnaires completed during each workshop, (2) observed team meetings, (3) adopted a method from psychology for measuring  awareness in decision-making situations called TD-SAGAT, and (4) evaluated the backlog data. 
% Results: Main finding including any meta-analysis results and sensitivity analyses. 
% 4. An outline of the proposed new solution. 
\textit{Results.} 
Practitioners preferred TD repayment and prioritization based on the system's evolution and cost calculations, i.e., repayment of so-called low-hanging fruits. Reminders in the backlog items, such as checkboxes or text templates, led to a sustainable rise in TD awareness.
% Conclusions: Implications for practice and future research.
% 5. A summary of how the solution was evaluated and what the outcomes of the evaluation were.
\textit{Conclusions.} 
We showed that a workshop-based approach is feasible and leads to sustainable process changes. New ideas for TDM applicable to other IT teams emerged, e.g., using a \textit{re-submission date}, using a \textit{Talked about TD} checkbox, and using visualizations for TD prioritization.

\end{abstract}


% \begin{highlights}
% \end{highlights}


\end{frontmatter}

%\linenumbers

\section{INTRODUCTION}
\label{sec:Introduction}

%introducing TD and ATD
    In 1992, Ward Cunningham~\cite{Cunningham1992} introduced the metaphor of technical debt (TD) to explain that code rewrites might be needed if the software is implemented under time pressure.
    Currently, a single instance of TD, called TD item, can be described as a construct that is beneficial in the short term but hinders future development~\cite{Avgeriou2016a, Kruchten2019}. 
    Multiple causes and consequences of TD have been identified and categorized~\cite{Rios2020, ramavc2022prevalence, wiese_it_2023}.
    Also, various TD types, like code, architectural, test, or infrastructure debt, have been uncovered~\cite{Li2015, Ernst2021}. 
    A conceptual model has been developed and gradually improved~\cite{Avgeriou2016a, Junior2022, wiese_it_2023}.
    
%Rationale & resulting  RQs
    Research on TD management (TDM) is manifold~\cite{Wiese2022, Guo2016d, Yli-Huumo2016, Ramasubbu2019a} and typically organized around so-called TD activities, such as TD identification, TD prioritization, or TD monitoring~\cite{Li2015}.
    However, two tertiary studies on TD management (TDM) found that holistic approaches for TDM, i.e., approaches including various TD activities and various TD types, are still missing~\cite{rios2018tertiary, Junior2022}.  
    In their vision paper on TDM, Avgeriou et al. specified that TD research currently focuses on single TD types, mostly code debt, but should include various TD types~\cite{avgeriou_technical_2023}.
    Additionally, research on systematically establishing a TDM process is missing.
    %Wiese et al.~\cite{wiese_it_2023} 
    Our previous study showed that IT managers considered TDM to be essential and their duty, but they were usually unable to establish a TDM process.
    Attempts to set up a TDM process were often half-hearted, ended prematurely, and led to ``cluttered backlogs''~\cite{wiese_it_2023}.
    Our goal was to analyze existing research efforts and, together with new research findings, tailor it to the needs of practitioners. 
    Moreover, we explored issues hindering a TDM establishment in practice. 
    These issues and intentions led to our first research question (RQ):
    
\vspace{0.2cm}
    \noindent
    \textbf{RQ 1: Which TDM approaches are useful for the practitioners of our case company?}    
    \begin{itemize} [itemsep=0pt]
            \item [] \textit{RQ 1.1.: Which TDM approaches do practitioners adopt?}
            \item [] \textit{RQ 1.2.: Which new approaches do practitioners develop?}
        \end{itemize}
        
    %Answering these questions allowed us to analyze existing research efforts and tailor further research to practitioners' needs.
    %Moreover, we explored issues hindering a TDM establishment in practice. 

    A frequently mentioned cause for TD is time pressure~\cite{Ernst2015, Martini2015a, Soliman2021a, Wiese2022}.
    Thus, establishing and maintaining a TDM process should not take much additional time.    
    The amount of time that has to be invested on top of the time for actually repaying TD  is also essential information. 
      For example, it is relevant when further IT teams negotiate a TDM process establishment with their IT management.
    This led us to our next RQ:
    
\vspace{0.2cm}
    \noindent
    \textbf{RQ 2: How much effort are the practitioners of our case company willing to invest in a TDM process?}
\vspace{0.2cm}
        % \begin{itemize} [itemsep=0pt]
        %     \item [] \textit{RQ 2.1.: How much effort is the team willing to invest in the initial establishment of a TDM process and the permanent maintenance of this TDM process?}
        %     \item [] \textit{RQ 2.2.: Which approaches are too complicated or take too much effort?}
        % \end{itemize} 
        
    %To answer this RQ, we identified the amount of time and effort a team invested in the TDM process. 

    Avgeriou et al. also identified that there is a need to study TD's ``social- tech\-ni\-cal and multi-dis\-ci\-pli\-nary aspects,'' which includes expertise from various domains such as ``human aspects of software engineering''~\cite{avgeriou_technical_2023}. 
    One of those social-techni\-cal aspects is a missing awareness of TD, a frequently mentioned cause for incurring TD~\cite{Martini2015a, Tonin2017, Rios2020,  Ramac2021, Wiese2022, wiese_it_2023}.
    %Numerous studies discuss the importance of TD awareness (TDA), e.g.,~\cite{}.
    In some studies, TD awareness is mainly mentioned as a necessity for TD management \cite{Holvitie2018}. 
    Further, multiple research studies report that establishing a TDM process led to a raised TD awareness but fall short in proving this claim, e.g.,~\cite{Russo_2022}.
    Other studies present a simplified perspective on what TD awareness entails, e.g., that TD awareness is simply the knowledge of the own system's TD items, e.g.,~\cite{Besker2017c, Ocker2019}.
    Some studies assess TD awareness, but they usually employ surveys asking the participant whether they are aware of TD in a one-time format lacking longitudinal observations, e.g.~\cite{Dong2019, Wiese2022}.
    However, to avoid and regularly repay TD, participants must be aware of TD in their daily routine, e.g., when making decisions on designs or architectures.
    Our goal was to analyze whether the establishment of a TDM process improved this kind of TD awareness, leading to RQ3: 
\vspace{0.2cm}
    \noindent
    \textbf{RQ 3: Is the team members' awareness of TD in decision-making situations improved sustainably by the TDM process's establishment? }
\vspace{0.2cm}
        % \begin{itemize} [itemsep=0pt]
        %     \item [] \textit{RQ 3.1.: Can TD be avoided by making more rational decisions?}
        %     \item [] \textit{RQ 3.2.: Is TD repaid when reasonable by making more rational decisions?} 
        % \end{itemize}
    
   % By answering this question, we aimed to identify the cognitive effects, i.e., socio-technical effects the establishment of a TDM process might have. 
    %Particularly, we researched the effect such a process and its establishment might have had on TD-related decision-making, i.e., whether to incur or repay a certain TD item.
    %After introducing a new topic, a raised awareness of this topic can be expected.
    %However, we analyzed the development of awareness in the long term to observe the sustainability of this change.
    % TDM Action Researches beschreiben und abgrenzen
    To analyze these three RQs, the action research method~\cite{staron_action_2020} provides a suitable research approach, combining an academic approach with practical insights.
    Five action research studies established TDM processes in companies~\cite{Oliveira2015, yli-huumo_developing_2016, borup_deliberative_2021, detofeno_technical_2021, detofeno_priortd_2022}.
    However, they each focused on a few TD activities and did not provide holistic approaches to establishing a TDM process that not only handled TD but also prevented it.
    Moreover, these studies rarely evaluated the long-term effects on the team's TD awareness during their decision-making regarding TD.

%Goal
    Our study aimed to establish a holistic TDM process and analyze the results in a long-term action research study.
    For this purpose, we created a workshop concept comprising the presentation of various TD research results to practitioners, structured by TD activities.
    Then, we let the practitioners decide which results and suggestions they wanted to adopt or adjust to their needs.
    This self-determined process aligns best with  the agile mindset of a self-organized team rather than providing a pre-defined TDM process.
    We employed the agile change management method~\cite{Fuhring2021a} from the economics domain because we identified that establishing a TDM process emulates an organizational change process.
    From this method, we adopted ideas such as conducting workshops to change a corporate mindset, employing agile change management sprints to support self-management, or diagnosing the current situation to identify the needed changes.
    %Additionally, we allowed the practitioners to specify further approaches.
    In addition, this approach fits well with the action research method, as both are iterative approaches, and the ``evaluation,'' ''learning,'' ''diagnosing,'' and ''action planning'' phases of an action research cycle can be represented within workshops of an agile change management.
    In this setting, the practitioners were also able to specify their own ideas for further approaches.
    They could also identify the effort each approach takes and whether they were willing to invest this effort to gain the resulting outcome.
    Such an agile workshop approach is highly flexible and takes the specifics of each team and company into account. 
    Accordingly, we assume that it is less context-specific than other approaches and can be adopted in multiple settings. 
    %To evaluate the long-term effects of this TDM establishment and to answer RQ3, we analyzed the team's TD awareness in decision-making situations. 
    %For this, we adopted a definition and measurement method from psychology that determines awareness in decision-making situations called \textit{Situation Awareness}.
    %Besides, we monitored the effort the participants had to invest in establishing and maintaining the TDM process.  
    
%Method
    During the workshop discussions, we evaluated which approaches developed or identified by academia were feasible in this practical environment and which research results were not applicable.
    We used questionnaires, interviews, and focus group discussions on the TDM topics during \ActionCycles action cycles.
    Moreover, we assessed the team's backlog and observed discussions for \StudyMonths months in \MeetingCnt meetings comprising \MeetingHrs meeting hours.
    Finally, we used the expertise of a psychology researcher to adopt a measurement method for \textit{Situation Awareness}, i.e., to determine TD awareness in decision-making situations. 
    This collaboration resulted in a new measurement method we call TD-SAGAT (Technical Debt's Situation Awareness Global Assessment Technique).
    We concluded the study when the team members were content with the resulting process and determined that the process did not need any further significant changes. % ToDo (number of meetings, number of meeting hours).

    By answering the research questions, we contributed insights into the long-term effects the establishment of a TDM process had in a practical context and which research ap\-proach\-es practitioners have chosen.
    Additionally, we identified new approaches supplementing a TDM process, i.e.,
    
    \begin{enumerate} [itemsep=0pt]
        \item using a \textit{re-submission date} for backlog items to recall them regularly
        \item adding a \textit{Talked about TD} checkbox to all backlog items to support TD prevention 
        \item using an \textit{educated guess} based on TD item attributes 
        \item calculating a \textit{return on investment} to optimize the \textit{educated guess} % but not as 
        \item using visualizations to support the prioritization of TD items, e.g., identifying \textit{low-hanging fruits}
       % \item 5 ??? Vervollständigen beim schreiben %ToDo
    \end{enumerate}

% study outline
\subsection{Outline}
    \Cref{sec:Background} introduces the background information that forms the basis for understanding this study and the workshop presentations. 
    We compare our study with related work in \Cref{sec:RelatedWork} % to explain the addressed research gap.
    and demonstrate our research methodology in \Cref{sec:Method}.
    To disclose the workshop design for replication, we explain the workshops in~\Cref{sec:workshops}.
    In~\Cref{sec:actioncycles}, we outline the progress of each action cycle in terms of actions taken and learnings.
    We present the research results in~\Cref{sec:Results}, which comprises the action research results to answer RQ 1 and RQ2 in~\Cref{sec:Results-RQ1} and \ref{sec:Results-RQ2} respectively, and the evaluation of the team's TD awareness to answer RQ3 in~\Cref{sec:Results-RQ3}.
    Afterward, we discuss the results in \Cref{sec:Discussion}, including a discussion of the TDM establishment with actionable recommendations for practitioners in~\Cref{sec:Discussion-RQ1}, a summary of the TD awareness effects in \Cref{sec:Discussion-RQ3} and a critical review of the study design in~\Cref{sec:discussion-studydesign}.
    We address the threats to the validity of our study in \Cref{sec:ThreatsToValidity}, and conclude our paper 
    in \Cref{sec:Conclusion}. 
    The additional material is available at Zenodo~\cite{AdditionalMaterial}. 
    It comprises all workshop slides and material, photographs of flip charts, evaluations (incl. questionnaire questions), and an introduction to the TD-SAGAT method. %and summarize the contributions for practitioners and researchers.

\section{BACKGROUND}
\label{sec:Background}

        This section introduces the TDM topics we similarly introduced to the practitioners during the action cycle workshops.

        \subsection{Technical Debt Metaphor and Definition} 
        \label{sec:Background_definition}
            Cunningham introduced TD as a metaphor for financial debt~\cite{Cunningham1992}.
            He described that prioritizing time over quality in software development can lead to the need for code rewrites. 
            The sub-optimal implementation is interpreted as \textit{debt}, the resulting problems as \textit{interest} rates, the refactoring as \textit{repayment}, and the refactoring cost as \textit{principal}.
            Later works described that the interest of TD %is not linear, as is the case with financial interest, but 
            is subject to a probability of occurrence, i.e., there is a certain \textit{interest probability}~\cite{Schmid2013}. 
            For example, when poorly written code needs extra time to be understood, the overall interest also depends on how often this code fragment must be read or how probable it is that it must be read. %, i.e., how often there is a need to change this code fragment (interest probability).

            According to the widely accepted ``Dagstuhl definition,'' TD items are constructs that provide short-term benefits but hinder the long-term development and maintenance of the software and mainly refer to internal software qualities~\cite{Avgeriou2016a}.
            A single instance of TD, i.e., one construct, is referred to as a \textit{TD item}~\cite{Kruchten2019}.
            Kruchten et al. introduced the ``colors in a backlog'' determining the category of a backlog item, i.e., ticket~\cite{Kruchten2012a}. % as functional, architectural, bug, or TD.
            They differentiate between whether objects are visible or invisible to customers and whether they increase a system's value or reduce its negative effects. 
            Functional and architectural items increase the system's value, but the results of architectural items are invisible to the customer.
            Bugs and TD items shall reduce negative effects; however, TD is invisible to the customer.
            Since it is still difficult to determine whether a backlog item is TD, Avgeriou et al. simplified the decision. They specified all constructs as TD that generate interest when being changed~\cite{avgeriou_technical_2023}.          

        \subsection{Technical Debt Types}
        \label{sec:Background_types}
            Various works addressed the challenge of describing different TD types based on the specific part of the software or aspect of its development that the technical debt item affected~\cite{Li2015, Ernst2021}.
            In their tertiary studies, Rios et al.~\cite{rios2018tertiary} found 15 different TD types, those being design, code, architectural, test, documentation, defect, infrastructure, requirements, people, build, process, automation test, usability, service, and versioning TD. 
            Moreover, new types like security debt~\cite{ahmadjee_assessing_2021} or update debt~\cite{wiese_it_2023} are still emerging.
            Sometimes, TD type names are not consistent across different works, e.g., Rios et al.~\cite {rios2018tertiary} mention ``people debt,'' which is similar to ``management and social debt'' defined by Ernst et al.~\cite{Ernst2021}.
            %  Architectural TD was identified as being one of the most dangerous types of TD~\cite{Martini2015a}. Nowadays, many studies focus on this TD type~\cite{Martini2015a, Besker2018b, Verdecchia2021}. 

        \subsection{Technical Debt Causes and Consequences}
        \label{sec:Background_CausesConsequences}      
        
            McConnell and Fowler distinguish between TD caused intentionally or unintentionally, i.e., unconsciously~\cite{McConnell2008a, Fowler2009}. 
            Intentional TD items result from a conscious decision, e.g., accepting a workaround to reach a deadline.
            Unintentional TD items result from missteps like poor design at the code level or architectural decisions that, over time, turn out to have been unfavorable.
            
            Regarding the consequences, Schmid presents the distinction between potential and effective TD \cite{Schmid2013}. 
            Effective TD is TD that is currently creating interest, e.g., poorly readable code that must be frequently changed. %, and with every change, the interest must be paid.
            Potential TD is TD that is rarely or never changed, e.g., poorly readable code in a legacy system that will be rewritten soon.
            
            The main causes and consequences of TD have been extensively studied and categorized by the InsighTD project~\cite{Rios2020, Ramac2021}.
            The InsighTD project describes eight categories of causes (e.g., ``lack of knowledge,'' ``external factors,'' ``development issues,'' or ``planning and management'') and six categories of consequences (e.g., ``external quality issues,'' ``internal quality issues,'' ``organizational,'' or ``planning and management'')~\cite{Ramac2021}.
            Our study on the IT manager's perspective on TD found that cause and consequence categories are often identical and form vicious cycles~\cite{wiese_it_2023}. 
            We found the following categories for causes and consequences equally: ``time and budget,'' ``business,'' ``management,'' ``technologies,'' and ``personnel/human resources.''
            % \begin{itemize}
            %     \item time and budget, e.g., taking a workaround to reach a given project deadline (cause) or not being able to keep a given deadline due to too many TD (consequence).
            %     \item business, e.g., changes in business strategies or changing legal requirements (cause), or customers leaving the company because new features are not delivered in time (consequence).
            %     \item management, e.g., difficult communication between different IT teams (cause), or not being able to keep project deadlines (consequence).
            %     \item technologies, e.g., still using outdated technologies / not updating outdated libraries (cause), or performance or deployment issues (consequence).
            %     \item personnel/human resources, e.g., too many inexperienced developers (cause), or experienced developers leaving the company due to too many TD items, e.g., old technologies (consequence).
            % \end{itemize}
            Besides, we identified two more categories for causes, i.e., ``de\-ci\-sion-making '' and ``awareness.''
            % \begin{itemize}
            %     \item decision-making, e.g., stakeholders making the decision to incur TD, e.g., to forego testing.
            %     \item awareness, e.g., stakeholders aiming for the fastest implementation because they are not aware that there might be consequences. 
            % \end{itemize}
            Moreover, causes and consequences often form causal chains, e.g., time pressure leading to bad decision-making (cause) or not keeping a given deadline, leading to customers leaving the company (consequence).
            % Causes associated with management dominate many lists of TD causes. 
            % Rios et al.~\cite{Rios2019d} found that in 34\% of the TD occurrences, TD is incurred due to a cause from the ``Planning and management'' category.
            % Ramač et al.~\cite{Ramac2021} conducted an extensive survey that found 725 ``Planning and Management'' cause occurrences, with the second most populated category being ``Development Issues'' (473 occurrences).
            % Vogel-Heuser et al.~\cite{VOGELHEUSER2021110809} identified that management decisions initiated 32\% of their researched TD items.
            
        \subsection{Technical Debt Management and Activities}
        \label{sec:Background_Activities}
            TDM has been intensively researched over the last decade, which resulted in two tertiary studies on this topic~\cite{Junior2022, rios2018tertiary} and two comprehensive management guidelines~\cite{Kruchten2019, Ernst2021} intended for software practitioners.
            Most TDM studies describe their approaches based on the so-called \textit{TD activities}, which were first described in a mapping study by Li et al.~\cite{Li2015}.

            \paragraph {TD identification} This comprises methods to detect TD such as (1)~static code analysis, e.g., SonarCloud~\cite{sonar_better_2025}, (2)~analyzing self-admitted TD, e.g., in ``todo'' tags in code comments~\cite{li_identifying_2022}, or (3)~recording TD in a TD backlog~\cite{Kruchten2019, Junior2022}.
            While automatic TD detection is repeatedly researched, automation tools often have limitations regarding the TD types, e.g., only identifying code debt but not architecture debt~\cite{avgeriou_technical_2023}. 
            Thus, manual TD identification is still frequently used, e.g.~\cite{Oliveira2015}. 

            \paragraph {TD documentation} This is the term for representing TD items and their attributes in lists, e.g., as a supplementary item type in a backlog of work items. 
            In backlog tools, work items are commonly referred to as tickets, i.e., one TD item is represented by one TD ticket.
            TD backlogs and lists lead to more visibility, make TD explicit, raise the perception of TD, and by this, raise the overall awareness for TD~\cite{Kruchten2012a, Eliasson2015, Guo2016a, Besker2019c}.
            %The backlog resulting from the adoption of this framework makes TD visible and explicit which enhances the overall awareness for TD \cite{Kruchten2012a, Eliasson2015, Guo2016a, Besker2019c}. %Treude2010,  %Overview=>Awareness
            
            \paragraph {TD measurement} This activity tries to quantify TD, e.g., by estimating costs for interest and principal or a cost-benefit ratio~\cite{Tom2013b, McConnell2008a, Ramasubbu2019a}.
            %Summarizing these costs might even lead to estimating a system's overall TD.
            In their recent systematic mapping study on quantification approaches, Perera et al. identified that most research focuses on the cost aspects, and benefit is rarely analyzed~\cite{perera_systematic_2024}. 
            Moreover, they found that research focuses on the cost retrospectively, i.e., on evaluating TD items after they are incurred. 
            Research rarely focuses on cost estimations to guide decisions on incurring TD~\cite{perera_systematic_2024}.
             
            \paragraph {TD prioritization} This activity ranks TD items depending on their importance. 
            %There are two variants of prioritization: 
            TD can be prioritized against functional requirements or each other~\cite{lenarduzzi_systematic_2020}.
            Prioritizing TD can be a one-time activity, but it is often an iterative activity, as a TD item's priority might depend on the current situation~\cite{lenarduzzi_systematic_2020}. 
            Multiple factors influencing a TD item's priority have been uncovered by academia, some of which we present in~\Cref{tab:approaches}. %, e.g., prioritizing dependent on the systems evolution~\cite{Schmid2013}, by business value of the affected system component~\cite{ReboucasDeAlmeida2018c}, by costs as calculated by TD measurement~\cite{Tom2013b}, or by the relevance of the affected quality attributes~\cite{rachow_architecture_2022}. 
            However, in their systematic literature review on TD prioritization, Lenarduzzi et al. uncovered 53 prioritization factors organized into the categories: business, customer, evolution, maintenance, system qualities, quality debt, productivity, project factors, social factors, other factors, and ``not specified''~\cite{lenarduzzi_systematic_2020}.

            \paragraph {TD prevention} This activity aims to avoid TD incurrence.  
            The InsighTD project identified various TD prevention strategies used in the industry~\cite{Rios2020, Freire2020a, Perez2021a}.
            We listed these and further prevention strategies from academia in~\Cref{tab:approaches}. % ``well-defined scope/requirements,'' ``code eval\-ua\-tion/\-stan\-dardization,'' ``following \-well-\-defined project processes,'' ``well-defined architecture/design,'' ``TD awareness/management,'' ``a\-doption of good practices,'' ``improving tests/coverage,'' ``good communication,'' and ``reviews'' \cite{Rios2020, Freire2020, Freire2020a, Perez2021a}.
            %In their comparison of encouraging, rewarding, forcing, and penalizing strategies for TD reduction, 
            %Besker et al. found some companies were using educational sessions on avoiding TD \cite{Besker2020d}.
            %Vogel-Heuser et al. advised using guidelines to prevent TD in embedded systems\cite{Vogel-Heuser2021}.
            %Wiese et al. present a framework to avoid TD by incorporating a TDM approach into project management~\cite{Wiese2022}.
            % Borowa et al. presented debiasing techniques to prevent cognitive biases that may lead to TD incurrence, such as ``challenging all decisions,'' ``explicitly gathering information about alternatives,'' ``documenting and passing on knowledge,'' ``explicitly registering all accounts of TD'' and ``defining and recording who is responsible'' \cite{Borowa2021}.
            % In their debiasing workshop tested on IT students, they focus on three techniques, i.e., explicitly searching for at least one solution alternative, at least one drawback per alternative, and at least one risk per alternative. Applying these techniques, they showed that the ``optimism bias,'' ``anchoring,'' and ``confirmation bias'' could be counteracted~\cite{borowa_debiasing_2022}. 

            \paragraph {TD monitoring \& visualization} This is concerned with observing a system's TD development over time~\cite{Li2015}.
            Visualizations of TD development based on lists of TD items can support monitoring TD.
            Various tools for TD visualization have been developed:
            DebtFlag~\cite{holvitie_debtflag_2013}, % acquires and visualizes self-admitted TD. 
            MultiDimEr~\cite{silva_multidimer_2022}, % analyzes and visualizes bug reports to identify emerging TD at an early stage. 
            VisminerTD~\cite{Mendes2019}, % combines TDM activities such as identifying and monitoring TD to visualize the evolution of TD items. 
            MIND~\cite{falessi_towards_2015}, and % quantifies and visualizes the interest associated with TD.
            TEDMA~\cite{fernandez-sanchez_open_2017}. % supports TDM by capturing static analysis metrics throughout software evolution.
            They visualize self-admitted TD, bug reports, static analysis metrics, or collections of TD items of a proprietary tool.

            % \begin{itemize}
            %     \item DebtFlag~\cite{holvitie_debtflag_2013} acquires and visualizes self-admitted TD.
            %     \item MultiDimEr~\cite{silva_multidimer_2022} analyzes and visualizes bug reports to identify emerging TD at an early stage.
            %     \item VisminerTD~\cite{Mendes2019} combines TDM activities such as identifying and monitoring TD to visualize the evolution of TD items.
            %     \item MIND~\cite{falessi_towards_2015} quantifies and visualizes the interest associated with TD.
            %     \item TEDMA~\cite{fernandez-sanchez_open_2017} supports TDM by capturing static analysis metrics over the course of software evolution.
            % \end{itemize}
             
            \paragraph {TD repayment} This is the term for resolving the TD item. 
            Busch\-mann et al. introduced three repayment categories: ignore, refactor, and rewrite~\cite{Buschmann2011}. 
            Various repayment methods are described in research as presented in~\Cref{tab:approaches}. 
            Ignoring includes only repaying TD if it impedes a functional requirement~\cite{FREIRE2023} or paying the interest~\cite{Buschmann2011}. 
            Rewriting involves waiting for the discontinuation or migration of a software system~\cite{Schmid2013, Schmid2013a}.
            Refactoring includes repaying TD (1)~if the cost-analysis proves the positive effect~\cite{McConnell2008a, Tom2013b}, (2)~by assigning a repayment quota~\cite{McConnell2008a, Wiese2022}, %, e.g., 10\% of each sprint, assigning every fifth sprint to TD repayment~\cite{McConnell2008a, Wiese2022}.
            %Additionally, refactoring comprises repaying TD 
            (3)~before a project if TD might hinder the development~\cite{Schmid2013}, or (4)~after a project if TD was incurred due to its project deadline~\cite{Wiese2022}.

    \section{RELATED WORK}
    \label{sec:RelatedWork}

        Our study aimed to establish a TDM process utilizing action research. %in an agile industry environment. 
        We analyzed not only the resulting process but also the change in practitioners' TD awareness.
        Hence, our related work section focuses on action research studies on TDM processes and the topic of TD awareness.

        \subsection{Establishing a Technical Debt Management process}
            
            Two recent tertiary studies on TDM by Rios et al.~\cite{rios2018tertiary}, and by Junior et al.~\cite{Junior2022} give an overview of the TDM research. 
            They both found that TDM is viewed as a set of TD activities and supporting strategies and tools, but they did not document any research on the systematic establishment of a TDM process.
            Both agreed that a holistic approach to TDM, including various TD activities and types, is missing. %, which makes it difficult for practitioners to profit from the research results. 
            Similarly, the vision paper by Avgeriou et al. identified that TD research needs to focus on multiple TD types instead of only focusing on code debt and should include more industrial datasets~\cite{avgeriou_technical_2023}.
            This coincided with our finding that IT managers lack suitable guidelines for establishing TDM~\cite{wiese_it_2023}.
            A general guideline for establishing TDM was presented in the book of Kruchten et al.~\cite{Kruchten2019}, mainly in the ``What can you do today?'' sections.
            Yet, the feasibility of this guideline has not been evaluated. % and may need optimization.

        \subsection{Action Research Studies on Technical Debt Management}
        \label{sec:RelatedWork_ActionResearchStudies}

            Five action research studies established TDM processes in companies~\cite{Oliveira2015, yli-huumo_developing_2016, borup_deliberative_2021, detofeno_technical_2021, detofeno_priortd_2022}.

            Detofeno et al. performed two action research studies to implement a TDM process in a Brazilian company~\cite{detofeno_technical_2021, detofeno_priortd_2022}. 
            The first study (two cycles in two years) established a TDM guild, i.e., a community of practice, on the TD topic~\cite{detofeno_technical_2021}.
            The guild contributed to the identification, monitoring, prevention, prioritization, and repayment of TD. 
            However, the actions were limited to code debt since they focused on static code analysis of PHP code.
            %The two action cycles over two years were connected to the company’s annual management cycle.
            In the second study (two cycles in two years and eight months), they developed a detailed TD prioritization mechanism but still limited their work to code debt~\cite{detofeno_priortd_2022}. 
            %This study comprised two action cycles in two years and eight months.

            Oliveira et al.'s study~\cite{Oliveira2015} (three action cycles in five mon\-ths) intended to confirm the feasibility of the TDM framework presented by Guo et al.~\cite{seaman_2011}. 
            %They performed three action cycles in five months.
            Their approach focused on identifying, documenting, and prioritizing existing TD items.
            Similarly, Yli-Huumo et al.~\cite{yli-huumo_developing_2016} focused their work on identification, documentation, and prioritization.
            They concentrated on uncovering the risks and benefits related to these TD activities.
            They performed five subsequent research steps: problem identification, TD identification, TD documentation, TD evaluation (i.e., measurement), and TD prioritization.
            However, they both missed the activities of TD  prevention and monitoring~\cite{Li2015} as well as the multiple ideas on TD repayment~\cite{Li2015, Buschmann2011}.
            
            Borup et al.'s study~\cite{borup_deliberative_2021} (two interventions in ten months) focused on TD deliberation, i.e., communicating about, weighing, and reflecting on TD.
            Therefore, they might have filled some of the gaps in the previous studies, particularly in TD prevention, but they did not focus on TD Management, e.g., documenting or prioritizing. 
            %They performed two interventions in ten months.
        
            These action research studies show shortcomings that we address in our study.
            First, our study focused on all major activities and did not differentiate TD types other than tagging backlog tickets with the respective TD type. 
            Second, we performed long-term research (\ActionCycles action cycles in \StudyMonths months) to evaluate whether processes are established and the team's TD awareness is improved sustainably. 
            %We evaluated the data of \StudyMonths months during \ActionCycles action cycles.
            Only the works of Detofeno et al. performed studies longer than ours; however, they had fewer action cycles and, therefore, fewer possibilities for intervention.
            Third, we observed team discussions as valuable supplemental data sour\-ces, which was not done by any other study.
            Finally, we focused on developing a TDM process together with practitioners, reinforcing an agile mindset rather than simply evaluating a pre-defined approach developed by researchers. 
            This is specifically relevant in agile teams to create acceptance for the TDM process among the practitioners.

        \subsection{Technical Debt Awareness}
        \label{sec:RelatedWork_Awareness}
            To analyze the current state of research on TD awareness and identify potential measures, we conducted a literature review and explored five established digital libraries: IEEE Xplore, ACM Digital Library, SpringerLink, ScienceDirect, and Wiley.
            %searching for the terms ``technical debt'' AND ``aware*'' in ACM Digital Library, IEEE Explore, Science Direct, and Springer
            We used \textbf{(``technical debt'') AND (aware OR awareness)} as a search string and searched titles, abstracts, and keywords to identify literature about TD Awareness.
            We did not do full-text searches as the terms ``aware'' and ``awareness'' are frequently used in everyday language. %in contexts unrelated to TD. 
            We may have missed some papers, but we aimed to gain an overview of the academic discussions on TD Awareness and assumed that we found all the literature focusing on TD Awareness. 
            After removing ten duplicates, we studied 59 unique references (see additional material~\cite{AdditionalMaterial}).
            
            We found out that TD awareness is an ambiguously used term. 
            Some studies interpreted TD awareness as knowing what TD is~\cite{Apa2020, Besker2017c, Holvitie2018, Baars2019a, zampetti_self-admitted_2021}; other studies used it as knowing the TD items in their own system~\cite{Besker2017c, Liu_2020}, and some interpreted TD awareness as not unintentionally incurring TD~\cite{Rocha2017b, Wiese2022}.  
            Many papers mentioned TD awareness as a prerequisite for TDM~\cite{Besker2017c, Kruchten2019}.
            Some papers have identified visualization~\cite{Eliasson2015}, gamification~\cite{Crespo2022}, and TDM~\cite{BALDASSARRE2020106377, Wiese2022} as feasible methods for raising TD awareness.
            Tonin et al. conducted a study specifically focusing on the effect of elevated TD awareness by regularly discussing this topic during IT students' development projects~\cite{Tonin2017}.

            % The need to study TD's ``social-technical and multi-dis\-ci\-pli\-nary aspects,'' which includes TD awareness, was an outcome of Avgeriou et al.'s vision paper~\cite{avgeriou_technical_2023}. 
            Unfortunately, few papers explicitly defined what they meant by the term TD awareness or described their measurements.
            We define TD awareness for decision-making situations in \Cref{sec:method-tdsagat}.
            %As of today, there is no common understanding of how to define and measure TD awareness.
            % Many papers interpreted TD awareness as knowing the definition of TD~\cite{Apa2020, Besker2017c, Holvitie2018, Baars2019a, zampetti_self-admitted_2021}. 
            % For fewer papers, TD awareness comprised knowing the TD items in a specific system~\cite{Besker2017c, Liu_2020} or avoiding unintentional TD incurrence~\cite{Rocha2017b, Wiese2022}. 
            To the best of the authors' knowledge, TD awareness in decision-making situations such as team meetings has not yet been researched.

            % Similarly, Avgeriou et al.'s vision paper~\cite{avgeriou_technical_2023} formulated the need for Data-driven research, i.e., using measurements to create data sets.
            To evaluate TD awareness, researchers commonly use questionnaires~\cite{berenguer_technical_2021, Wiese2022}, which have the disadvantage of being subjective measurements.
            Others deduced TD awareness from indicators derived by static analysis tools~\cite{Crespo2022}, which only indirectly point to TD awareness.
            Kruchten et al.~\cite{Kruchten2019} introduced five levels of TD awareness (from not knowing the term TD to having control over their TD). 
            However, these levels indicate the maturity of the TDM process, and likewise, can only indirectly point to a team's TD awareness.

            In our study, we addressed these shortcomings by supplementing meeting observations, which, to the best of the authors' knowledge, have not been used previously.
            Additionally, we identified potential measuring methods from the field of Psychology.
            %investigated the Psychology domain to identify potential measuring methods and found 
            We found conceptual similarities to TD among three awareness concepts: contingency awareness, bounded awareness, and situation awareness.
            We finally adopted a measuring method for situation awareness, which we further explain in~\Cref{sec:method-tdsagat}. 
            %In our study, we addressed these shortcomings by supplementing meeting observations of team meetings and TD-SAGAT, a me\-thod adopted from psychology, as measurements for TD awareness.
            This allowed us to provide more objective measurements and quantitative data in addition to the qualitative data.

\section{METHOD}
\label{sec:Method}

%Justification of action research
Our study aimed to create a TDM process in cooperation with our industry partner based on previous academic findings.
From a business perspective, this resembled a small change management project as work processes were changed~\cite{Fuhring2021a}.
For such a change process, it was essential to include the employees who have to work with these processes~\cite{Fuhring2021a}.
Moreover, the agile work concepts adopted in most IT companies heavily rely on a team's self-organization and do not promote process changes that are enforced by higher hierarchical levels~\cite{meyer_agile_2014,maximini_scrum_2018}.

Action research~\cite{staron_action_2020}, an approach in which both the practitioners from a company and the research group work together to understand a problem and develop a solution, seemed especially suitable for the purpose.
Therefore, we chose this method to answer RQ 1\&2, determining the adopted approaches and the time required to establish the TDM process. 
We collected data from the workshops and subsequent discussions in the action cycles, workshop questionnaires, and, on rare occasions, interviews. Additionally, we analyzed the backlog development.

Along with the action research, we analyzed the effect the TDM establishment had on the participants' awareness of TD to answer RQ 3.
For this purpose, we collected three kinds of data throughout the whole action research.
The participants completed questionnaires during the workshops, we ob\-served the team during their meetings, and we adopted an approach from psychology to assess the participants' TD situation awareness during meetings, which we call TD-SAGAT.
%Additionally, we analyzed the development of the team's backlog to provide the team with evaluations of their work and to observe the effect the TDM establishment has on the number of TD tickets and their quality.

In the following sections, we introduce our industry partner, outline the basics of the action research, explain the analysis of the team's backlog, and introduce the interviews we used.
Subsequently, we introduce the three data-gathering methods used to analyze the participants' awareness of TD.
The questions used for the action research and awareness analyses were included in the same questionnaires and are, thus, explained together in~\Cref{sec:method-surveys}. 

\subsection{Industry partner}
\label{sec:company}

    % What were we looking for
    % How did we find them
    % What did we get
    For this research, we were cooperating with a German company with over 2,000 employees that had been in operation for over 70 years.
    The company produces signal-processing products, which means the software development of this company is located in the embedded, i.e., the mechatronic domain. 
    The company chose not to disclose its name, so the information in this article and the additional material were anonymized.

    \subsubsection{Team}
    
    For our study, we worked particularly with one IT team that was developing and maintaining a central framework used by other IT teams in the same company. 
    Informally, the team was divided into two subteams, the \textit{algorithm subteam} focusing on the signal processing algorithms and the \textit{framework subteam} on the supporting framework.
    While the algorithm subteam did not further differentiate different roles, the framework subteam had a team architect, a quality assurance specialist, and one specialized developer with an engineering background to bridge the gap between the disciplines. 
    As the algorithm subteam did not suffer from much TD, we reported on their insights during the action cycles (RQ1, \Cref{sec:Results-RQ1}) but focused our awareness evaluation on the framework subteam (RQ3, \Cref{sec:Results-RQ3}).
    %An overview of all team members can be seen in~\Cref{tab:participants}.
    \Cref{tab:participants} shows an overview of all team members.
    
    The team was newly assembled but worked on an existing framework with lots of TD, which, therefore, were the first focus of this team. 
    During the study, this focus shifted to implementing more new functionalities.
    Further, the team continuously improved the processes it had just established. 

% 
%     As proposed by Staron~\cite{staron_action_2020}, we additionally divided the team into an \textit{action team} to execute the one-time actions (e.g., implementing a new ticket type, discussing its attributes, and selectable values of combo-boxes) and a\textit{ reference team} to monitor the soundness of the action team's work and to counteract the bias of the action team.
% 
    
   % It used an adapted Scrum process with three-week sprints ending with a retrospective.
        	\begin{table}
    	    \centering
    	    \footnotesize
        	\begin{tabular}{llccrc}%p{1cm} p{4cm} p{1,4cm} p{1,4cm}  p{1,4cm} p{1,4cm} p{1cm} p{1,1cm} p{1,1cm}} % p{2,1cm}
    			\toprule
    		Partici- & Role & Sub- & AR & Years & Gender \\ 
    		pant &  & team & team & in IT &  \\ 
    			\midrule 
    			P1 	& Team manager           & - & A   & 3-5  & f \\ %SaS
    			P2 	& Team architect         & F & A   & 6-10  & m \\ %AB
    			P3 	& Developer and engineer & F & R   & \textgreater 10 & m \\ %AH 
    			P4 	& Quality assurance      & F & A   & 0-2  & m \\ %SH
    			P5 	& Developer              & F & R   & \textgreater 10  & m \\ %TR 
    			P6 	& Developer              & A & A   & 6-10 & m \\ %ER
    			P7 	& Developer              & A & R   & 6-10 & f \\ %DF
    			P8 	& Developer              & A & R   & 0-2  & m \\ %M
     			\bottomrule
    		\end{tabular}
    		\caption{Study participants with their role, subteam assignment to Framework (F) or Algorithms (A), and Action research (AR) team assignment to action team (A) or reference team (R)}
    		\label{tab:participants}
    	\end{table}

    \subsubsection{Meetings \& Backlog}

    The team used an adapted Scrum process with three-week sprints ending with a retrospective.
    The team had daily stand-up meetings and organized its work tickets in the backlog tool Microsoft Azure DevOps~\cite{microsoft_azureDevops_2025}. %\footnote{\url{https://azure.microsoft.com/en-us/products/devops/}}.
    Their backlog was configured as an agile process based on a ``portfolio backlog'' defined by Azure DevOps~\cite{ microsoft_devops_portfolio_2023,microsoft_Azure_Backlog_2024}. %\footnote{\url{https://learn.microsoft.com/en-us/azure/devops/reference/add-portfolio-backlogs}}. %; \url{https://learn.microsoft.com/en-us/azure/devops/boards/backlogs/backlogs-overview}}. 
    The team used the following hierarchically ordered work item types for functional requirements defined by that process: initiatives, epics, features, user stories, and tasks. 
    Additionally, they employed a particular ticket type for bugs. 
    During the study, they added another specialized type for TD.
    Their tickets could have had six different statuses, i.e., their Scrum board comprised six columns typical for Scrum processes: backlog, in proposal, planned, in progress, blocked, testing, and done.

    At the beginning of our study, the team had three meetings to organize their work:
    \begin{itemize} [itemsep=-2pt]
        \item \textit{Grooming}: Recording necessary attributes for new tickets, e.g., description, effort, or acceptance criteria. After being ``groomed,'' the tickets were not given a specific status other than ``backlog.''
        \item \textit{Refinement}: Selecting and preparing the tickets for the next sprint and finalizing attribute information. After being refined, the tickets transitioned from the status ``back\-log'' to ``in proposal.''
        \item \textit{Review \& planning}: Reviewing the tickets of the past sprint and planning the tickets for the next sprint. After being ``planned'' the tickets transitioned from the status ``in proposal'' to ''planned.''
    \end{itemize}

    During our work with the team, the meeting structure has changed. 
    The two subteams decided to separate their meetings from each other and reduce the number of meetings. 
    The framework subteam decided on two meetings: A refinement meeting for analyzing tickets and recording their attributes (leading to the ticket status ``in proposal'')  and a planning meeting to review the old and organize the next sprint (leading to the ticket status ``planned''). 
    The algorithm subteam decided on one meeting combining refinement \& planning.

\subsection{Action Research}
\label{sec:method-actionresearch}

    Multiple research studies identified the adoption of TDM processes in the industry as insufficient, e.g.,~\cite{Martini2018d, wiese_it_2023}.
    Introducing a TDM process aims to change work processes, and action research is a research method that aims to create and improve work processes.  
    The research is conducted in iteratively repeating action cycles, which comprise five phases each:  
    \begin{itemize} [itemsep=0pt]
        \item Diagnosing: identifying which problem should be solved
        \item Action planning: identifying a set of actions to solve the problem
        \item Action taking: execution of the identified actions
        \item Evaluating: analyzing the actions' effects and whether they solved the problem
        \item Learning: Identifying improvements to enhance the action's efficacy in effectively addressing the problem.
    \end{itemize}
    The action research terminates when the problems are solved, which makes it crucial to define a termination goal before starting the action research.
    We decided to end our action research when the participants noted their satisfaction with the developed process in a questionnaire. 
    
\begin{figure*}%[H]
            \centering
            \includegraphics[width=0.9\textwidth]{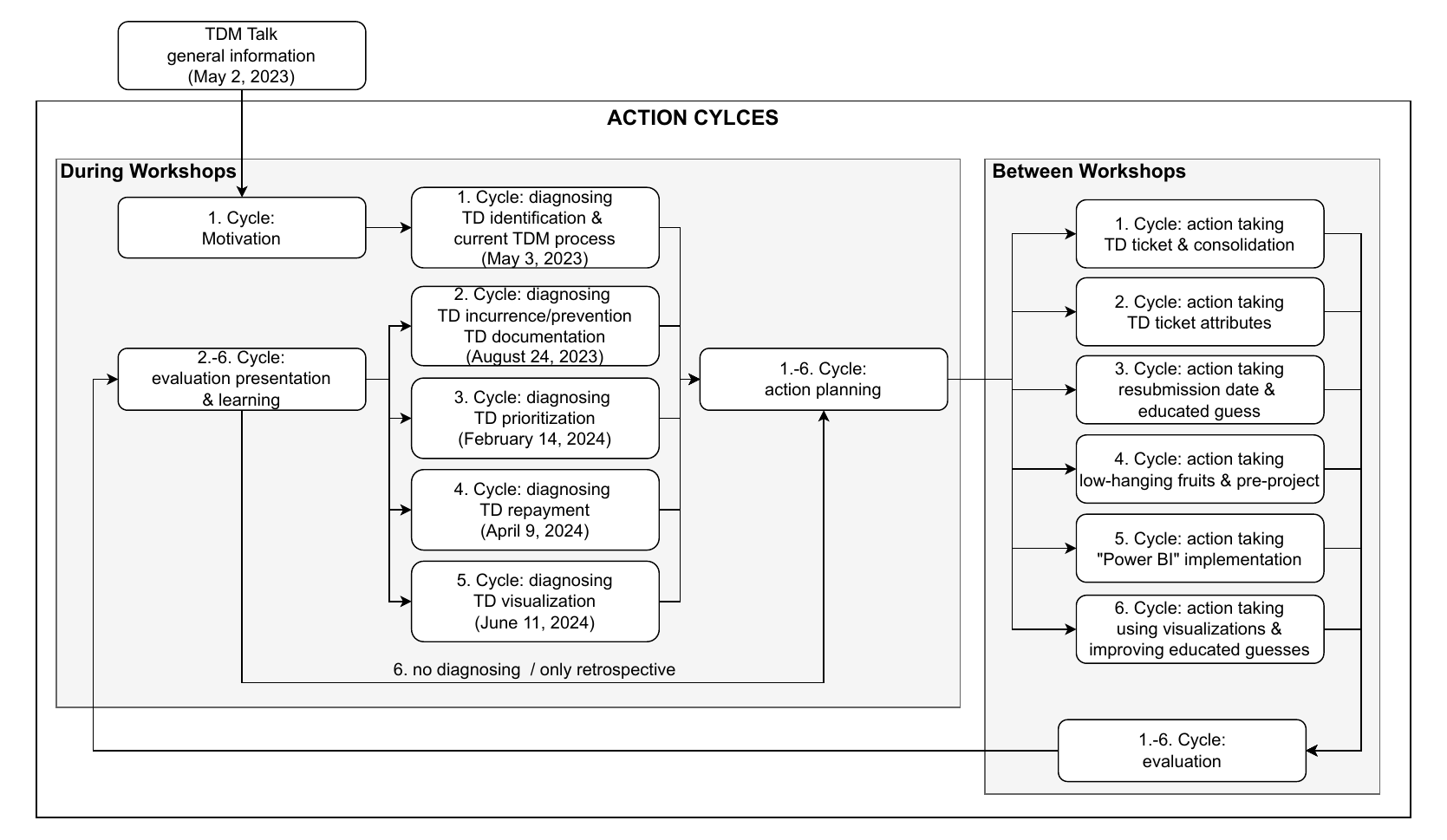}
            \caption{Study Design - iterating action cycles including five workshops on TD activities}
            \label{fig:Study_Design}
            %\vspace{-5mm}
        \end{figure*}

    Our iterative action research design, including all five workshop topics and major actions taken, is presented in \Cref{fig:Study_Design}. 
    A detailed overview of how we conducted our action cycles is presented in the following section. 

    %reflexivity, credibility, resonance, usefulness, and transferability

    Action research allows the researcher to interact with and influence the study's participants~\cite{staron_action_2020}. %, which distinguishes this method from other research methods.    
    Accordingly, action research is interpreted as a field experiment with regard to the ABC framework~\cite{Stol2018}, which means the study is conducted in a natural, pre-existing setting. 
    In this regard, it is similar to design science research. 
    Yet, in contrast to changing work processes, the primary aim of design science research is to develop an artifact that is suitable for practitioners. %, the goal of action research is to change work processes, which makes it suitable for our purpose.
    %Therefore, we found the action research method to be suitable to identify TDM approaches that are useful for practitioners by establishing a TDM process.
    The downside of field experiments is that the precision of measurement is affected by ``confounding contextual factors,'' and measurements cannot be statistically generalized~\cite{Stol2018}.
    For action research, \textit{ACM Special Interest Group on Software Engineering}'s empirical standards accept this lack of generalizability or reliability in favor of, e.g., reflexivity, credibility, resonance, usefulness, and transferability~\cite{sigsoft_acm_2025}. 
    %\footnote{\url{https://www2.sigsoft.org/EmpiricalStandards/docs/standards?standard=ActionResearch}} 
    Nonetheless, we took measurements to counteract biases.
    For example,  we followed Staron's~\cite{staron_action_2020} proposition to separate the team into an \textit{action team} to execute the actions and a\textit{ reference team} to monitor the soundness of the action team's work (see~\Cref{tab:participants}).
    We informed the team members about their roles and their roles' tasks at the beginning of each workshop. 
    %An overview of each team member's role is provided in~\Cref{tab:participants}.
    To further reduce biases, the researchers did not decide on the actions that were taken. 
    For example, researchers could hint that the current amount of meetings and their structure lead to irritations, but it is the team's task to choose if and in which way to change the meeting structure.
    %As such, the researchers are not responsible for the selection of actions but report on the actions taken.

    In the following, we present our action cycles, the accompanying backlog analysis, and the two interviews we employed during the action research. 
    We present the questionnaires used during the workshops in~\Cref{sec:method-surveys} together with the questionnaires and questions for the awareness analysis.

    \subsubsection{Action Cycles}
    \label{sec:method-cycles}
    
    We conducted our action research in action cycles, each consisting of the five phases as outlined in the previous section. An overview of the cycles is presented in~ \Cref{fig:Study_Design}.  %by Stol et al.~\cite{staron_action_2020}: diagnosing, action planning, action taking, evaluating, and lear\-ning.

    Each cycle started with a one-day workshop. 
     A more detailed agenda of the workshops and the employed workshop methods are explained in~\Cref{sec:workshops}. 
    In the morning, the participants were presented with the results of the last cycle's evaluations (results of ``evaluating''). 
    They discussed these results and developed ideas for improving the process (``learning'' and ``action planning'').    
     The first workshop started with a general introduction and motivation for TD in the morning instead of the ``evaluating'' and ``learning'' parts (see~\Cref{sec:workshops}). 
    In the afternoon, the re\-sear\-ch\-ers presented a new TD activity with its respective research results and approaches to the participants. 
    The team discussed these approaches (``diagnosing'') and decided which approach to adopt, adapt to their needs, or dismiss (``action planning'').
    Between workshops was the time for ``action taking.'' 
    Some actions were ongoing, e.g., documenting TDs when they occur, while others were one-time actions, e.g., consolidating the backlog by tagging TD tickets or reviewing the existing TD tickets to record new attributes.
     The action team performed one-time actions, including making decisions on details, e.g., the content of attribute combo boxes. 
    The results were presented to the complete team during the regular meetings or the following workshops and discussed if necessary. 
   
    The first five workshops each had a focus on one or two TD actions, these being:
    TD identification \& definition, TD prevention, measurement \&  documentation, TD prioritization, TD repayment, and TD visualization \& monitoring.
    After that, the workshops were reduced to retrospectives that focused on the ``evaluating,'' ``learning,'' and ``action planning'' parts.
    The cycles and, therefore, the study ended when all participants found the process sufficiently defined and working except for minor changes, which we analyzed with a progress questionnaire (see~\Cref{sec:method-surveys}).

    \subsubsection{TD Backlog}
    \label{sec:method-backlog}
    
    In the following, we explain how the team changed their backlog during the action research by introducing a TD ticket type and corresponding attributes.
    These changes were part of the action research, as employing a TD backlog is a significant step in establishing a TDM process. 
    The analysis of the TD backlog was part of the ``evaluating'' phase of each action cycle.
    The illustrations helped the participants gain insights into their backlog data and evaluate the usefulness of having and monitoring TD items and their attributes.
    As such, the changing backlog is primarily part of the actions (and explained in~\Cref{sec:Results-RQ1}) and only secondarily a method to analyze the action research's success. 
    Furthermore, we used the number of TD items as an indicator to support RQ3.
    %Thus, these changes are references in the 
    %Further, we will outline our approach to analyzing this data.
    %We evaluated the backlog data to present them to the practitioners during the ``evaluation'' phase of each action cycle. 
    %The purpose of this presentation was to provide the participants with insight into their backlog.
    %To support RQ3, we also analyzed opened and closed TD tickets. 
    %However, the purpose of the evaluations was not to prove the success or failure of the action research and the establishment of the TDM process but to use them to improve the associated TDM process.

        % \subsubsection{Technical Debt Backlog}
        % \label{sec:method-backlog}
        
            \paragraph{ Data Collection}

            For each workshop, we gathered the TD tickets from the backlog by exporting and analyzing them using Microsoft Excel~\cite{microsoft_excel_2025}. %footnote{\url{https://www.microsoft.com/en-us/microsoft-365/excel}}.
            In the consolidation phase, the team tagged \TDitemsTagged items as TD.
            For the last cycle, i.e., the retrospective, we used the visualizations directly from the visualization tool developed during the action research (see~\Cref{sec:Results-Cylce5}). 
            
            The TD ticket type was established after the first workshop.  
                % These newly introduced attributes were interest, interest probability, and contagiousness on a five-step scale from very low to very high, a breaking change checkbox, and a hierarchy for the affected component.
            It was extended by the following attributes after the second workshop, which we used for our evaluations:
                \begin{itemize} [itemsep=0pt]
                    \item \textit{Interest}: 
                    The five elements of the interest scale were: ``very low (less than a minute),'' ``low (several minutes),'' ``medium (one hour),'' ``high (one day),'' and ``very high (more than one day).'' 
                    %We assumed several minutes to be 15 minutes and more than one day to be two days and normalized all values to ``required minutes.''
                    \item \textit{Interest probability}: 
                    The elements of the interest probability scale were: ``very low (less than once a year),'' ``low (once a year),'' ``medium (once a month),'' ``high (once a week),'' and ``very high (once a day).'' 
                    %We normalized these values to ``times per month.''
                    \item 
                    \textit{Contagiousness}: 
                    Contagiousness, i.e., whether the principal might grow over time, was initially measured on a five-step scale from very low to very high. In the fifth cycle, the team decided to simplify this measurement to a simple ``yes/no'' check box.
                    \item \textit{Breaking change}: 
                    A simple ``yes/no'' checkbox recorded whether the implementation incurred breaking changes, i.e., effects on other components.
                    \item \textit{Component (Area)}: 
                    Components were structured in a hierarchy utilizing Azure DevOps' standard attribute ``area.''  
                    \item \textit{Effort (principal)}: 
                     This value ranged from one to five, where one to four means one to four person days, and five means more than five days.          
                    In the fifth action cycle, the team decided to denote the effort on a continuous scale in person-days. 
                    %We assumed 10 days for a given five. 
                    %We calculated the ``effort in minutes'' assuming an eight-hour workday.
                \end{itemize}

                \paragraph{ Data Analysis}
                For the first workshop, we did not evaluate the backlog items as TD had not been tagged yet.
                
                For the second and third workshops, we evaluated the number of opened and closed TD tickets per month and illustrated the number of TD versus functional and bug tickets.
                
                %The TD ticket type and its attributes were established during the third cycle. 
                %These attributes were interest, interest probability, and contagiousness on a five-step scale from very low to very high, a breaking change checkbox, and a hierarchy for the affected component.
                %Thus,  
                From the fourth workshop on, we evaluated the new attributes as follows:
                \vspace{-0.2cm}
                \begin{itemize} [itemsep=-2pt]
                    \item number of opened and closed TD tickets per month,
                    \item share of tickets per TD type,
                    \item share of tickets per estimated priority,
                    \item share of tickets per \textit{interest risk}, i.e., the interest ID (from 0 to 4) times the interest probability ID (from 0 to 4),
                    \item share of tickets per estimated contagiousness,
                    \item share of breaking changes, and
                    \item share of tickets' \textit{interest risk} versus their effort.
                \end{itemize}
                \vspace{-0.1cm}

                For the fifth workshop, we added a calculation based on the idea of a \textit{return on investment} (ROI) value to experiment with the idea of calculating a priority value. Due to the experimental setting, we made some assumptions: % using the definitions in the teams' wiki:
                \begin{itemize} [itemsep=0pt]
                \item %\textit{Interest}: 
                %     The five elements of the interest scale were, among other things, explained as: ``less than a minute'' (very low), ``several minutes'' (low), ``one hour'' (medium), ``one day'' (high), and ``more than one day'' (very high). 
                For the interest, we assumed ``several minutes'' to be 15 minutes and ``more than one day'' to be two days and normalized all values to \textit{interest in minutes}. 
                This led to the values of 1, 15, 60, 480, or 960 minutes (very low to very high) for the interest.
                \item %\textit{Interest probability}: 
                %     The elements of the interest probability scale were, among other things, explained as: ``once a year or less'' (very low), ``once a year'' (low), ``once a month'' (medium), ``once a week'' (high), and ``once a day'' (very high). 
                For interest probability, we normalized the selected values to \textit{times per month}, leading to the values of 30, 4.5, 1, 0.0027, or 0.0013 times per month (very low to very high).
                %\item \textit{Effort (principal)}: 
                %     This value ranges from one to five, where one to four means one to four days, and five means more than five days.
                \item 
                For the effort, we assumed 10 days for a given five days, as long as the team did not estimate their effort on a continuous scale. 
                After this change, we used the given effort for calculations. 
                We multiplied the effort by 60 to determine the \textit{effort in minutes} (\textit{EFFORT}).
                \end{itemize}
                
                %     We calculated the ``effort in minutes'' assuming an eight-hour workday.
                % \end{itemize}
                
                Using these values, we identified the \textit{interest in minutes per month} (\textit{INTEREST}) for each TD item by multiplying \textit{interest in minutes} and \textit{times per month}.
                We set the \textit{interest in minutes per month} against the \textit{effort in minutes} to calculate the number of months it takes to achieve an ROI (ROI in months -- \textit{ROIM}) as follows:
                 \[\text{\textit{ROIM}} \enspace \text{months}= \frac{\text{\textit{EFFORT}}  \enspace \text{min}}{\text{\textit{INTEREST}} \enspace \text{min/months}}\]
                We acknowledge the limitations of this calculation, i.e., (1)~that we made assumptions, (2)~the underlying values were all estimations, and (3)~the values on the scale were not equidistant, i.e., the distance between ``once a year'' and ``once a month'' is not the same as the distance between ``once a week'' and ``once a day.''
                %This might lead to imprecise results.
                We separated the results into the following categories so that the items were distributed evenly for the most part, and large gaps between the ordered single values also led to different categories: 
                \begin{itemize}[itemsep=-2pt]
                    \item \(\text{\textit{ROIM} } \textless \text{ 1 month} \to \text{very high, }\)
                    \item \(\text{\textit{ROIM} } \textless \text{ 2 months} \to \text{high, } \)
                    \item \(\text{\textit{ROIM} } \textless \text{ 1 year} \to \text{medium, }\)   
                    \item \(\text{\textit{ROIM} } \textless \text{ 3 years} \to \text{low, } \)
                    \item \(\text{\textit{ROIM} } \geq \text{ 3 years} \to \text{very low.}\)
                \end{itemize}
                
                We only used these values experimentally as a basis to discuss a calculated priority with the participants as part of the action cycle.
                The limitations and ways to counteract these limitations, as well as correct assumptions, were part of the discussion as presented in~\Cref{sec:Results-Cylce5}.

                For the sixth workshop, i.e., after implementing the backlog's visualization in the fifth cycle, we used the visualization directly, and no additional analysis was necessary. 

        \subsubsection{Interviews}
        \label{sec:method-interviews}
        %In two situations, we also used one informal and one semi-formal interview to support the action research.
        % However, we did not regularly use interviews.
        We used an informal interview to get initial insight into the team's members and processes. Further, we conducted semi-formal interviews to develop the visualizations. 

        % Instead, we primarily combined team discussions in the workshops and questionnaires to gather insights.
        
           \paragraph{ Data Collection} 
            At the beginning of the study, we consulted the team's IT manager for information about the current TDM status, the current working processes, including ticket types and status, and the stakeholders, including other IT managers and team members.
    
            For the visualization, we used semi-structured interviews to gather use cases for a visualization. 
            We interviewed the action team members,
            %and %the team member responsible for the quality assessment.
            %We 
            %asked for use cases, why these use cases would be valuable, and whether any specific TD ticket attributes should be visualized. 
            %Finally, we asked for non-func\-tio\-nal requirements regarding a visualization tool, e.g., security.
            % what the participants wanted to be visualized, 
            asking the following questions: %todo -> Anastasia 
            \begin{itemize} [itemsep=-2pt]
                \item Could you please name two to three use cases in which the visualization tool may be useful?
                \item Do you have ideas for the content of potential visualizations? 
                \item Why do you want to see this specific content?
                \item Which ticket attributes would be particularly useful for a visualization? Which ones would not?
                \item Are there any non-functional requirements, e.g., security, for the application regarding your team/company?
                \item Do you have any other comments on the tool or the visualizations?
            \end{itemize}

            \paragraph{ Data Analysis}
            The interviews were analyzed informally.
            The first interview was used for understanding and reporting the team's processes.
            The results of the semi-formal interviews on potential visualizations were assembled to create the mock-ups discussed during the ``diagnosing'' phase of the fifth workshop as presented in~\Cref{sec:Results-Cylce5} and~\Cref{tab:visualization}. The answers to the question regarding non-functional requirements were used to specify the visualization environment and supported the decision to use Microsoft Power BI~\cite{microsoft_powerbi_2025}.

    \subsection{Evaluation of TD Awareness}

     To answer RQ3, we continuously evaluated the team's awareness of TD in decision-making situations using three different measurement methods: Questionnaires during the workshops for a self-assessment, observations of team meetings by researchers for an external assessment, and TD-SAGAT for an immediate assessment in decision-making situations.
     For comparability, all three methods used the same ten questions presented in~\Cref{sec:SAGATquestionnairedevelopment}.
     However, additional questions were evaluated for observations and questionnaires. 
     As mentioned previously, the additional questions were related to the action research.
     An overview of all questions is presented in~\Cref{tab:questions}.
     
     We explain each method in the following three sections.

        \subsubsection{Situation Awareness Global Assessment Technique for TD (TD-SAGAT)}
        \label{sec:method-tdsagat}

                %In RQ3, we ask for the effects the TDM process might have on the participants' TD awareness.
                As TD awareness is ambiguously used in TD research (see~\Cref{sec:RelatedWork}), we searched for a way to define TD awareness more precisely for our study and determine a fitting measurement method.
                Together with a researcher from the psychology domain, we searched a database of scientific papers that is used predominantly in psychology (i.e., PsycInfo~\cite{american_psychological_association_apa_2025}) for awareness concepts and subsequent measuring methods.
                We identified 17 distinct awareness terms and their corresponding definitions. 
                We dismissed 14 terms, e.g., ambient or phonemic awareness, that did not conceptually overlap with TD awareness (see~\cite{AdditionalMaterial}).
                We identified conceptual similarities to contingency awareness, bounded awareness, and situation awareness.
    
                \textbf{Contingency awareness} refers to the relationship between two events, which appears to fit the TDM problem, where the causes and consequences of TD are often not linked by stakeholders~\cite{wiese_it_2023}.
                Within the context of affective learning, contingency awareness is defined as the relation between conditional and unconditional stimuli~\cite{dawson1987human}.
                This refers to whether people are aware that one stimulus (e.g., a neutral image, such as a balloon) has been influenced by a simultaneous stimulus (e.g., an electric shock).
                However, this measurement method is poorly adaptable to TD Awareness because the stimuli (TD incurrence and TD consequences) have to be shown to participants, yet are too abstract in the case of TDM. 
    
                \textbf{Bounded awareness} refers to the fact that people often focus on a limited portion of the information available and make decisions based on that rather than considering it in full~\cite{chugh_bounded_2007}. 
                This bias often stems from the information people are requested to focus on. 
                This means that if a developer's goal is to finish an assignment within a certain time frame, they will focus on software that works, not whether they have optimized it to limit TD.  
                Unfortunately, there is no metric specifically designed to measure bounded awareness. 
                Instead, measurements have been designed for specific examples~\cite{Bazerman2016}.

                We found that \textbf{Situation Awareness (SA)}~\cite{Endsley1988} fits our concern as it focuses on decision-making, which we consider to be the underlying concern of TDM. 
                IT stakeholders must make two essential decisions in TD: Whether to incur TD or not, and whether to repay it or not~\cite{wiese_it_2023, Buschmann2011}.
                Consequently, we deemed the corresponding Situation Awareness Global Assessment Technique (SAGAT)~\cite{Endsley2000} to be the most appropriate research method for our concern.
                Moreover, this was the approach that most objectively measured awareness.
                
                \paragraph{Situation Awareness (SA)}
                
                \textit{Situation awareness (SA)}, as proposed by Endsley~\cite{Endsley1988}, ``refers to the perception of the elements of the environment within a volume of time and space, the comprehension of their meaning and the projection of their status in the near future''~\cite{zhou2023research}. % (Zhou et al., 2023, p. 203).  
                Going with this definition of SA, we defined TD awareness:
                \begin{framed}
                \textbf{TD awareness} is the (1)~\textbf{perception} of the prerequisites (e.g., alternatives, benefits, drawbacks) of a decision (e.g., incurring TD) during decision-making situations (e.g., refinement meetings), (2)~the \textbf{comprehension} of these prerequisites' meaning, and (3)~the \textbf{projection} of their status in the future (e.g., TD's consequences).
                \end{framed}
                \vspace{-0,1cm}
                With the workshops and information we give the participants, they get a primary perception of the prerequisites~(1). 
                By researching this TD awareness over the course of the study, we want to analyze whether the participants comprehend those prerequisites~(2) and whether they are able to use them in de\-ci\-sion-making, i.e., for the projection of future effects~(3).

                \paragraph{Situation Awareness Global Assessment Technique (SAGAT)}
                
                The idea behind the \textit{Situation Awareness Global Assessment Technique (SAGAT)} is to interrupt participants in their work or a simulation of their work and ask them questions about their situation~\cite{Endsley2000}.
                Interrupting participants with questions is still a subjective assessment. 
                Yet, participants asked about their current thought processes are more likely to provide objective answers as opposed to answering retrospectively (e.g., at the end of the week)~\cite{schwarz2007science}. 
                
                Moreover, this approach includes a process for systematically breaking down the specific awareness to its prerequisites, i.e., pieces of information needed to make informed decisions.  % following the term used in the definition.
                %The prerequisites were derived from 
                The main \textit{goals} and \textit{subgoals} for the situation lead to \textit{decisions} that have to be made to achieve the (sub)goal. 
                The \textit{SA prerequisites} are then derived for these decisions. 
                Finally, the appropriate \textit{queries} are designed from these prerequisites, questioning the participants about whether they considered the prerequisites during the decision-making before interruptions~\cite{Endsley2000}. 
                In the original papers on this technique, these ``prerequisites'' are also called ``requirements.'' 
                To avoid misunderstandings regarding ``(non-)functional requirements,'' we use the term ``prerequisites'' in this paper.

                %Questionnaire development
                \paragraph{TD-SAGAT Questionnaire Development}
                \label{sec:SAGATquestionnairedevelopment}

                We adopted the SAGAT approach and named this specific method \textit{TD-SAGAT}. 
                First, we utilized the systematic query-creation aspect of the SAGAT approach to collect TD awareness prerequisites and subsequent queries.
                We defined the following as the main goal and inferred subgoals from the TD definition from Avgeriou et al.~\cite{Avgeriou2016a} and the most common cause for TD according to the InsighTD project, which is time pressure~\cite{Rios2019}: 

                \begin{framed}
                \noindent
                \textbf{GOAL:} \\
                Have a system that is easy to change and fulfills the users' (functional) requirements. \\
                \noindent
                \textbf{SUBGOALS:}\\
                (1) Develop a new system that is easy to change within a given timeframe. \\
                (2) Implement a new functionality in a system so that it is still easy to change within a given timeframe.\\
                (3) Refactor (incl. architectural refactorings) a system so that it is easy to change after the refactoring with the least amount of time.\\
                (4) Maintain a system (update its components) so that it is easy to change with the least amount of time and effort.
                \end{framed}
                \vspace{-0,1cm}
                
                From these subgoals, we derived various decisions and respective prerequisites (see additional material~\cite{AdditionalMaterial}), e.g.:
                \textit{Could the preferred solution option incur TD?} 
                The resulting SA prerequisites were knowing (1)~alternative solutions, (2)~the benefits, and (3)~the drawbacks of each solution.
                Another decision was: \textit{Should we incur this TD item or not?}
                The resulting SA prerequisites were knowing (1)~the effort for the assumed best and the TD-incurring solution, (2)~the principal and interest, (3)~the risks of both solutions, and (4)~the competing functional requirement. 
                Finally, this led us to the following ten prerequisites:
                
                \begin{framed}
                \noindent
                \textbf{PREREQUISITES:}\\
                (1)~solution alternatives, (2)~solution's benefits, (3)~solution's drawbacks, (4)~solution's risks, (5)~effort for implementing a solution, (6)~TD item's principal, (7)~TD item's interest, (8)~component affected by TD, (9)~functional requirement potentially competing with a TD item, and (10)~quality attributes affected by TD.
                \end{framed}
                \vspace{-0,1cm}
                %For TD, these prerequisites can be viewed as the available information on a TD item. 
                %The participants should consider these prerequisites to make conscious decisions rather than acting on unconscious \textit{gut feelings}. 

                To validate the resulting prerequisites, we surveyed six experts gathered from our personal network and had them create their own hierarchy of goals, subgoals, decisions, and prerequisites. 
                Three experts were software engineers with over 20 years of IT experience, each from a different domain. 
                The other three experts were researchers who had all published peer-reviewed papers on TD.
                We did not discuss the topic of TD awareness with them prior to this study. 
                Comparing their input with our initial design, we found that the goals and subgoals were sometimes defined slightly differently. %, e.g., by referencing quality attributes like maintainability. 
                However, we did not receive further input regarding the prerequisites, which we interpreted as confirmation of our prerequisites.
                % Regarding the \textbf{(sub)goal}, four participants declared conscious decision-making regarding TD and avoiding unconscious TD incurrence as a goal.
                % Two participants each (1)~aligned the goals with quality attributes (i.e., maintainability, evolvability, reliability, and security), (2)~focused the goals on stakeholder satisfaction, or (3)~mentioned establishing a TDM process as a goal.
                % Regarding the \textbf{decisions} derived from the goals, five participants mentioned TD incurrence (including TD prevention) as a relevant decision.
                % Three participants specified decisions regarding TD repayment.
                % One participant suggested the decision of whether to maintain a system versus rebuilding it. 
                % On the level of \textbf{prerequisites}, two participants mentioned the importance of knowing and comparing alternative solution options together with their benefits, drawbacks, consequences, and effort required. 
                % Two participants also acknowledged the uncertainty associated with some decisions, e.g., if developers lack know-how.
                % Two participants specified the knowledge of the system's or components' details as prerequisites, e.g., knowledge of the technologies, the dependencies, and the evolution path.

                % Data gathering - Interrupting meetings
                
                \paragraph{ Data Collection}
                We identified meetings in which tickets were discussed, evaluated, and estimated, such as agile grooming/re\-fine\-ment meetings, as situations when to use our TD-SAGAT method. 
                We applied this method starting in August 2023 and used the TD-SAGAT interruptions only once a month to minimize the Haw\-thorne effect, i.e., observing behavior can influence the behavior~\cite{MCCAMBRIDGE2014, chiesa_making_2008}.
                We interrupted the participants at two to three randomly chosen points in time during a one-hour meeting using a random alarm clock app~\cite{randomticker_random_2025}. %\footnote{\url{https://shorturl.at/NDcyN}}. 
                The participants were then invited to complete the TD-SAGAT questionnaire (see additional material~\cite{AdditionalMaterial}), asking whether they considered the prerequisites during the previous ticket's discussion with ``yes'' or ``no'' as answer options. 
                All questions are presented in~\Cref{tab:questions} (also see~\Cref{sec:method-surveys}). 
                %The whole questionnaire is part of the additional material~\cite{AdditionalMaterial}.
                We ordered all questions randomly to avoid systematic and contextual effects, e.g., that subsequent questionnaire items are more likely to be answered similarly~\cite{royal_impact_2016}.
                %Instead, we used the interruption only once a month to obtain the data in a setting as natural as possible. 
                Only participants familiar with the discussed topic answer\-ed the TD-SAGAT questionnaire, which sometimes resulted in few answers. 
                However, we assume the overall results are still valid, as we conducted multiple interruptions per meeting and only evaluated the results as a sum.
                Overall, we used the TD-SAGAT interruptions during \TDSAGATMeetingCnt meetings, resulting in \TDSAGATMeetingAnswersCnt datasets, i.e., an average of \TDSAGATMeetingAnswersAvg per meeting.
                A step-by-step introduction to the TD-SAGAT method is part of our additional material\cite{AdditionalMaterial}.

                %Data analysis
                \paragraph{ Data Analysis}
                For each prerequisite, we counted the number of participants considering this prerequisite per meeting,  i.e., we summarized the ``yes'' answers across all inter\-rup\-tions, i.e., tickets. 
                We then calculated the percentage of ``yes'' answers relative to all answers. 
                We arranged this on a timeline to see whether the number of participants per prerequisite increased during the action cycles.
                In the end, we compared these results with the observations' results to see whether there is a difference in what participants think about and what they talk about.

                We presented these results in each workshop's ``evaluation'' phase, comparing them with the observation results.
                For this paper, we furthermore compare them to the results of the workshop questionnaire.
                To enable a better overview, we split the prerequisites into three categories: 
                \begin{itemize} [itemsep=0pt]
                    \item \textit{comparing solutions:} comprising solution alternatives, solution's benefits, solution's drawbacks, and solution's risks
                    \item \textit{evaluating costs:} comprising effort for implementing a solution, TD item's principal, and TD item's interest
                    \item \textit{considering potential consequences:} comprising the af\-fec\-ted components, a functional requirement potentially competing with a TD item, and the affected quality attributes
                \end{itemize}

        \subsubsection{Observations}
        \label{sec:method-observation}

                \paragraph{ Data Collection}
                \label{sec:method-observation-collection} 
                
                For the first three months, we observed the review \& planning meetings. % because we were unaware of the other meetings.
                Initially, many tickets were still debated in these meetings. 
                However, discussions became less frequent as the structure of the meetings became more established over time, and only tickets for the upcoming sprint were selected.
                Thus, from August on, we also participated in refinement and grooming meetings, during which much more discussion on the single backlog items occurred.
                Consequently, we focused our observation on these two meeting types for the rest of the study.
                
                For the observations, we used a predefined observation protocol (see~\cite{AdditionalMaterial}) where the researchers marked the following observations per ticket, i.e., discussion item:
                \begin{enumerate} [itemsep=-2pt]
                    \item Ticket ID and subject to enable discussion in the next workshop's ''learning \& evaluation'' phase
                    \item Ticket type given by the team (e.g., TD, bug, user story)  to reveal potential deviations between backlog and discussion
                    \item Checkboxes to indicate whether the team versus the researchers consider a ticket to be a TD repayment to determine if there were any different assessments on what is considered a TD item
                    \item Checkbox to indicate whether the team runs the risk of incurring TD unconsciously, e.g., when someone was already implementing a ticket even though it has not been refined and, thus, alternatives have not been analyzed, or when a ticket was planned for the next sprint even though crucial information for the implementation was still missing  %ing TD,
                    \item Checkbox to indicate  whether TD was incurred intentionally to determine instances of conscious TD incurrence  
                    \item Checkboxes for each TD-SAGAT prerequisite to indicate whether it was discussed %: alternatives, drawbacks, risks, effort/principal, interest, contagiousness, affected components, competing functions, and quality attributes. 
                \end{enumerate}
                We explained the researcher's evaluation in a \textit{remark} column for discussion in the next workshop's ``learning'' phase.
                
                The first and second authors of this paper performed the observations. 
                We analyzed \DoubleObservation meetings with both researchers to align our working methods. 
                Then, we alternated the observations between us.
                Overall, we observed the discussion of \ItemsObserved items in \MeetingCnt meetings watching \MeetingHrs hours of video recordings.
                %one researcher and the other.
                % We marked whether a topic was discussed, but did not analyze its frequency due to the time limitations when filling out the protocol during an ongoing discussion. 
                % Additionally, the discussions were typically very short, and one requirement was rarely mentioned twice.
                
                \paragraph{ Data Analysis}
                For data analysis, we counted the number of tickets (i.e., ticket discussions) for which the respective observations occurred per meeting.
                For May to July 2023, these numbers were based on the review \& planning meetings. 
                For the rest of the study, the values are based on the refinement and grooming meetings.
                To better compare the results with other methods, we calculated the percentage of counted tickets relative to the total number of discussed tickets per meeting. 
                For the TD-related values (interest, contagiousness, and competing functional requirements), we used the number of TD tickets as the basis.
                
                We presented these results in each workshop's ``evaluation'' phase and this paper, comparing the prerequisites (observation no. 5) with the TD-SAGAT results.
                To enable a better overview, we split the prerequisites into the same categories as mentioned in~\Cref{sec:method-tdsagat}. 

                The additional observation items were presented separately on a timeline, focusing on the risk of unconscious TD  incurrence (observation no. 3) and dispute on whether an item is TD or not (observation no. 2).
                % \begin{itemize} [itemsep=-2pt]
                %     \item \textit{comparing solutions:} comprising solution alternatives, solution's benefits, solution's drawback, and solution's risks
                %     \item \textit{evaluating costs:} comprising effort for implementing a solution, TD item's principal, and TD item's interest
                %     \item \textit{evaluating consequences:} comprising the affected components, a functional requirement potentially competing with a TD item, and the affected quality attributes
                % \end{itemize}
                
                Finally, we clustered the ``remarks'' by common topics  and used the single items as examples to support the discussion in the following workshop's ``learning'' phase.
                Similarly, information on observations no. 1 (backlog deviation) and no. 4 (intentional TD incurrence) was presented as discussion topics with examples. 
                We did not present the values in a figure, as the instances were extremely rare, e.g., only two instances of intentional TD incurrence during the study's duration.
                
        \subsubsection{Questionnaires}
        \label{sec:method-surveys}

    \begin{table*}[t]
        \centering
        \footnotesize
        \begin{tabular} {p{0.3cm} p{3cm} p{9cm} p{1cm} p{1cm} p{1cm}  }
            \hline
            \textbf{No.} & \textbf{Question} & \textbf{Sub-Question or answer-options if applicable} & \textbf{Answer} & \textbf{Ques-} &  \textbf{Used to} \\
            &  & & \textbf{type} & \textbf{tionnaire}   & \textbf{answer} \\
            \hline
            1a &  \multirow[t]{10}{3cm}{When I decide for/against an architecture,  a design, or a way to implement a requirement, I know ... (questions: 1, 2, 3, 4, 5, 8)
            OR
            When I decide for/against a refactoring (TD repayment), I know ... (questions: 6, 7, 9, 10} 
                &  what is an alternative solution option to my favorite solution option?
            & y/n & KP & RQ3 \\
            2a &  & what are the advantages of my favorite solution option?
            & y/n & KP & RQ3 \\
            3a &  &  what are the disadvantages of my favorite solution option?
            & y/n & KP & RQ3 \\
            4a &  &  what are the risks of the supposedly faster solution?
            & y/n & KP & RQ3 \\
            5a &  &  how much effort is required for a supposedly ideal solution?
            & y/n & KP & RQ3 \\
            6a &  &  how much will the repayment (the refactoring) cost?
            & y/n & KP & RQ3 \\
            7a &  &  how much effort will it take to keep the current solution?
            & y/n & KP & RQ3 \\
            8a &  &  what competing functional (technical) requirements are there?
            & y/n & KP & RQ3 \\
            9a &  &  which components are affected by the repayment (the refactoring)?
            & y/n & KP & RQ3 \\
            10a &  &  what are the requirements for the quality attributes (e.g. availability, performance, usability, maintainability) for the affected component?]
            & y/n & KP & RQ3 \\
            \hline
            1b  &   \multirow[t]{10}{3cm}{Regarding the topic currently under discussion, I considered ... (Please answer spontaneously! If you have to think long and hard about whether you have (just) thought about something, the answer is probably -- No)}  
                   &  what alternatives are there to the solutions discussed?
            & y/n & S    & RQ3 \\  
            2b  &  & what are the advantages of the solutions discussed?
            & y/n & S & RQ3 \\  
            3b  &  & what are the disadvantages of the solutions discussed?
            & y/n & S    & RQ3 \\  
            4b  & & what are the risks of the solutions discussed?
            & y/n & S    & RQ3 \\  
            5b  &  & how much effort is required for the solutions discussed?
            & y/n & S    & RQ3 \\  
            6b  &  & how much effort (if any) will it take to convert the discussed solutions into the ideal solution (repayment of TS)?
            & y/n & S    & RQ3 \\  
            7b  &  & how high, if any, is the (additional) effort required to permanently maintain the solutions discussed? (Interest from TS)
            & y/n & S    & RQ3 \\  
            8b  &  & what competing functional (technical) requirements are there?
            & y/n & S    & RQ3 \\  
            9b  &  & which components are affected by the solutions.
            & y/n & S    & RQ3 \\  
            10b  &  & which quality attributes are influenced.
            & y/n & S    & RQ3 \\  

            \hline
            12 &  \multicolumn{2}{p{12cm}}{How much time (in person-hours) did you invest in initializing the technical debt management process (e.g., workshop participation or backlog consolidation) in the last cycle?} 
            & number & P & RQ2\\         
            \hline
            13 & \multicolumn{2}{p{12cm}}{How much time (in person-hours) did you invest in the process of managing technical debt in the last cycle (excluding initial efforts such as workshop participation or backlog consolidation)?} 
            & number & R & RQ2 \\
            \hline
            
            14 & %\multirow[t]{3}{5cm}{Please indicate how you assess the work with the backlog and the number of technical debts in the backlog.} 
            \multicolumn{2}{p{12cm}}{Do you consider the amount of technical debt in the current backlog to be a consistently appropriate amount?}
            & y/n & P & RQ1 \\
            15 & \multicolumn{2}{p{12cm}}{Do you think the amount of technical debt taken on in the last cycle was appropriate?}
            & y/n & P & RQ1 \\
            16 & \multicolumn{2}{p{12cm}}{Do you think the amount of technical debt repaid in the last cycle was appropriate?}
            & y/n & P & RQ1  \\
            17 & \multicolumn{2}{p{12cm}}{Are you satisfied with the process, backlog and technical debt situation? Or should we further optimize the process?}
            & y/n & P & RQ1  \\            
            \hline
            
            18a & \multicolumn{2}{p{12cm}}{What are your goals related to the process? What should change in your everyday life? How should technical debt be handled at the end of the project?}
            & text & K & supports action\\
            18b & \multicolumn{2}{p{12cm}}{What are your goals related to the process? What is going “really well” now? What should we improve in the next step?}
            & text & P & research (RQ1) \\         
            \hline
            19a & \multicolumn{2}{p{12cm}}{What expectations do you have for the project? How should the project work? What shouldn't be forgotten? What should be given less importance? What should definitely not happen?}
            & text & K & supports action \\
            19b & \multicolumn{2}{p{12cm}}{What expectations do you have for the project? What is going “really well” in the project? What should change?}
            & text & P & research (RQ1)\\
        \end{tabular}
        \caption{Overview of all questions from the kickoff (K), progress (P), and TD-SAGAT questionnaire (S), their answer types, and their reference to the RQs }
        \label{tab:questions}
    \end{table*}

            Throughout the study, we used three different questionnaires: one questionnaire for the first, i.e., kickoff workshop, one to analyze the progress made so far during the following workshops, and one during the TD-SAGAT interventions.
            
            \paragraph{ Data Collection}
            
            Each of the three questionnaires comprised the ten questions about the TD-SAGAT prerequisites in a ``yes'' or ``no'' format.
            However, for the kickoff and process questionnaire, we separated this evaluation into the three situations where they (a)~might potentially incur TD, (b)~discuss repaying TD, and (c)~discuss updating the existing system, e.g., libraries or infrastructure (resulting in 30 questions).
            
            The kickoff questionnaire additionally included questions on whether the participants had heard of TD and its related concepts, whether they were familiar with the various TD types, and how they currently manage TD. 
            All these questions had predefined values to simplify the completion of the questionnaire.
            We did not add these questions to this paper as they were not used for scientific evaluation, but to get to know the participants and team. Still, we added them to the additional material~\cite{AdditionalMaterial} as they might help other researchers when replicating this study.
            
            The process questionnaire comprised additional questions on the progress of the study, enabling us to analyze the termination point for the action research.
            Moreover, we asked for the time invested in the management of TD, which was separated into the initial effort and the continuous effort to answer RQ2.
            Finally, the kickoff and the process questionnaire additionally asked for the goals and expectations of the participants in two open questions to give participants an opportunity to express their opinions and take these into account in the further development of the TDM process.
            
            All questions that were evaluated for this study are presented in~\Cref{tab:questions}.
            In some cases, a slightly different wording of the questions was necessary. 
            In these cases, we added both questions and numbered them as a/b.
            For creating the questions on the prerequisites, we followed the SAGAT guidelines presented in~\Cref{sec:SAGATquestionnairedevelopment}.
            The questionnaires were created using LimeSurvey~\cite{limesurvey_limesurvey_2025}. %\footnote{\url{https://www.limesurvey.org/}}.          
            They were completed electronically via smartphones or laptops, depending on each participant's preferences.     
            All questionnaires were pre-tested with two senior developers who were not part of the research team.
            In addition, the questionnaires were all completed while the researchers were present, allowing any uncertainties regarding the wording of the questions to be discussed immediately.

                \paragraph{ Data Analysis}
                During the evaluation, we recognized the complexity of asking for the prerequisites in three different scenarios and the hardship of comparing them with the results from other methods (also see~\Cref{sec:discussion-studydesign}).
                Therefore, we merged the separated results to reduce them to the ten TD-SAGAT prerequisites.
                For this, we used the results from the questions on TD incurrence (a) if the prerequisites were not TD-related, i.e., alternatives, benefits, drawbacks, risks, effort, and competing functional requirement (1a -5a and 8a of~\Cref{tab:questions}).
                We used the answers to the TD repayment questions (b) if the prerequisites were TD-related, i.e., principal, interest, affected component, and quality attributes. (6a, 7a, 9a, and 10a  of~\Cref{tab:questions}).  
                We created visualizations for the prerequisites' results and presented the progress in each workshop's ``learning'' phase and in this paper, comparing the questionnaire's answers to the results of the observations and TD-SAGAT.

                We used the questions from the kickoff questionnaire regarding the current TDM process to understand and report the team's current processes in~\Cref{sec:initialSituation}. 
                We presented the results to the team in the second workshop. 
                However, we did not analyze these questions scientifically, nor did we present the visualizations in this paper due to their limited significance compared to the textual description.

                We utilized questions 12 and 13 (\Cref{tab:questions}) from the progress questionnaire to answer RQ2.
                Questions 14 to 17 indicated the right time to terminate the action research.
                Furthermore, we compared these backlog-related questions with the actual amount of TD in the backlog.
                Finally, the open-ended questions (18 and 19) gave qualitative feedback on the TDM process and the project's progress.
                We used the answers to address emerging topics in the following workshop or separate meetings, e.g., special issues the algorithm subteam identified (see~\Cref{sec:Results-Cylce2}).

    \begin{table*}[t]
        \centering
        \footnotesize
        \begin{tabular} {p{2,2cm} p{8cm} p{0.5cm} p{5cm}  }
            \hline
            \textbf{TD activity} & \hypertarget{approachesTbl}{\textbf{Presented approaches}} & \textbf{Used} &  \textbf{Remark}  \\
            \hline
            TD organization               & establishing a TD champion~\cite{jaspan_defining_2023}
                                            & y & called TD manager\\
            \hline 
            TD identification             & automatically by static analysis tools            
                                            & (y) & only as supporting method \\ 
            \& definition                 & manually by Dagstuhl definition~\cite{Avgeriou2016a}
                                            & (y) & but other TD causes exist \\ 
                                          & manually by backlog colors~\cite{Kruchten2012a} %, i.e., a quadrant for visible or invisible to customers and increasing a system’s value or reducing its negative effects
                                          %~\cite{Kruchten2012a}
                                            & n & no external customers exist \\ 
                                          & manually by identifying whether TD interest occurs~\cite{avgeriou_technical_2023}
                                            & y & \\ 
            \hline 
            TD prevention    	          & well-defined requirements, code eval\-ua\-tion/\-stan\-dardization, following \-well-\-defined project processes, well-defined architecture/design, using good practices, improving tests, and reviews \cite{Rios2020, Freire2020a, Perez2021a}
                                            & (y) & the team generally optimized their work processes; not all strategies were followed\\ 
                          	         & educational sessions on how to avoid TD \cite{Besker2020d} 
                                            & n & workshop worked as an educational session, but no other sessions were planned\\ 
                          	         & incorporating a TDM approach into project management~\cite{Wiese2022} 
                                            & n & \\ 
                          	         & debiasing decision-making by listing at least one solution alternative, one drawback, and one risk per alternative~\cite{Borowa2021, borowa_debiasing_2022} 
                                            & y & \\ 
                          	         &  better decision-making by considering quality attributes~\cite{iso_25010_2023, bass2012software} 
                                            & n & \\ 
                          	         &  counteract TD inheritance between disciplines in mechatronics~\cite{Dong2019, Vogel-Heuser2021}
                                            & n & \\ 
            \hline 
            TD measurement 	              & interest, interest probability, effort/principal for prioritization~\cite{Tom2013b, McConnell2008a} 
                                                    & y & to identify \textit{low-hanging fruits} \\ 
                                          & interest, interest probability, effort/principal for decision-making~\cite{perera_systematic_2024} 
                                            & y &  \\
            \hline 
    TD documentation & recording TD in a backlog~\cite{Kruchten2019, Junior2022}             
                                            & y & \\ 
            \hline 
            TD prioritization             & prioritizing TD against functional requirements~\cite{lenarduzzi_systematic_2020}       
                                            & n & \\ 
                          	         & prioritizing TD against each other~\cite{lenarduzzi_systematic_2020}        
                                            & y & \\ 
                          	         & prioritizing once~\cite{lenarduzzi_systematic_2020}          
                                            & n & \\ 
                          	         & prioritizing constantly~\cite{lenarduzzi_systematic_2020}         
                                            & y & by using a \textit{re-submission date}\\ 
                          	         & based on the systems evolution~\cite{Schmid2013}       
                                            & y & \\ 
                          	         & based on the business value~\cite{ReboucasDeAlmeida2018c}              
                                            & n & \\ 
                          	         & based on cost attributes (TD measurements)~\cite{Tom2013b}              
                                            & y & to identify \textit{low-hanging fruits}\\ 
                          	         & based on the affected quality attributes~\cite{rachow_architecture_2022}             
                                            & n & \\ 
            \hline 
            TD repayment 	              & repay TD if it impedes a functional requirement~\cite{FREIRE2023}       
                                            & y & similar to refactor if it impedes development\\ 
                          	         & pay the interest~\cite{Buschmann2011}              
                                            & y & \\ 
                          	         & wait for discontinuing or migrating of the affected software system~\cite{Schmid2013, Schmid2013a}              
                                            & y & use \textit{re-submission date} to recall TD items \\ 
                          	         & repay TD if the positive effect is proven by analyzing the costs~\cite{McConnell2008a, Tom2013b}               
                                            & y & repay \textit{low-hanging fruits} if time allows \\ 
                          	         & repay TD by assigning a quota~\cite{McConnell2008a, Wiese2022}              
                                            & n & \\ 
                          	         & repay if TD might hinder the system's evolution in the near future~\cite{Schmid2013}              
                                            & y & \\ 
                          	         & repay if TD was incurred due to a project deadline~\cite{Wiese2022}              
                                            & n & \\
            \hline 
            TD monitoring \& visualization& DebtFlag~\cite{holvitie_debtflag_2013}, MultiDimEr~\cite{silva_multidimer_2022}, VisminerTD~\cite{Mendes2019}, MIND~\cite{falessi_towards_2015}, TEDMA~\cite{fernandez-sanchez_open_2017}
                                            & n & not for backlog visualization, or only for self-developed tools \\   
                          	         & self-developed Microsoft PowerBI TD visualization (\Cref{sec:Results-Cylce5})       
                                            & y & interfaces to Azure DevOps and Jira exist \\    
            \hline
        \end{tabular}
        \caption{Overview of presented approaches with references and their adoption/use}
        \label{tab:approaches}
    \end{table*}

\section{WORKSHOPS}
\label{sec:workshops}
    
    In this section, we describe the five workshops in detail. 
    Each workshop comprised the ``evaluation,'' ''learning,'' ''diagnosing,'' and ''action planning'' phases of the respective action research cycle.
    The first workshop followed a unique structure as there was no ``evaluation \& learning'' part but a need to get to know each other, analyze the team's current TD processes, and organize the upcoming work. 
    Workshops two to five followed a similar structure, with a slight deviation for the second workshop, as two activities were introduced.
    The first and second workshops were done on-site, while the subsequent workshops were held online following the team's preferences.
    
\subsection{Recurring Intervention Formats}
    We repeatedly employed two intervention formats typical for workshops: the ``Love it -- Change it -- Leave it'' and the ``1-4-all'' format.
    
    In the retrospective ``Love it -- Change it -- Leave it'' format, participants are asked to write down what they like about the current process, what needs to be changed, and what should be abandoned. 
    Each participant writes down each thought on one moderation card and presents their results one after the other by sticking them on a flipchart or whiteboard.
    The results of all participants are then clustered by the team. 
    In the end, possible actions are derived, discussed, and potentially taken.

    In the ``1-4-all,'' questions given by the moderator are thought through briefly by each participant on their own (1). 
    Then, the questions are discussed in detail in groups of four members each (4). 
    Finally, each group presents its result to the other groups to find commonalities and discuss discrepancies until a common answer is formed (all).

\hypertarget{WS1}{
\subsection{First Workshop / Kickoff Workshop}}
\label{sec:workshops_1}
    We started the first workshop by motivating the TD topic.
    Throughout the entire first workshop, we collected topics relevant to the practitioners on an additional flip chart (i.e., a topic cache), which we then used at the end of the study to determine whether all topics were adequately addressed.
    
    First, we informed the team about the details of the upcoming study.
    After that, we analyzed the team's motivation using an adaptation of  Führing's ``organizational energy template''~\cite{Fuhring2021a} where each participant was asked to identify their opinion on the topic's importance versus their willingness to invest effort into that topic.
    The participants marked their self-assessment, their assessment of their team, and of their company, on a matrix with four quadrants on a flip chart (see~\cite{AdditionalMaterial}).
    We combined this with a short introduction of each participant, including their name, position, IT experience, time of company affiliation, and hobby.
    Additionally, we motivated the TD topic by playing a game developed to enable discussions on TD~\cite{wiese_techdebt_2025}.
    
    %new TD activity discussion
    After a lunch break, the participants completed the workshop (kick-off) questionnaire before the actual action cycles started. 
    We diagnosed the current TD management process in three subsequent steps:
    \begin{enumerate}[itemsep=0pt]
        \item We collected concrete examples of TD, i.e., TD stories of TD incurrence/causes, TD consequences, and TD repayment, to get a first insight and to give the participants some room to present their experiences with TD. 
        This also supported changing a potentially frustrated basic mood into a more constructive mood.
        \item We identified the current approaches to managing TD by discussing the following questions in a ``1-4-all'' format.
        \vspace{-0,6cm}
            \begin{itemize}[itemsep=0pt]
                \item What are you currently doing after you have identified a TD item? (e.g., document, communicate, repay, ignore) 
                \item How do you currently identify TD items?
                \item In which situation do you consciously incur TD? On which information do you base this decision?
            \end{itemize}
       
        \item We gathered requirements for process improvement using a ``Love it -- Change it -- Leave it'' format
    \end{enumerate}
    % First, we asked the participants to describe concrete instances, i.e., examples, of when they dealt with TD regarding its incurrence, consequences, and repayment.
    % Second, we requested the participants to elaborate on how they discover TD items, how they react when they discover a TD item, and in which cases they are willing to incur TD.
    % Third, we invited the participants to discuss their wishes and requirements for a TDM process by asking for things that already work appropriately, things they want to change, and things they plan to discontinue.
    Eventually, we discussed the organization of the change project and action research.
    We identified meetings where TD items were typically discussed to observe these meetings for evaluation, and we named one team member to fulfill the role of a \textit{TD manager}, similar to a ``TD champion'' mentioned in~\cite{jaspan_defining_2023}. 
    The tasks for the TD manager were:  
    \begin{itemize}  [itemsep=-2pt]
        \item to be part of the action team and function as the link between the research team and practitioners, 
        \item to implement the approved actions in cooperation with the researchers and potentially other teams, e.g., backlog administrators, 
        \item to support other teams of the company in adopting and establishing a TDM process, and
        \item to be the "voice of reason" in meetings, continuously reminding everyone of the TD topic, i.e., to ensure the process would still be followed after the study ended.
    \end{itemize}
    We ended the workshop by reusing the organizational energy matrix to identify changes in motivation and allowing the participants to give feedback on the first workshop.

    \begin{table*}
        \centering
        \footnotesize
        \begin{tabular} {p{0,4cm} p{0,8cm} p{4cm}  p{0,5cm} p{7cm} p{1,2cm} p{1,7cm}  }
            \hline
            No. & Dash- & Use Case  				& Chart & Implemented visualization & Fig. 	& Adoption  \\
                & board & Question to be answered 	& type  &					&  		& / Usage \\
            \hline 
            %\hline 
                V1 & M + T & Which TD tickets can influence the development of a specific user story?     
                    & list & We list the TD tickets in the component (areaPath) and its sub-components where the user story will be implemented. (The original mock-up was not adopted but changed to a list.)
                    & \ref{fig:influence}
                    &  used \\ 
            %\hline 
                V2 & M + T & How big is the current amount and effort of all TD tickets?     
                    & text & We display the number of TD tickets and the sum of the effort as text. 
                    & \ref{fig:AmountEffort}
                    & rarely used  \\ 
            %\hline 
                V3 & M & Which TD item's repayment will be quickly amortized?     
                    & list & We list the 20 TD tickets with the lowest ROI in months. (The original mock-up was not adopted but changed to a list.) %(\textit{Low Hanging Fruits})
                    & \ref{fig:profitable}
                    & usage planned \\ 
            %\hline 
                V4 & M &  How many costs can we save by repaying this TD item?    
                    &  line & We show the accumulating interest costs (calculation see~\Cref{sec:method-backlog}) per TD item. 
                    & add. mat.
                    &  not used\\ 
            %\hline 
                V5 & M &  How does the amount of TD develop over time?     
                    & line &  We display the effort sum of all TD tickets per month that were still open in the respective month. 
                    & \ref{fig:openovertimeeffort}
                    &  rarely used \\ 
            %\hline 
                V6 & M &  Which TD tickets can only be repaid by person X?    
                    &—& No visualization: From the works council's point of view, it is critical to present the work performance of a single person.
                    & --
                    & not used  \\
            %\hline  
                V7 & T & How accurately are the priorities estimated?     
                    &  bar & We show the \textit{educated guess} and the calculated \textit{ROI-based} priority (calculation see \cref{sec:method-backlog}) as overlaid bars.
                    & \ref{fig:discrepancy}
                    & used \\ 
            %\hline 
                V8 & T & How many TD tickets will be resubmitted for a given meeting?    
                    & dot &  We show the number of TD tickets per \textit{re-submission date} versus the \textit{educated guess} or the calculated priority (customizable). % The upper left region is dyed green and marked with \textit{Do}. The lower right region is dyed red and marked with \textit{Wait}.
                    & add. mat. 
                    &  not used \\ 
            %\hline 
                V9 & T &  Which tickets are \textit{low-hanging fruits} (i.e., high priority with low effort/principal)?     
                    & dot & We show the number of TD tickets per effort versus the priority. The upper left and lower right regions are dyed green/red and marked with \textit{Low-hanging fruits}/\textit{Wait}.
                    & \ref{fig:lowhanging}.
                    &  used \\ 
            %\hline 
                V10& T &  Which tickets are contagious and should be repaid soon?    
                    & list & We represent contagious TD tickets as a column containing a red exclamation mark in the detailed table views.
                    & \ref{fig:profitable}
                    & used  \\ 
            \hline
        \end{tabular}
        \caption{Overview of visualizations with underlying uses cases/questions, assignment to a dashboard (M - manager, T - team), final implementation details, the corresponding figure (add. mat. = additional material) and their adoption by the team }
        \label{tab:visualization}
    \end{table*}
\subsection{Second to Fifth Workshop}
\label{sec:workshops_2_5}

    The second to fifth workshops were structured in similar ways.
    Each workshop started by reminding the participants of the workshop's goal and their role (action/reference team) in the change process and study. 
    Then, the researchers presented the evaluation results of the previous action cycle (``evaluating''). 
    During this presentation, the team discussed the findings if needed. 
    Before lunch, the participants were asked for their evaluation of the actions taken during the previous cycle in a ``Love it -- Change it -- Leave it'' format (``learning'').
    After the break, the participants completed the workshop questionnaire. 
    Then, the researchers introduced the new topics, namely TDM approaches, for the respective workshops in presentations that each lasted no more than 45 minutes.
    An overview of all workshop topics is given in~\Cref{tab:actioncycles}, and the respective research approaches are presented in~\Cref{tab:approaches}.
    The details of each workshop's topics are explained in the following ``topics'' subsections.
    After the introduction, we asked the participants to discuss the new topics and decide which approaches to adopt (``diagnosing'' and ``action planning'') utilizing the ``1-4-all'' format. 
    %The exact questions for each topic can be derived from the workshop slides in the additional material~\cite{AdditionalMaterial}.
    %The introduced and discussed topics are further presented in~\Cref{tab:actioncycles} and the respective ``Diagnosing'' Subsections of~\Cref{sec:actioncycles}.
    At the end of each day, we determined the topics that were raised during the day and the measurements that the team decided on. 
    We assigned at least one responsible participant for each measurement.
    
    While all workshops were planned this way, slight deviations regarding the breaks were possible. 
    %For example, sometimes, the evaluation's presentation lasted longer due to rising discussions, and thus, the retrospective was postponed until after the lunch break.
    A generic agenda for these workshops is part of the additional material~\cite{AdditionalMaterial}.
    
    The third to fifth workshops were done online utilizing Microsoft Teams~\cite{microsoft_teams_2025} due to the participants' preferences. %\footnote{ \scriptsize\url{https://www.microsoft.com/en-us/microsoft-teams/}}
    In these cases, we used breakout rooms for the ``1-4-all'' discussion in ``groups of four''  and Mural~\cite{mural_work_2025}for the ``Love it -- Change it -- Leave it'' part, employing its ``Retrospective Template~\cite{mural_retrospective_2025}. %\footnote{\url{https://www.mural.co/templates/retrospective}}.''     \footnote{\url{https://www.mural.co/}} 
    
       \begin{figure*} [t!]
        \begin{tabular}{@{}cc@{}}
                \subfigure[V1: TD tickets (TD ID) that influence a given user story (US ID) with component or sub-component (areaPath) and the sub-component's level]
                {	\label{fig:influence}
                    \includegraphics[width=0.5\textwidth]{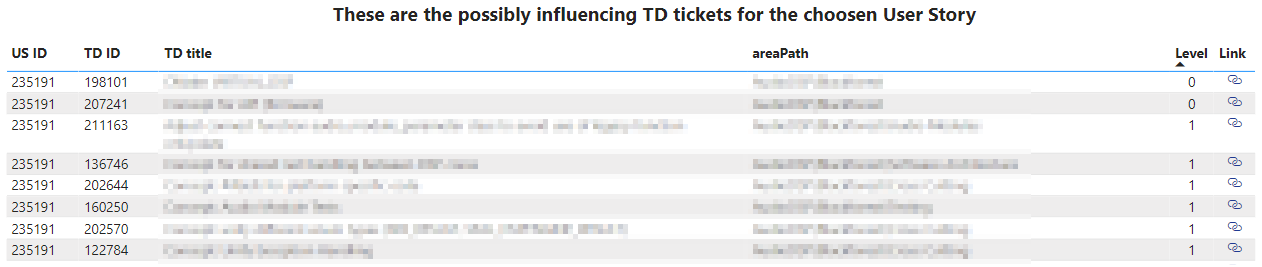}
                } &
                \subfigure[V3: TD tickets (ID) ranked by the ROI in months, with \textit{educated guess} (given) and \textit{ROI-based} (calculated) priority and the information on contagiousness (V10)]
                {	\label{fig:profitable}
                    \includegraphics[width=0.5\textwidth]{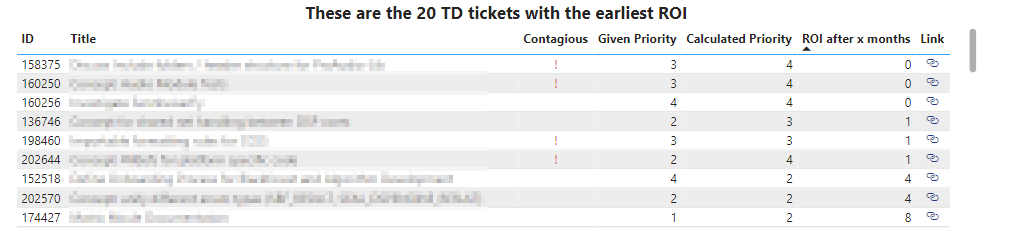}
                }\\
                % \subfigure[V1: User stories that are highly influenced by TD items (not used, changed to a list)]
                % {	\label{fig:influence}
                %     \includegraphics[width=0.48\textwidth]{img/influence.png}
                % }\\
                % \subfigure[V8: TD items per priority and \textit{re-submission date} (not used)]
                % {	\label{fig:resubmission}
                %     \includegraphics[width=0.48\textwidth]{img/resub_given.png}
                % }&
                % \subfigure[V4: TD items, whose repayment is highly profitable, i.e., with an ROI in a few months (not used, changed to a list)]
                % {	\label{fig:profitable}
                %     \includegraphics[width=0.48\textwidth]{img/profitable.png}
                % }\\
                \subfigure[V2: Current amount and effort over all TD tickets and for very high and very low priority tickets]
                {	\label{fig:AmountEffort}
                    \includegraphics[width=0.35\textwidth]{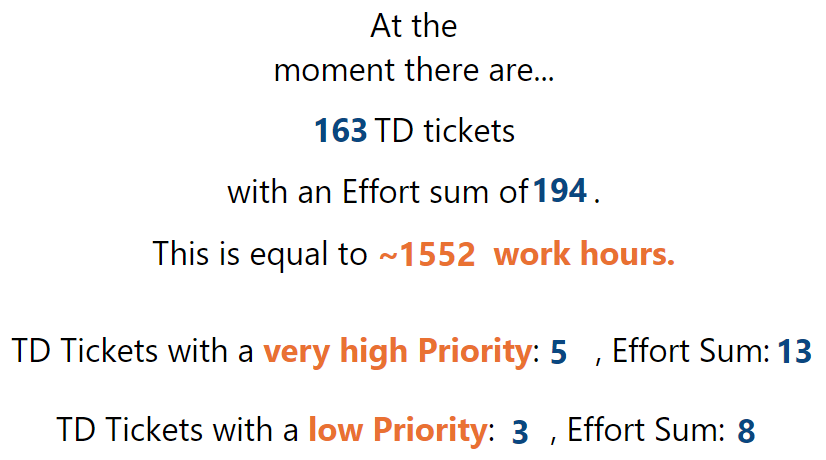}
                } &
                \subfigure[V5: Effort summarized for open TD items per month over time]
                {	\label{fig:openovertimeeffort}
                    \includegraphics[width=0.5\textwidth]{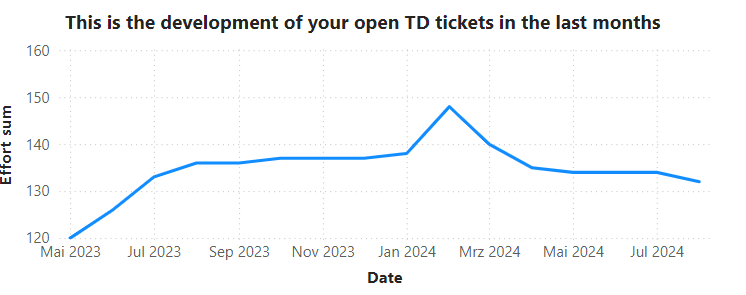}
                } \\
                \subfigure[V7: Priority discrepancies between \textit{educated guess} and \textit{ROI-based} priority]
                {	\label{fig:discrepancy}
                    \includegraphics[width=0.45\textwidth]{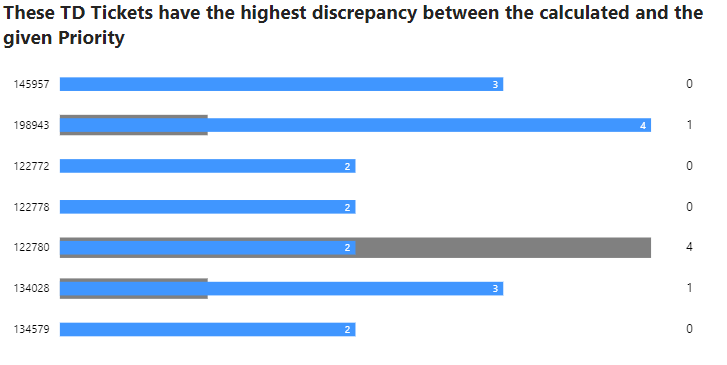}
                } &
                \subfigure[V9: TD items per priority and effort to identify \textit{low-hanging fruits}, where the size of the dot represent the amount of TD tickets]
                {	\label{fig:lowhanging}
                    \includegraphics[width=0.45\textwidth]{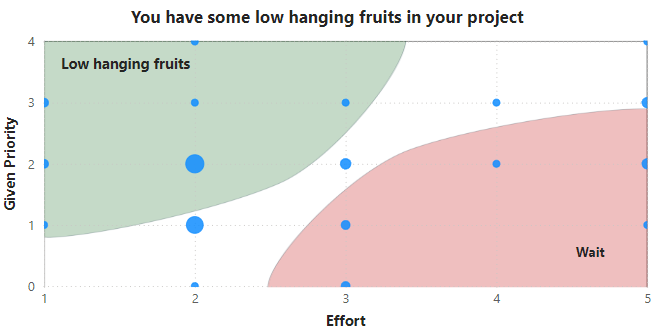}
                }\\
        \end{tabular}
    
        \caption{Visualizations created to manage TD items in the team's backlog and their adoption/usage by the team}
        \label{fig:visualizations}
    \end{figure*}

\hypertarget{WS2}{
\subsubsection{Second Workshop -- Topics}} 
\label{sec:2WS_topics}
    
    The second workshop marginally deviated from this procedure as two activities were introduced (TD prevention and TD measurement \& documentation), and two discussions were conducted:
    We presented the works shown in~\Cref{tab:approaches} about TD in mechatronics~\cite{Dong2019, Vogel-Heuser2021}, quality attributes~\cite{iso_25010_2023, bass2012software} and cognitive biases~\cite{zalewski_cognitive_2017, Borowa2021, borowa_debiasing_2025} and their impact on decision-making.
    Using the ``1-4-all'' format, the participants explored optimizations to the TD ticket by adding TD attribute fields to provide the respective vital information for de\-ci\-sion-making of managers and developers. 
    Further, we asked them to determine key performance indicators (KPIs) and ways to calculate them. 
    For the (additional) discussion on quality attributes, each participant was asked to rank the attributes in a priority that would be typical for their system. 
    Everyone presented their ranking, and the team was then asked to discuss and find a ranking that each participant was satisfied with.  
    The resulting discussion about trade-offs and the hardship of deciding between quality attributes was on purpose and one of the learning goals. 
    In the ``1-4-all'' discussion part of the second workshop, we asked the participants to explore optimizations to the TD ticket, i.e., the TD documentation, by adding TD attribute fields. % to provide vital information for de\-ci\-sion-making. 

\hypertarget{WS3}{
\subsubsection{Third Workshop -- Topics}}
\label{sec:3WS_topics}
    We informed the practitioners that TD could be prioritized against functional requirements or each other and that prioritization could be a one-time activity or iterative~\cite{lenarduzzi_systematic_2020}. 
    Additionally, we introduced various prioritization factors uncovered by aca\-demia~\cite{Schmid2013, ReboucasDeAlmeida2018c, Tom2013b, rachow_architecture_2022}, including the comprehensive list from the SLR on TD prioritization by Lenarduzzi et al.~\cite{lenarduzzi_systematic_2020} (see~\Cref{sec:Background_Activities}). 
    %new TD activity discussion
    Using the ``1-4-all'' format, we asked the participants to discuss the prioritization factors and formulate additional ones if appropriate.
    The participants discussed which factors are most relevant to their work and which ones they can evaluate or estimate in a reasonable amount of time.
    Finally, we wanted to know the participants' opinions on whether a priority calculation might be feasible or if priorities should be set by the participants' \textit{gut feelings}.

\hypertarget{WS4}{
\subsubsection{Fourth Workshop - Topics}} 
\label{sec:4WS_topics}
    We introduced the repayment categories: ignore, refactor, and rewrite~\cite{Buschmann2011} and the methods explained in~\Cref{sec:Background_Activities} and referenced in~\Cref{tab:approaches}~\cite{Yli-Huumo2016, FREIRE2023, Buschmann2011, Schmid2013, McConnell2008a, Tom2013b, Wiese2022}.
    For the discussion part, we asked the participants to examine if they could uncover further repayment methods, and the participants discussed in which contexts they might use which approach. 

\hypertarget{WS5}{
\subsubsection{Fifth Workshop - Topics}} 
\label{sec:5WS_topics}

    In contrast to the other workshops, we did not use the related work on visualizing TD as inspiration, as they do not visualize backlog tickets but bugs, self-admitted TD, or static analysis metrics, or they visualize the tickets from a proprietary tool.
    However, we presented five TD visualizations found in research studies, as shown in~\Cref{tab:approaches}.
    
    Instead of using the visualizations from research, we developed a TD visualization based on Microsoft PowerBI~\cite{microsoft_powerbi_2025} that can easily be adapted to individual backlog needs. 
    Moreover, PowerBI provides pre-im\-ple\-men\-ted interfaces for Azure DevOps~\cite{microsoft_azureDevops_2025} and Jira~\cite{Jira2022}, which are commonly used backlog tools in the industry. %\footnote{\url{https://www.atlassian.com/de/software/jira}}
    For this purpose, we interviewed the action team, i.e., the two TD managers (P2, P6) of both subteams, the team manager (P1), and the person responsible for quality assurance (P4) before the fifth workshop. % to be able to create a viable visualization of the backlog data. 
     From these interviews, we gathered use cases for visualization and structured them into two categories, resulting in dashboards for the IT management and the development team.
    We found that the questions presented in Table~\Cref{tab:visualization} should be addressed through visualizations.

    We introduced the concept of storytelling with visualizations to exemplify how data-driven storytelling might support decision-making~\cite{gershon_what_2001, kosara_storytelling_2013, airaldi_data-driven_2021}.
    Subsequently, we presented the visualization mock-ups designed by a researcher based on the previous interviews and their resulting use cases.
     We included one visualization idea based on calculating an ROI value as explained in \Cref{sec:method-backlog}.
    %new TD activity discussion
    With this ROI approach, we intended to get feedback from the team on whether calculating a priority would be suitable, assuming the team would make the assumptions rather than the researchers. 
    %We also wanted to get feedback on how to improve this approach if it is found to be unsuitable.
        
    In the discussion part, we questioned whether the presented mock-ups answered the questions comprehensibly and sufficiently. 
    The team also discussed additional use cases and whether some questions or use cases and their corresponding visualizations were superfluous.
    Finally, the participants discussed in which situations, e.g., meetings, they might use visualizations.
    The discussion about and final decision on the adoption of those visualizations is presented in the result~\Cref{sec:Results-Cylce6}.
    In~\Cref{fig:visualizations}, we illustrate all visualizations that were evaluated as suitable after a four-week testing period.

\subsection{Workshop Material}

    More details on each workshop, including the researcher's presentation, the actions taken, and the learnings, are discussed in~\Cref{sec:actioncycles}.
    For reproduction purposes, we have added all slides prepared for the workshops and the initial talk, as well as photographs of flip charts, to the additional material~\cite{AdditionalMaterial}. 
    They include further details on concrete evaluations, actions, measurements, and learnings, and can be exploited to replicate this study.

\section{ACTION CYCLES}
\label{sec:actioncycles}    
    In the following, we introduce the action cycles, which provide details on the adoption of the approaches and the development of new approaches to manage TD.
    Each action cycle starts with a workshop day, which includes the ``evaluation's presentation,'' ``learning,'' ``diagnosing,'' and ``action planning'' phases of the action research method, as shown in~\Cref{fig:Study_Design}. 
    The workshop day is followed by the ``action taking'' and ``evaluation'' phases between two workshops.
    
    In this section, we briefly reference the \textit{diagnosing} phase, i.e., the discussed topics during the workshops as explained in~\Cref{sec:workshops_2_5} for each cycle.
    Subsequently, we focus on the \textit{actions taken} phase in terms of changes to the backlog and changes to the processes.
    Moreover, we present the \textit{learnings \& emerging issues} of each cycle, providing research insights that did not lead to actions. 
    %In the end, we give an additional overview of the TD backlog's numerical development during the action cycles
    The workshop topics, actions taken, and learnings are outlined in~\Cref{tab:actioncycles}.
    The table can be used to get an overview of all cycles and facilitates navigating to the most engaging topics.

    \begin{sidewaystable*}
      %  \centering
      \footnotesize
      %  \landscape
        \begin{tabular} {lllllll}%p{0.5cm} p{0.7cm} p{4cm}  p{4cm} p{4cm}  }
            \hline
            No. & Date & Workshop topics 				& \hyperlink{approachesTbl}{Approaches} from  & Actions taken - backlog & Actions taken - processes & Learnings \& Emerging issues \\
             &  & 			&  TD activity &  &  \\
            \hline
             Intro- & May 2, 2023
                    & TD definition \& TD types     %WS topic
                    & TD definition      
                    &       % Actions backlog
                    & 
                    &  \\
            duction & 
                    & Causes \& consequences     %WS topic
                    & TD identification   
                    &       % Actions backlog
                    & 
                    &  \\
            Talk    & 
                    & Cost \& benefit estimation     %WS topic
                    & TD measurement    
                    &       % Actions backlog
                    & 
                    & \\
            (1hr.)  & 
                    & Prioritization factors     %WS topic
                    & TD prioritization 
                    &       % Actions backlog
                    & 
                    & \\
                    & 
                    & TD backlog \& TAP framework     %WS topic
                    & TD repayment
                    &       % Actions backlog
                    & 
                    & \\
                    
            \hline 
                1   & \hyperlink{WS1}{May 3, 2023} 
                    & Study introduction %WS topic
                    & TD identification 
                    & \hyperlink{tickettype}{New ticket type}       % Actions backlog
                    & \hyperlink{identification}{TD identification by customer visibility}
                    & \hyperlink{tradeOffs}{TD vs. architecture trade-offs} \\ 
                    & 
                    & Recap of the previous presentation  %WS topic
                    &  \& definition
                    & \hyperlink{consolidation}{Ticket consolidation}      % Actions backlog
                    & \hyperlink{TDManager}{TD manager}  
                    &  \\
                    &
                    & TD game     %WS topic
                    & (Multiple for game) 
                    &      % Actions backlog
                    & 
                    &  \\
                    &
                    & Previous TD instances     %WS topic   %of TD incurrence, consequences, repayment
                    &        %Record unrecorded TD items 
                    &            % Actions backlog
                    &
                    &  \\
                    & 
                    & Current TDM process      %WS topic  % actions for identifying TD, dealing with TD, incurring TD 
                    &    
                    &       % Actions backlog
                    &
                    &  \\
                    & 
                    & Wishes and requirements for a TDM process      %WS topic %(Love it -- Change it -- Leave it)
                    & 
                    &      % Actions backlog
                    & 
                    &  \\
                    &
                    & TD manager \& meetings      %WS topic
                    & 
                    &      % Actions backlog
                    & 
                    &  \\
                    
            \hline 
                2   & \hyperlink{WS2}{August 24, 2023} 
                    & TD in mechatronics     %WS topic
                    & TD prevention
                    & \hyperlink{attributes}{TD ticket attributes }      % Actions backlog
                    & \hyperlink{qualityrequirements}{Quality requirements' relevance}
                    & \hyperlink{backlogadmin}{Backlog administration}\\ 
                    &
                    & Quality attributes     %WS topic
                    &
                    & \hyperlink{TalkedaboutTD}{Talked about TD} % Actions backlog
                    & %\hyperlink{queriedbycustomer}{TD repayment queried by the customer }      
                    & \hyperlink{algorithmsteam}{TD in the algorithm subteam} \\
                    &  
                    & Cognitive biases     %WS topic
                    & 
                    &       % Actions backlog
                    &
                    &  \\%\hyperlink{bla}{Separation into two subteams }\\
                    &
                    & TD tickets attributes     %WS topic
                    & TD measurements 
                    &        % Actions backlog
                    &
                    & \\
                    &
                    &      %WS topic
                    & \& documentation
                    &        % Actions backlog
                    &
                    & \\

            \hline 
                3   & \hyperlink{WS3}{February 2, 2024}
                    & Prioritization vs. TD or functional requirements      %WS topic
                    & TD prioritization
                    & \hyperlink{resubmission}{Re-submission date}% Actions backlog
                    & \hyperlink{educatedguess}{Educated guess for priority}
                    & \\
                    &
                    & Prioritize one-time vs. recurring     %WS topic
                    & 
                    &       % Actions backlog
                    & \hyperlink{reorganize}{Reorganize meetings }
                    &  \\ %- separate framework from algorithm team \\
                    &
                    & Prioritization factors      %WS topic
                    & 
                    &       % Actions backlog
                    &
                    & \\ % \\Reorganize meetings - reduce the number of meetings \\ 
                    &
                    & Calculate a priority vs gut feeling
                    &
                    &       % Actions backlog
                    & 
                    &  \\
                    
            \hline 
                4   & \hyperlink{WS4}{April 9, 2024}
                    & TD repayment strategies      %WS topic %- Ignore, Refactor, Rewrite
                    & TD repayment
                    & \hyperlink{BenefitsEtc}{Benefits, drawbacks, risks} % Actions backlog
                    & \hyperlink{pay}{Pay interest}
                    & \hyperlink{centraltendency}{''Central tendency bias''}  \\
                    &
                    &     %WS topic
                    & 
                    &        % Actions backlog
                    & \hyperlink{lowhangig}{Low-hanging fruits}
                    &  \\
                    &
                    &     %WS topic
                    & 
                    &        % Actions backlog
                    & \hyperlink{before}{Repay before a project}
                    &  \\
                    &
                    &      %WS topic
                    & 
                    &       % Actions backlog
                    & \hyperlink{after}{Repay after a project}
                    &   \\
            \hline 
                5   & \hyperlink{WS5}{June 11, 2024}
                    & Visualization from research      %WS topic
                    & TD visualization
                    & \hyperlink{ValuableVis}{Valuable visualizations}      % Actions backlog
                    & \hyperlink{VisSituations}{Situations to use visualizations}
                    & \hyperlink{workProcesses}{Work processes are TDM's basis} \\
                    & 
                    & Story-telling with visualization      %WS topic
                    & \& monitoring
                    &  % Actions backlog
                    & \hyperlink{ROIprio}{ROI-based priority}
                    & \\ %\hyperlink{otherTeams}{Working with other teams}\\
                    &     
                    & Visualization ideas based on interviews   %WS topic
                    &
                    &     % Actions backlog
                    & \hyperlink{guessFactors}{Factors for the \textit{educated guess}}     
                    & \\
                    &     
                    & Visualization ideas based on ROI  %WS topic
                    &
                    &       % Actions backlog
                    & \hyperlink{worstcase}{Worst-case scenarios}  
                    &  \\
                    &     
                    &  %WS topic
                    &
                    &       % Actions backlog
                    & \hyperlink{continuousEffort}{Estimate effort on a continuous scale }    
                    &  \\
                    &     
                    &  %WS topic
                    &
                    &       % Actions backlog
                    & \hyperlink{silos}{Knowledge silos.}    
                    &  \\
            \hline 
            6 Retro-& August 8, 2024
                    &      %WS topic
                    &    
                    & \hyperlink{Revision}{Revising the template}   % Actions backlog
                    & \hyperlink{futureExpectations}{Expectations for the future}
                    & \hyperlink{satisfaction}{Satisfaction with the backlog }\\
            spective&    
                    &     %WS topic
                    &
                    &  \hyperlink{relevantVis}{Relevant visualizations}       % Actions backlog
                    & 
                    & \hyperlink{simlyfiedVis}{Simplyfied visualisations} \\
            \hline
        \end{tabular}
        \caption{%\centering 
        	Systematic overview of action cycles, including workshops, approaches, actions taken, and learnings. %\\
        	(The hyperlinks to the workshops', actions', and learnings' explanations in the text can be used to simplify navigating this paper's results.)}
        \label{tab:actioncycles}
    \end{sidewaystable*}

 \subsection{Initial Situation}
    \label{sec:initialSituation}
    %Zusammenfassung der Anfangssituation
    The team started their TDM establishment without having followed a structured approach beforehand. 
    %In the questionnaire, all participants agreed they did not manage TD. 
    %One participant had not even heard of TD before the study's start.
    According to the questionnaire, all participants agreed they did not manage TD; they had no approaches to avoiding, monitoring, prioritizing, or repaying TD in a structured manner.
    
    In the kickoff discussion, the participants explained various situations where TD consequences hit them.
    For instance, the company opted to use an outdated CPU because waiting for the new version would have delayed the project by two additional months. 
    Meanwhile, the demand in the market for the new feature was intense. 
    Consequently, the team is now facing numerous performance and maintenance issues due to the choice of CPU. 
    The team mentioned that they had set up SonarQube to identify code debt, but they had no process for dealing with the results. 
    Still, most of the TD identification was done by the developers who were stumbling across TD-related issues,  e.g.,  \textit{``P2: \ldots during implementation or when a complaint or a request actually comes in from some project. \ldots I have a ticket, and I'm supposed to expand or implement something, and \ldots it's working, but then I look again, [and] I'd have to incur TD to get it working, and I don't like doing that. And then I look at why, and then I might realize, yes, there is [already] TD somewhere.''} %, whether consciously or unconsciously.''}
    
    In these cases, they either repaid the TD immediately or discussed it with colleagues and created tickets in their backlog, but they had no structured process for TD documentation.

    P8 noted: \textit{``That's a general question about how big the TD is. If it's a small problem that you can fix easily, then you would probably fix it yourself directly, but if it's really something bigger, then it should definitely be brought up for discussion.''}
     The manager mentioned: \textit{``So I think we do discuss it and maybe even write a ticket, but we don't do it systematically. We don't have a specific type of ticket where we say what alternative there is, how much effort would it take to remove it and so on.''}
     Finally, P5 remarked: \textit{`` \ldots P2 has already done that somewhere. \ldots We discussed this in the group, that this is somehow recorded separately [by each participant].''}
     P3 and P6 complained about TD prioritization: \textit{``P3: `Prioritization is not always ideal, but\' P6: `I also find it really difficult when it comes to prioritization because I don't think we can prioritize a lot of TD on our own. So, a lot of TD that comes to us may or may not have an impact on others, and then you have to talk to other teams first.''}
    
     Regarding conscious TD incurrence, the manager explained: \textit{``so without judging it negatively now - so that we then have the pressure from other project teams from outside \ldots we have to see: Ok, do we have the time to build it in properly and build it in properly together with them and do we have to take on this technical debt, because there's this time pressure from outside.''}
    Concerning the wishes for the project, the team particularly mentioned the importance of communication and better documentation of TD. 
    P8: \textit{``We had just discussed in our team that communication is always a very important point, and it's also very important here because you're often so deep in your framework and then discuss with your colleagues what really makes sense or are you actually already too deep in your mess  \ldots ''}
     P2: \textit{``I would also like detailed descriptions [of TD], I already mentioned that earlier, that you can really... that you can then see what the consequences are if you can estimate them at the time; that you give a bit of context around why this is technical debt at all.''}
    The topic cache on the flip chart stated similar topics, e.g., general TD decision-making (incl. a guideline for management decisions), the TD ticket quality, and continuous communication, e.g., with colleagues that function as their customers, enterprise architecture management, and other TDM stakeholders.

    %However, those usually were ignored when deciding on future development.
    %Moreover, they were still figuring out their general work processes.
    %The meeting structure and ticket status transitions repeatedly led to misunderstandings.

 \subsection{First Cycle - TD Identification \& Current TDM process}
    \label{sec:Results-Cylce1}

    The interventions started on May 2nd, 2023, with a one-hour talk to inform the team about the basic intricacies of TD. 
    This talk included inspiration on TD from various studies~\cite{Cunningham1992, Avgeriou2016a, Li2015, wiese_it_2023, Wiese2022, Schmid2013, Tom2013b, ReboucasDeAlmeida2018c, rachow_architecture_2022, Kruchten2012f} introduced in~\Cref{sec:Background}. 
    The talk was directly followed by the first one-day workshop on May 3rd, 2023, which  started the action research and  served as the kick-off regarding the change management process.
    The agenda of the first workshop is explained in~\Cref{sec:workshops_1}.  
    % Additionally, throughout the first workshop, we collected topics to analyze and discuss during the whole study in a topic cache (see~\cite{AdditionalMaterial}.
    % This cache included topics like TD decision-making, TD identification, a guideline for management decisions, or defining stakeholders for TDM.  
    % We reviewed the cache at the end of the study.
     
    \paragraph{Diagnosing} $~$
    
    In this workshop, we diagnosed the current TD management process in three subsequent discussions: 
    We asked the participants to describe concrete TD instances, elaborate on how they react to TD items, and discuss their wishes and requirements for a TDM process as explained in~\Cref{sec:workshops_1}.
    %one-time Actions taken
    
    \paragraph{Actions Taken -- Backlog} $~$
    
    \hypertarget{tickettype}{\textit{New ticket type.}}
    The team created a new ticket type for TD tickets in the backlog.
    This ticket type did not yet incorporate any TD-specific attributes.
    Initially, it was simply a means of identifying and tagging TD items in the backlog.
    
    \hypertarget{consolidation}{\textit{Ticket consolidation.}}
    The team identified all TD items in the backlog and created TD tickets for all TD items they had in mind but had not yet recorded.
    The advantage of this approach was that at least all TD items that had been hindering the team at one point, i.e., all effective TD items~\cite{Schmid2013}, were recorded.
   % Regarding ongoing actions, they decided to record all TD items they came across in the future as tickets in the backlog.
    At this point, 127 tickets were in the backlog, of which 109 had the status ``new,'' i.e., they were not yet refined.
    The action team identified 65 tickets as TD items.
    %At the end of July, there were 74 TD tickets in the backlog, 14 of which were not in the status ``new'.

    \paragraph{Actions Taken  -- Processes} $~$

    \hypertarget{TDManager}{\textit{TD Manager.}}
    During the first workshop, we established a TD manager, as explained in~\Cref{sec:workshops_1}.

    \hypertarget{identification}{\textit{TD identification by customer visibility.}}
    We tried to distinguish TD tickets from functional or bug tickets using the four colors of the backlog established by Kruchten et al.~\cite{Kruchten2012a}.
    However, the team created software for other IT teams within the company. 
    Thus, the question arose whether an item is TD when it is not visible to the other teams or only if it is not visible to the end customer.
    
     For example, the following discussion with their manager arose: \textit{``P1: `But if it's not working as expected right now, isn't that a bug, or am I misunderstanding something here?'' P2: `Well, it's a bug in their [the other team's] application then, but I mean, we have made life difficult for them; why aren't we doing that [avoiding the bug] already? We can do that implicitly when we set up our controller  \ldots ' P1: `Ah, and they have to make a workaround somehow so that it works properly for them, right?'''} 
    The team first decided to adopt the product owner's viewpoint and interpret the other teams as their customers.
    Yet during the study, the team embraced a simplified method by inspecting whether the TD items created interest for the team as described by Avgeriou et al.~\cite{avgeriou_technical_2023}.
    Furthermore, they considered whether negotiations with management are necessary for the implementation of this ticket.

    % \begin{figure}
    %     \centering
    %     \includegraphics[width=0.3\textwidth]{img/before_consolidation.png}
    %     \caption{Backlog shares before consolidation}
    %     \label{fig:before_consolidation}
    % \end{figure}
    % \begin{figure}
    %     \centering
    %     \includegraphics[width=0.3\textwidth]{img/after_consolidation.png}
    %     \caption{Backlog shares after consolidation}
    %     \label{fig:after_consolidation}
    % \end{figure}
    
    % \begin{figure}
    %     \begin{tabular}{@{}cc@{}}
    %         \subfigure[before consolidation]
    %             {	\label{fig:before_consolidation}
    %                 \includegraphics[width=0.22\textwidth]{img/before_consolidation.png}
    %                 %\Description{A significant difference can be seen.}
    %             }&
    %         \subfigure[after consolidation]
    %             {	\label{fig:after_consolidation}
    %                 \includegraphics[width=0.22\textwidth]{img/after_consolidation.png}
    %                 %\Description{A difference between groups can be seen, but that difference is not significant.}
    %             }
    %     \end{tabular}
    %     \caption{Backlog shares before and after the ticket consolidation}
    % \end{figure}

    \paragraph{Learnings \& Emerging Issues} $~$

%         \textit{Impact of existing processes.} The team was newly set up, and many processes were not yet established. 
%         This became particularly apparent during the various meetings, as some team members were unclear on these meetings' purposes, which also overlapped. 
%         % This led to subsequent problems: 
%         % (1)~Some tickets were discussed multiple times. 
%         % (2)~Other tickets, particularly old ones, vanished from view. Many TD tickets were among these tickets. 
%         % (3)~
%         Moreover, some tickets were created after the implementation had already started. %, e.g., during the planning meeting. 
%         This means that the implementation was not discussed among the team, creating a risk of TD incurrence due to bad decision-making.
%         Consequently, the team decided to avoid implementing a task without first creating and refining a suitable ticket for it.
% %        \item 

        \hypertarget{tradeOffs}{\textit{TD vs. architecture trade-offs.}}
        Initially, the practitioners had problems distinguishing between TD and architecture trade-offs and identified all trade-offs as TD. 
        We explained the difference, and the team decided to classify an issue as TD only if they had a more suitable solution alternative in mind.

 \subsection{Second Cycle - TD Documentation and Prevention}
    \label{sec:Results-Cylce2}
    
        \begin{figure*}[t]
        \centering
        \includegraphics[width=\textwidth]{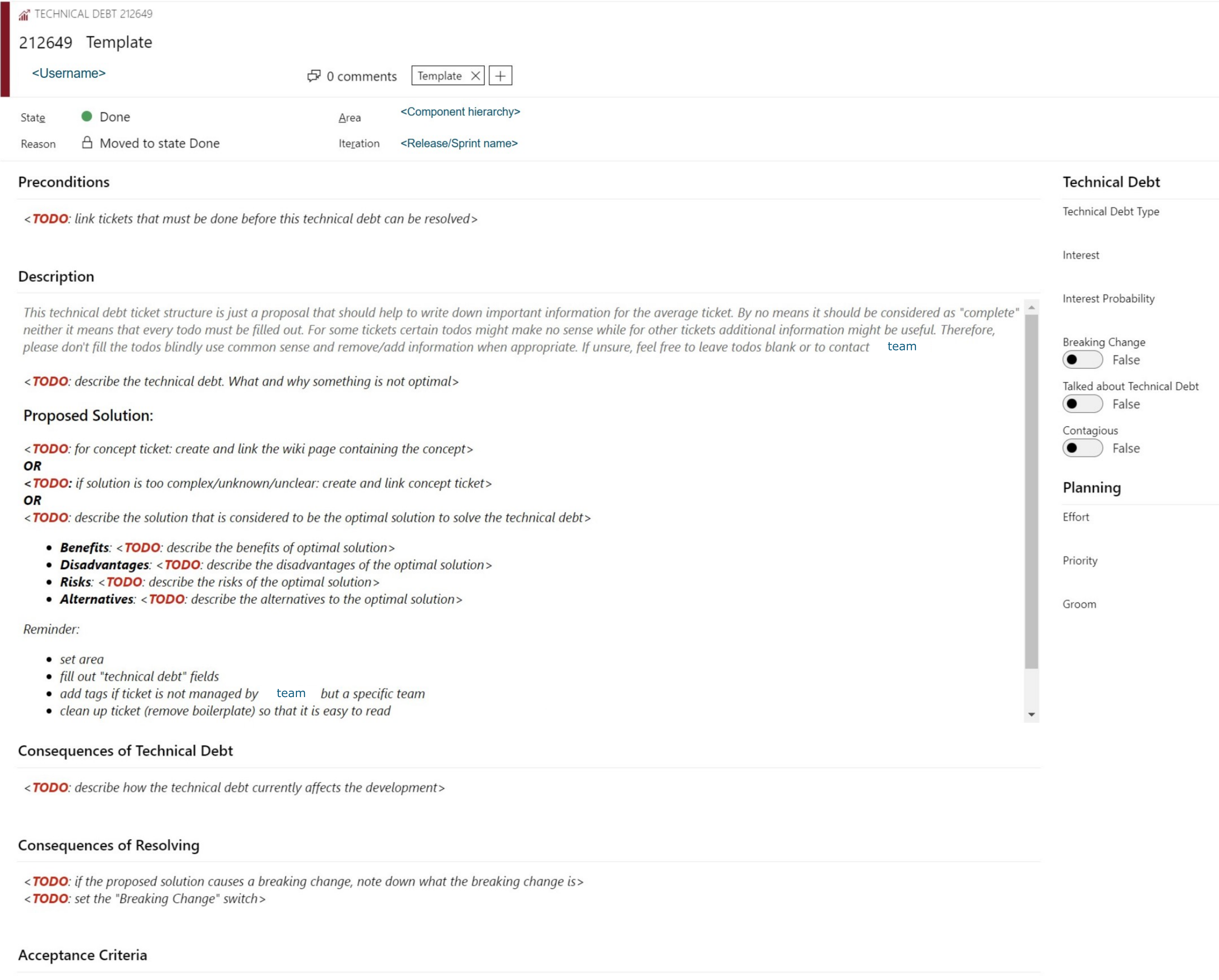} % 0.8textwidth bringt etwa 3-4 Zeilen
        \caption{Screenshot of the team's TD ticket template}
        \label{fig:TD_Ticket}
        %\vspace{-5mm}
    \end{figure*}
    
  %  \paragraph{Workshop/Diagnosing} $~$
    %new TD activity info
    The second workshop on August 24th, 2023, was on the topics of TD incurrence \& prevention by optimizing TD decision-making and TD documentation to improve the TD ticket. 
    
    \paragraph{Diagnosing} $~$
    
    In this workshop, we discussed methods to prevent TD (e.g., use of quality attributes, cognitive biases) and goals for TD documentation discussions as explained in~\Cref{sec:2WS_topics}.

    \paragraph{Actions Taken  -- Backlog} $~$
    
    \hypertarget{attributes}{\textit{TD ticket's attributes.}}
    The TD ticket was enhanced by adding attributes for the TD item's interest, interest probability, contagiousness, whether the repayment results in a breaking change, the affected component, the consequences of the TD item, and the consequences of repaying it as explained in~\Cref{sec:method-backlog}. 
    The final version of the team's TD ticket can be seen in~\Cref{fig:TD_Ticket}.
    This figure and an excerpt from the team's wiki describing these attributes are included in the additional material~\cite{AdditionalMaterial}.
    
    \hypertarget{TalkedaboutTD}{\textit{Talked about TD Checkbox.}}
    The practitioners determined that creating a TD ticket type is not sufficient.
    Instead, changing all ticket types and adding necessary information for deci\-sion-making is essential to avoid TD. % debt, particularly.
    As a first approach, they added a \textit{Talked about TD} checkbox to all ticket types, i.e., user story tickets, bug tickets, features, epics, etc.
    This checkbox should remind the team to evaluate the risk of incurring TD when implementing the respective ticket. %alternative solutions and possible risks of the preferred solution during the refinement meetings. 
    %CUT candidate
    %To the authors' knowledge, this idea is a novelty in academic TD knowledge even though it is similar to the idea of architecture decision records~\cite{VANHEESCH2012795}.

    \paragraph{Actions Taken  -- Processes} $~$

    \hypertarget{qualityrequirements}{\textit{Quality requirement's relevance.}}
    %For the algorithms team, the quality requirements were essential but generally not provided from the outset.
    The workshop high\-light\-ed the relevance of quality attributes, especially for the algorithms subteam.
    For them, the quality requirements were essential but generally not provided from the outset.
    Instead, they were assumed by single developers, which led to misunderstandings with the customers and, consequently, to TD in previous implementations.
    The team chose to discuss quality requirements with each other and the customers before starting the development process from now on.
    % Third, it became evident that implementing one algorithm ticket usually took weeks, which made it impossible to overlook the whole implementation process. 
    % Consequently, the developers started splitting their tickets into various steps and utilizing the hierarchy of initiatives, features, epics, user stories, and tasks provided by their backlog tool.

    % \hypertarget{queriedbycustomer}{\textit{TD repayment queried by customers.}}
    % We discussed tickets that might be interpreted as TD but were also queried by the customers, e.g., performance optimizations or the utilization of new technologies, like artificial intelligence.
    % These tickets did not face the typical problems of TD items, i.e., their implementation did not have to be argued with customers. 
    % Thus, the team decided to keep these tickets as functional requirements.

    \paragraph{Learnings \& Emerging Issues} $~$
        
        \hypertarget{backlogadmin}{\textit{Backlog administration.}}
        The team had to discuss changing the ticket design, e.g., adding the attributes, with the team responsible for the backlog software.
        Particularly, adding the \textit{Talked about TD} checkbox to all ticket types took another two months of organization and negotiation. 
        Our previous research showed that not adding another tool for TDM is important~\cite{wiese_it_2023}.
        Yet, adapting existing tools to TDM might be a problem that many companies face. 
        If the whole company uses the backlog tool, there might be restrictions on its adaptation.
        Besides, adding value calculations or rule-based pre-filling of attributes might be challenging for proprietary tools.
    
  %      \item 
        % \textit{The team's separation into two subteams.}
        % The framework subteam mainly led the TDM process establishment and TD ticket development because they faced the most TD.
        % As a result, the algorithm developers had trouble identifying their TD and relating their work to the developed TDM approaches. 
        % We noticed that every subteam needed initial guidance on TD identification and TDM attributes (e.g., cost or prioritization factors) and that the description in the wiki was not self-ex\-plan\-atory.
        % Therefore, an extra meeting to discuss TD in the realm of the algorithm team was scheduled.
        % In this meeting, questions similar to those of the framework team arose. %, e.g., the difference between architecture trade-offs and TD.
        
        \hypertarget{algorithmsteam}{\textit{TD in the algorithm subteam.}}
        The algorithms were developed in an unrestricted environment, often as part of research and development work, e.g., by experimenting and developing prototypes.
        %Comparing solution alternatives, i.e., algorithm implementations, beforehand was complicated. 
        Accordingly, the amount of effective TD in the algorithm team seemed very low. 
        We decided to focus the research study on the framework subteam and use the algorithm team's insights as supplementary input.
        However, we also established a dedicated TD manager for this subteam.
        %In discussions with the algorithm developers, it became apparent that

 \subsection{Third Cycle - TD Measurement \& Prioritization}
    \label{sec:Results-Cylce3}
    
    The third workshop took place on February 14, 2024. 
    We initially planned to focus this cycle on TD visualization (now the fifth cycle).
    However, prioritizing TD was already a frequent discussion topic in the previous cycles, which led us to adjust the initial plan according to the practitioners' needs.
    
    \paragraph{Diagnosing} $~$
    
    %new TD activity info 
    We informed the practitioners about TD prioritization factors and discussed the relevant factors and processes for the team as presented in~\Cref{sec:3WS_topics}. 
    
    \paragraph{Actions Taken  -- Backlog} $~$

    \hypertarget{resubmission}{\textit{Re-submission date.}} 
    The team recognized that tickets got lost in the backlog, i.e., some tickets, particularly TD tickets, have a low priority and stay at the end of the backlog list until they are forgotten. 
    The team decided to add a \textit{re-submission date} to all ticket types, which represents an iterative prioritization approach~\cite{lenarduzzi_systematic_2020}. 
    This date ensured no ticket was lost in the backlog and also indicated the ticket's urgency.
    Urgent tickets would be rescheduled soon, while less urgent tickets would be postponed for longer periods.
    %CUT candidate:
    This also had the advantage of separating priority from urgency, following the idea of the Eisenhower matrix~\cite{baer_eisenhower_2014,wikipedia_eisenhower_2025, miro_eisenhower_2025}, creating a quadrant of urgency and importance to decide a task's future. %\footnote{\url{https://miro.com/strategic-planning/what-is-an-eisenhower-matrix/}}
    The team really valued this approach as mentioned in one of the meetings: \textit{``P4: `One really notices that there's at least a bit of a flow now, right? It's getting better every time.' P3: `And as a result, nothing is forgotten, which I think is nice because they [TD tickets] come up every now and then. P4: `There's still room for improvement, but we're on the right track.'''}

    \paragraph{Actions Taken  -- Processes} $~$

    \hypertarget{educatedguess}{\textit{Educated guess for priority.}}
    In the workshop's discussion regarding TD prioritization, we asked the team whether they would like to calculate the priority based on the recorded attributes. 
    The team, particularly the algorithm developers, who all had a mathematical background, did not think it was possible to calculate a priority that builds on all relevant attributes.
    Some attributes, e.g., the developers' availability and the requirements of the teams using their framework, change quickly or cannot be recorded.
    The team decided the priority should be an \textit{educated guess}, i.e., an estimation based on knowledge of the underlying ticket's attributes.
    This should be sufficient and much better than the mere \textit{gut feeling} they used before.
    Fittingly, the \textit{resubmission date} was used for continuous re-evaluation of this guess.

    \hypertarget{reorganize}{\textit{Reorganize meetings.}}
    In the ``learning'' phase, it became apparent that many TD-related problems originated from poorly designed work processes, such as communication issues or the number and rationale of meetings.
    Consequently, another significant change was to reduce the three meetings (planning, grooming, and refinement) and separate the meetings of both subteams starting in March 2024. 
    The framework subteam decided to have two meetings: refinement and planning. 
    Initially, they had three half-hour refinement meetings each week to review the TD backlog and discuss each ticket and its attributes, including the new \textit{re-submission date}.
    The algorithms subteam decided to have one meeting that combined refinement \& planning.
    They also decided not to use the \textit{re-submission date}, as their TD backlog was smaller and organized hierarchically.

   % \paragraph{Learnings \& Emerging Issues} $~$
    
    % \textit{Avoid micro-management.}
    % The team discussed how to make ticket recording easier.
    % As a result, they decided to change the contagiousness five-step scale to a checkbox.
    % They did not want to add additional attributes to the TD ticket for prioritization.
    % Nonetheless, to better overview all tickets, the team decided to emphasize creating meaningful ticket titles.
    % They also planned to improve overall ticket quality, e.g., by better explaining the requirements, advantages, and disadvantages.

 \subsection{Fourth Cycle - TD Repayment}
    \label{sec:Results-Cylce4}

    %\paragraph{Workshop/Diagnosing} $~$
    
    %new TD activity info 
    In the fourth workshop, held on April 09, 2024, we informed the participants about the different methods of addressing TD.
    
    \paragraph{Diagnosing} $~$

    We introduced the repayment categories:  ignore, refactor, and rewrite~\cite{Buschmann2011} and the respective approaches and discussed the adoption of these as explained in~\Cref{sec:4WS_topics}. 
    
    \paragraph{Actions Taken  -- Backlog} $~$

    \hypertarget{BenefitsEtc}{\textit{Benefits, drawbacks, and risks. }}
    Our observations uncovered that the \textit{Talked about TD} checkbox led the team to discuss alternative solutions. 
    However, discussions about the benefits, drawbacks, and risks of the solutions were rare. 
    In contrast, the TD-SAGAT evaluation showed that these aspects affected many team members' minds (see~\Cref{fig:alternatives}).
    As the discussion might help share all information across a team, they decided to add a reminder to each ticket type to discuss these attributes by creating a text template  providing space to elaborate on each attribute (functionality provided by the backlog tool).
    %Unfortunately, implementing this reminder took them until the fifth cycle because they had to negotiate this with the manager and the backlog administrators.  

    \paragraph{Actions Taken  -- Processes} $~$
    
    %one-time Actions taken
    \hypertarget{pay}{\textit{Pay interest.}}
    The participants decided to pay the interest if the priority and urgency were low and the effort was high. However, they did not want to close these tickets but to keep them open in the backlog with a far-off \textit{re-submission date}.

    \hypertarget{lowhangig}{\textit{Low-hanging fruits.}}
    The team planned to repay \textit{low-hang\-ing fruits}, i.e., TD tickets with a high priority and low effort, as bridging tickets, i.e., if a developer found some spare time to bridge between functional requirements.
    
    \hypertarget{before}{\textit{Before a project.}}
    The participants decided to repay TD before a project, which was their most valued repayment method.
    In their backlog's ticket hierarchy, they considered a feature, epic, or user story ticket to be a project.
    They planned to use this method by reviewing the TD tickets related to the respective ticket's affected component. 
    
    \hypertarget{after}{\textit{After a project.}} 
    The team intended to adopt the project-related TD repayment after a project. 
    As the team does not utilize projects, they considered initiatives (see~\Cref{sec:method-backlog}) as projects for this repayment method.
    This means the team planned to repay all TD tickets incurred during an initiative's implementation before it was closed. 
    However, as initiatives were rarely used, they were not able to use this method before the study's end.

    %No one-time actions were planned as the team decided which methods to use during the workshop and directly took action.

    \paragraph{Learnings \& Emerging Issues} $~$

        \hypertarget{centraltendency}{\textit{``Central tendency bias.''}}
        We presented a calculated priority based on \textit{interest risk} as explained in \Cref{sec:method-backlog}. 
        Comparing this priority with the \textit{educated guess} priority demonstrated that team members seldom gave very low or very high priorities. 
        We attribute this to the ``central tendency bias'' of preferring medium values instead of extreme ones~\cite{pimentel2019some}.
        We discussed this observation with the team members, who determined to use ``very low'' and ``very high'' more boldly and frequently.
        However, the tendency remained until the end of the study.

\subsection{Fifth Cycle - TD Monitoring \& Visualization}
    \label{sec:Results-Cylce5}

    The fifth workshop took place on June 11, 2024, and was on the topic of visualization. 
   %  In contrast to the other action cycles, we did not use the related work on visualizing TD as inspiration, as they do not visualize backlog tickets but bugs, self-admitted TD, or static analysis metrics, or they visualize the tickets from a proprietary tool.
   %  Instead, we developed a TD visualization based on Microsoft Power BI that can easily be adapted to individual backlog needs. 
   %  Moreover, PowerBI provides pre-im\-ple\-men\-ted interfaces for Azure DevOps and Jira\footnote{\url{https://www.atlassian.com/de/software/jira}}, which are commonly used backlog tools in the industry. 
    
   %  \paragraph{Diagnosing - Interviews} $~$
    
   %  To create a viable visualization of the backlog data, we first interviewed the action team, i.e., the two TD managers of both subteams, the team manager, and the person responsible for quality assurance, before the fifth workshop. 
   %  From this, we gathered use cases for visualization and structured them into two categories, resulting in dashboards for the IT management and the development team.
   % % The following use cases for a visualization of TD tickets were identified:
   %  We identified that the questions listed in~\Cref{tab:visualization} should be answered by visualizations.

    \paragraph{Diagnosing} $~$
    
    %new TD activity info 
    We presented TD visualizations found in research studies as shown in~\Cref{tab:approaches} and visualization mock-ups, which we created for the use cases that emerged from the interviews.
    In the discussion part, we examined the suitability of the mock-ups as explained in~\Cref{sec:5WS_topics}. 

    \paragraph{Actions Taken  -- Backlog} $~$

    \hypertarget{ValuableVis}{ \textit{Valuable visualization.}}
     We presented mock-ups for all visualizations mentioned in~\Cref{tab:visualization} (visualizations see additional material~\cite{AdditionalMaterial}).
    %the visualization's implementation that the participants identified as feasible and valuable in \Cref{tab:visualization}  (``Adoption/Usage'' column) and in~\Cref{fig:visualizations}. %even though they finally only used a few.
    The team decided that a button to  switch between \textit{educated guess} and \textit{ROI-based} priorities should be included for all visualizations that are based on priority.
    Notably, they discarded V6 (TD tickets that must be repaid by a specific person) from the outset because a visualization for a single team member would not be compliant with the works council.
    Based on the outcomes of the discussion, a researcher further improved all visualizations to make them better understandable. 
    The implementation was completed on July 23, 2024, and the visualizations have been used in team meetings since then to assess their suitability. 
    
    \paragraph{Actions Taken  -- Processes} $~$
    
    \hypertarget{VisSituations}{\textit{Situations to use visualization.}}
    The algorithm subteam decided not to use visualization as they had too few TD tickets (two at that time) for a visualization to be valuable.
    The framework subteam decided to use the visualization at the beginning of each refinement meeting.
    However, they changed their decision during the ``action taking'' to using the visualization in more specific situations, e.g., when looking for \textit{low-hanging fruits} or TD items that affect a functional requirement. 

    \hypertarget{ROIprio}{\textit{ROI-based priority.}}
    As intended, the \textit{ROI-based} priority was intensely discussed, and its limitations were laid out.
    While the team members initially were reluctant, they acknowledged during the discussion that a calculated value might have benefits,  e.g., summarizing at least some relevant attributes. 
    However, the underlying values, e.g., interest and interest probability, had to be chosen more carefully, keeping in mind their effects on the priority. 
    They did not consider these effects during the previous refinements. 
    Thus, they planned to re-evaluate the tickets with a significant difference between the \textit{educated guess} and the \textit{ROI-based} priority. 
    They decided to use visualization V7, i.e.,~\Cref{fig:discrepancy},  for this purpose. 
    
    Analyzing the differences between \textit{educated guess} and \textit{ROI-based} priorities led to three additional learnings and changes in their processes to improve their \textit{educated guesses}:  
    
    \hypertarget{guessFactors}{\textit{(1)~Factors for the \textit{educated guess}. }}
    %\textit{Including effort in \textit{educated guess}. }
    While the calculation of the \textit{ROI-based} priority included the effort of a ticket, the team members usually did not include the effort in their \textit{educated guess}. 
    They decided that including the effort was unnecessary as the visualization V9 of \textit{low-hanging fruits} would give them the required result without changing their estimations. 
    %\textit{Contagiousness and breaking change.}
    In contrast to that, the team included the information on contagiousness and breaking change in the \textit{educated guess} while these factors were not included in the calculation of the ROI.
    Moreover, the team discussed that attributes such as interest and interest probabilities might change over time, e.g., due to a component being used more frequently.
    They included these expectations for the future in their \textit{educated guess}.

    \hypertarget{worstcase}{\textit{(2)~Worst-case scenarios.}}
    Initially, the team always estimated a worst-case scenario for interest probability, which might have led to higher values in the calculated priority. 
    However, they did not specify whether the \textit{educated guess} should be based on a median or worst-case scenario.
    We identified this issue by observing the team comparing the \textit{educated guess} and \textit{ROI-based priority} using visualization V7 (priority discrepancies).
    P2 and P3 discussed a TD item where the documentation should be updated, including an example code snippet: \textit{``P2: Is that really painful? P3: \ldots Yes, it's actually painful - especially when people are new to it, \ldots it's not easy to understand: why we do it and why-- P2: It can take a day if people don't understand it, can't it? P3: Yes, it's complicated. P2: And that happens a lot? P3: \ldots That's the daily work when developing algorithms. \ldots P2: Yes, but is that high or medium? Because once they've done that, it's no longer so painful for them. That's the worst-case scenario here. [pause] P2: Yes, we can set the high [previously medium] for my sake. I mean, at least then it's done. Otherwise, we'll just push this [TD item/backlog item] around.''}
    In this case, the team reprioritized the TD item based on the worst-case scenario aligning it with the worst-case scenario used for the interest.  
    Confronted with these observations, the team decided to align their estimations and (re-)evaluate all values using medium expectations. % Also, they decided that probability should be interpreted more like a frequency.

    \hypertarget{continuousEffort}{\textit{(3)~Estimate effort on a continuous scale.}}
    The participants realized that the effort scale, where an effort of five meant anything above five, was too coarse and unsuitable for a calculation.
    This problem was supported by the meeting's observations: \textit{``P5: I've still added ‘5’ to the effort, I don't know exactly how long it will actually take. But it will probably take a week or so. I don't know exactly what the '5' is somehow.''} 
    They decided to still interpret the numbers for the effort as person-days, but to allow any values, including values above five. % from one to five and, from then on, in steps of five. 
    %The bigger steps for higher values were chosen due to the rising uncertainty of estimations.

    \hypertarget{silos}{\textit{Knowledge silos.}}
    Comparing TD-SAGAT results with the observations, it became apparent that participants considered many prerequisites but did not talk about them. 
    The participants explained that this might have been because only persons knowledgeable about a ticket completed the survey. 
    They conceded that knowledgeable persons tend to expect other participants to know the same or consider knowledge sharing unimportant, which is known as the ``curse of the knowledge bias''~\cite{zalewski_cognitive_2017}.
    Thus, they did not discuss prerequisites while still considering them.
    One striking example was the moderator of most of the meetings, i.e., the TD manager, who wrote down every piece of relevant information in the backlog item and, by this, learned details about every ticket and became a kind of a \textit{knowledge silo}.
    In contrast, the non-moderating team members might not have put enough thought into tickets. 
    Thus, the team decided to rotate the moderation of Scrum meetings to counteract the creation of such \textit{knowledge silos}.  % such as refinement or planning. 

    %one-time Actions taken

    \paragraph{Learnings \& Emerging Issues} $~$

        \hypertarget{workProcesses}{\textit{Work processes are TDM's basis.}}
        During all cycles, it became increasingly apparent that all work processes influenced the establishment of a TDM process, and, in turn, establishing TDM affected all work processes. 
        %Furthermore, all team members must have the same understanding of these processes, which was not always the case in this team.
        On the downside, establishing a TDM process required much effort, which included identifying and discussing all work processes,  e.g., to avoid TD, they improved the quality of all their backlog items by adding reminders to discuss alternatives, drawbacks, and risks. 
        On a very positive note, however, the team improved all their work processes,  not only TD-related process artifacts, e.g., they optimized their meeting structure and avoided \textit{knowledge silos}.

    \begin{figure} [t]
        \begin{tabular}{@{}c@{}}
                \\
                \subfigure[Number of participants satisfied  with the TD items in the backlog]
                {	\label{fig:BacklogSatisfaction}
                    \includegraphics[width=0.49\textwidth]{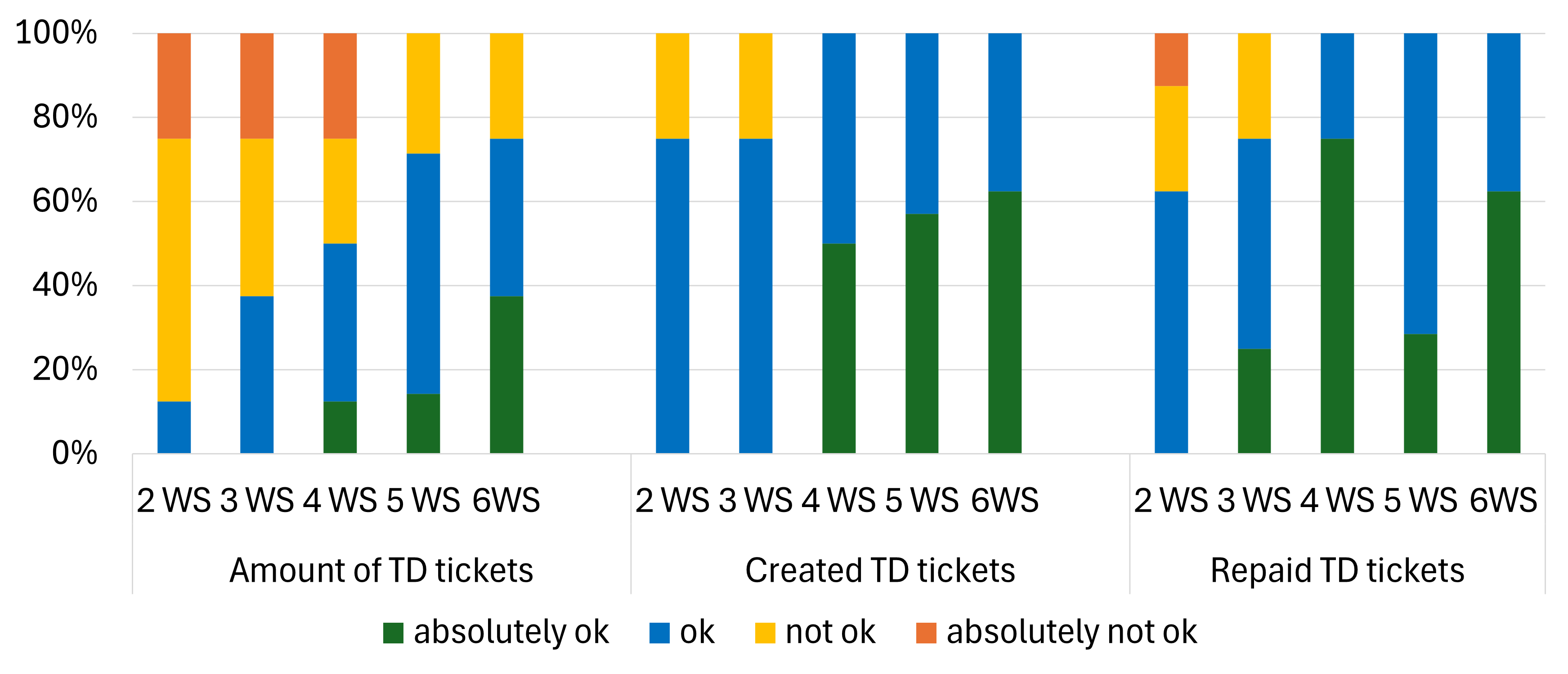}
                }\\
                \subfigure[Number of participants satisfied  with the TDM process]
                {	\label{fig:ProcessSatisfaction}
                    \includegraphics[width=0.40\textwidth]{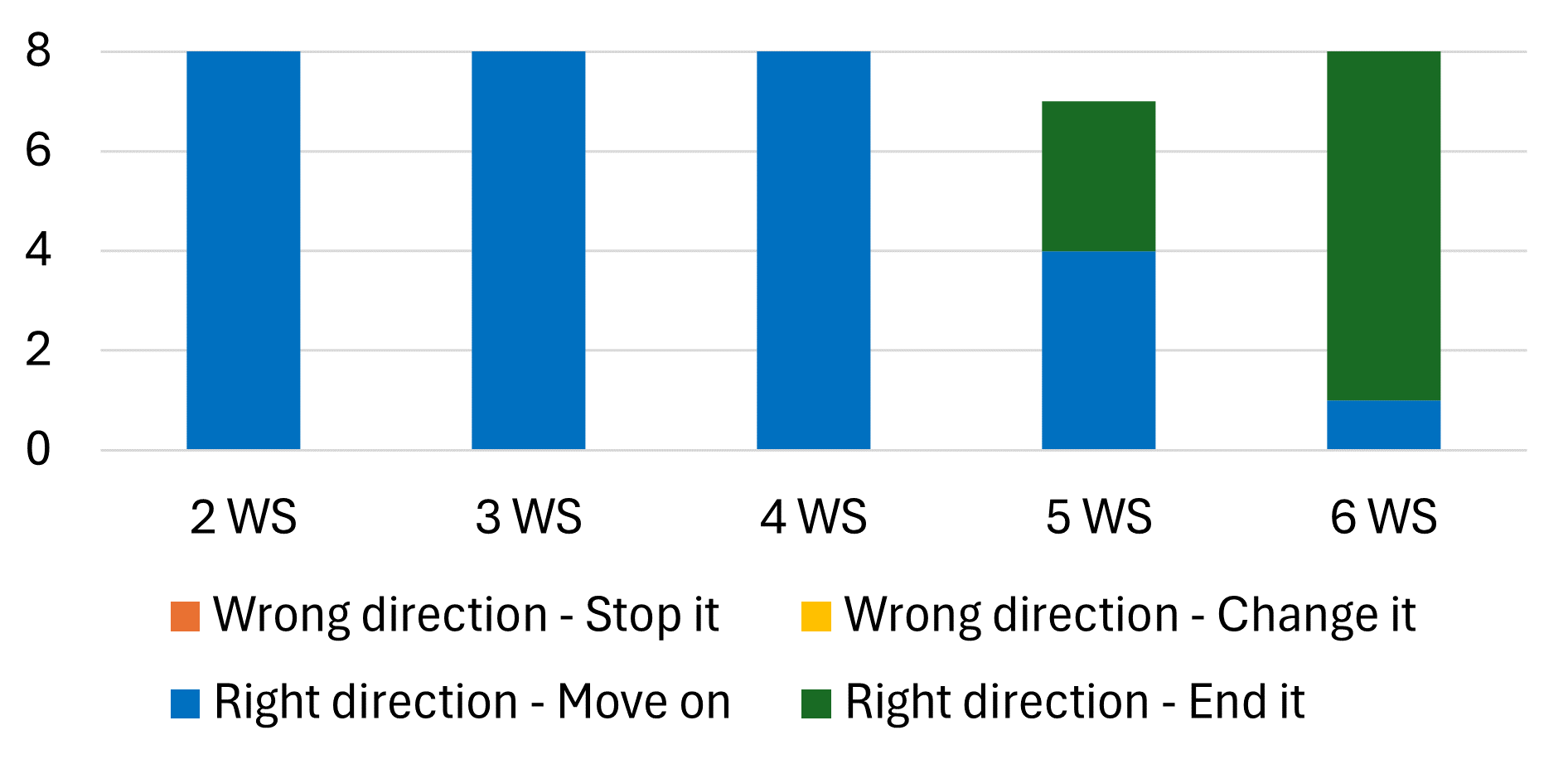}
                }
        \end{tabular}
        \caption{ Participant's satisfaction with the study's results per workshop}
        \label{fig:satsifaction}
    \end{figure}
    
 \subsection{Sixth Cycle - Retrospective}
    \label{sec:Results-Cylce6}

    The first and last retrospective meeting took place on August 8, 2024. 
    In this format, we did not present new ideas but focused on discussing the evaluations from the previous cycle. %observations and   

    \paragraph{Actions Taken  -- Backlog} $~$

        \hypertarget{Revision}{\textit{Revising the template.}}
        The reminder for “alternatives, benefits, drawbacks, and risks” was valuable for the discussion, but filling out the information in the ticket was deemed micro-management.
        The benefits of a ticket were usually obvious and part of the regular description.
        Evaluation of alternatives was already done when using the \textit{Talked about TD} checkbox or as part of a concept ticket.
        Similarly, explaining the effort was deemed micro-management:
        \textit{``P3: `But [we're] getting into micromanagement now.' P2: `Yes, that's the case now -- I don't know -- I wouldn't actually write such obvious things as effort. That it's an effort.'''} 
        Thus, the team decided to focus the discussion on drawbacks and risks as they provide essential information often underestimated or ignored, e.g.,  \textit{``P2: The disadvantage is that we'll probably end up with dependencies. Because you would probably come to the conclusion that you want to take [tool name] command line arguments to do that.''}

        \hypertarget{relevantVis}{\textit{Relevant visualizations.}}
        After the team had used the improved visualization for four weeks, we discussed each visualization's suitability.
        The team identified that V1 (\Cref{fig:influence}  -- TD tickets influencing a user story) and V9 (\Cref{fig:lowhanging}  -- Low Hanging Fruits) were the most valuable visualizations. 
        However,  the team had difficulties understanding the original V1 visualization (see additional material~\cite{AdditionalMaterial}) and, thus,  decided to use a list of TD tickets for a given user story with a filter for the user story instead (see \Cref{fig:influence}).
         The original versions of V3 and V4 were similarly hard to understand (see additional material~\cite{AdditionalMaterial}). 
        Again, the team decided to use a simplified visualization, i.e., a list of TD items whose repayment is highly profitable (see \Cref{fig:profitable}). 
        This visualization has not yet been used, but the manager plans to use it for argumentation in the future. 
        %The manager informed us that this is only due to the short usage time and that they plan to use this visualization in suitable situations for argumentation in the future. 
        %Again, the participants determined that a simplified visualization was sufficient. 
        %We changed the visualization to a list of tickets with a column showing the number of months in which the TD is profitable. % (Grafik folgt). 
        %The figures of the filtered list (V1 and V3) are part of the additional material~\cite{AdditionalMaterial}).
        V7 (\Cref{fig:discrepancy} -- Priority discrepancies between \textit{educated guess} and \textit{ROI-based priority}) was used extensively to optimize the \textit{educated guess} but might not be used in the long term.
        V8 (TD items per priority and re-submission date)  was not beneficial at all and was removed from the dashboard.
         V2 (\Cref{fig:AmountEffort} -- Amount and effort over all TD items) and V5 (\Cref{fig:openovertime}  -- Effort of TD tickets over time) are used for orientation but not as a part of any TDM process steps.
        %All other visualizations were still considered intriguing but will not be used regularly.

    \paragraph{Actions Taken  -- Processes} $~$

        \hypertarget{futureExpectations}{\textit{Expectations for the future.}}
        We confronted the team with our thought that including future expectations in their \textit{educated guess} contradicts the basic idea of the \textit{re-submission date}.
        When using a \textit{re-submission date}, it seems more reasonable to base all attributes on the current situation, set an appropriate \textit{re-submission date}, and change attributes when the \textit{re-submission date} is reached.
        We noticed this issue while watching the team compare the \textit{educated guess} and \textit{ROI-based priority} using visualization V7 (priority discrepancies). 
        For example, the team discussed whether a component should be able to use more than two kernels, which is currently not possible. P2 and P3 rediscussed the priority: \textit{``P2: In any case, I think we've prioritized it correctly. It needs to be high because we're going to need it soon. Yes? P3: Hmmm P2: That means we're not going to reprioritize it now.''} 
        In this case, the team included the future expectations in their educated guess.
        In the same meeting, they also discussed the inclusion of a block size parameter in a component. In this case, a workaround exists that all developers are aware of. Again, P2 and P3 discussed this: \textit{``P2: Yes, so it doesn't hurt at the moment, because people are used to setting it for every module. P3: Yes, yes, exactly. Otherwise, we usually did it \ldots with a wrapper \ldots  P2: Well, yes, I would still say that it remains medium. I wouldn't downgrade it now or anything. P3: No, we still have the other priority with the groom date [= re-submission date] \ldots''}
        In this case, the team used the \textit{re-submission date} to account for the future expectations. 
        The team regarded our thoughts as valuable, revised their previous decision, and included future expectations only through a shorter \textit{re-submission} period.

    \paragraph{Emerging Issues/Learning} $~$

%CUT candidate:
        % \textit{Identifying the affected component.}
        % The team informed us that it is sometimes hard to identify the proper affected component as the architecture is less than optimal and is going through a re-design.
        % Thus, they used more general components (e.g., ``system components'' or ``cross-cutting concerns'') in some cases, which made it harder to identify TD affecting these functional tickets. 
        % This problem should be resolved when the architecture is optimized.

        \hypertarget{satisfaction}{\textit{Satisfaction with the backlog.}}
        While \Cref{fig:openovertime} and \ref{fig:closedopen_count} show that the number of TD items increased over time,  the satisfaction with the backlog improved (\Cref{fig:BacklogSatisfaction}).
        %One interesting discussion topic was whether the satisfaction with the backlog means that more or less TD are incurred or repaid.
        The participants realized that the improved satisfaction values and the success of the change process are not rooted in the amount of TD but in the ability to manage the backlog and distinguish relevant from irrelevant TD.

        \hypertarget{simlyfiedVis}{\textit{Simplyfied visualisations.}}
        We identified that visualizations, even though they were built on the basis of the participant's requirements, were sometimes hard to understand.
        In multiple situations, the participants preferred more straightforward visualizations, e.g., simple lists for V1 and V3 and a simple pictogram for the contagiousness of a ticket in the list presentation. It would be interesting to further analyze whether this is due to the stakeholders being technical stakeholders and whether business stakeholders might prefer other visualizations.

 \subsection{Terminating the Action Research}
    \label{sec:terminating}
    \Cref{fig:ProcessSatisfaction} shows that the participants wanted to end the action research after the fifth cycle, i.e., the sixth workshop in August 2024.
    At this point, the participants were content with the incurrence and repayment of TD (\Cref{fig:BacklogSatisfaction}).
    The two participants who were dissatisfied with the overall TD amount in the backlog reported that they saw  no reason to change the process further but to utilize the process to reach an acceptable amount of TD in the long term.
    Furthermore, we revisited the topic cache from the kickoff workshop and examined whether relevant topics remained unresolved. 
    The team concluded that all significant topics had been addressed.

\section{RESULTS}
\label{sec:Results}
In the following subsections, we present the answers to our three RQs on the adopted and newly created approaches (\Cref{sec:Results-RQ1}), the effort establishing a TDM process takes (\Cref{sec:Results-RQ2}), and the effects the establishment has on TD awareness (\Cref{sec:Results-RQ3}).

\subsection{RQ1: Which TDM approaches are useful for the practitioners of our case company?}
\label{sec:Results-RQ1}
\noindent

    The action research determined the useful approaches, i.e., the approaches adopted in the ``action taking'' phases. 
    We presented the details on and rationales for the adoption of the approaches in the previous~\Cref{sec:actioncycles}.
    %An overview of all introduced approaches with their adoption and remarks is presented in~\Cref{tab:approaches}.  %answering RQ 1.1. 
    \Cref{tab:approaches} gives an overview of all approaches and their adoption (usage) or rejection by the practitioners for each TD activity answering RQ1.
    % To answer RQ 1.2., using a \textit{Talked about TD} checkbox and a \textit{resubmission date} in the backlog were approaches newly developed and employed by the participants (answering RQ 1.2.). 
    % Furthermore, using a \textit{low-hanging-fruits} visualization and employing an \textit{educated guess} for prioritization were new additions to their TDM process.
    % The utilization of a return on investment calculation (\textit{ROI-based} prioritization) was a new approach but only adopted partly, i.e., to optimize the team's process but not as part of the process. 
    Below, we summarize the useful approaches (\Cref{sec:Results-RQ1-approaches}) and provide an overview of the quantitative development of the team's backlog (\Cref{sec:Results-RQ1-backlog}).
    
\subsubsection{Adopted and Developed Approaches}
\label{sec:Results-RQ1-approaches}
    \paragraph{TD Identification}   
        %The TD identification did not change much, and most TD identification was done manually by the developers.
    %Partly adopted 
    % Not adopted
        The team struggled to identify what items should be considered TD based on their visibility~\cite{Kruchten2012a}. 
        Also, the ``Dagstuhl definition''~\cite{Avgeriou2016a} did not entirely fit their needs. 
    %Adopted
        The most feasible definition was to identify an item's interest and specify it as TD if interest occurred~\cite{avgeriou_technical_2023}.
    %New approach
        In addition, they used the question of whether this ticket's implementation needs to be negotiated with the management as guidance.
        While the team had set up a SonarQube installation, they only used this as a supplementary tool for TD identification.
    
    %Adopted
    \paragraph{TD Documentation / TD Backlog (qualitative)}  
        For TD documentation, the team used its existing backlog. 
        TD items were recorded as tickets using a newly created ticket type for TD as proposed by Kruchten et al.~\cite{Kruchten2012a} and Rubin~\cite{scrum2012practical}.
        We found that improving the templates for backlog items and adding new ones is a favorable way to establish a TDM process, as the team already organized all its work around these backlog items.
        Thus, this approach achieved a high level of acceptance among all team members.
        %The item types, i.e., templates, were used to remind the team members of the information, i.e., ticket's attributes, needed for successful decision-making and, thus, to keep the awareness of certain topics high.
        Additionally, the backlog items can be used to enable communication with business stakeholders by providing information necessary for their decision-making and subsequently visualizing the items in an aggregated way. 
       % The attributes of the TD ticket type, as well as other ticket types, were added to support other TD activities, such as prevention and prioritization.
    % A major issue was      
    
        % We identified that the restricted use of a company's backlog tool was a significant issue.
        % Moreover, adding value calculations or rule-based pre-filling of attributes was challenging for the proprietary tool.
        % We assume that this might be a reason generally impeding TDM approaches by practitioners. % on a team level.
        % A solution would be for the backlog tool vendors to integrate TD ticket types by default.
        
    %New approach
        %Tickets of all types were discussed and refined in a weekly refinement meeting. 
        %A new process was set up to refine newly found TD items as well as all backlog ticket types by using a \textit{re-submission date}. 
        A new approach was using a \textit{re-sub\-mis\-sion date} to iteratively adapt the attributes to new circumstances. 
        By this, the team ensured that no items got lost in the backlog. 
        They could safely set \textit{re-submission dates} far away, e.g., when using the repayment method ``waiting'' when they expected a potential TD item to sort itself out over time. 
        %The \textit{re-submission date} is a new approach created by the practitioners and not based on suggestions by the researchers.

    \paragraph{TD Prevention}
        Regarding TD prevention, the team's first action was to record and refine tickets before starting their implementation, i.e., they were ``following well-defined project processes'' as mentioned by the InsighTD project~\cite{Freire2020a, Perez2021a, Rios2020}. 
        %We did not present this information from the outset but brought this to the team's attention during the ``learning'' phase. 
        %This aligns with the TD prevention method ``following well-defined project processes'' uncovered in the InsighTD project~\cite{Freire2020a, Perez2021a, Rios2020}. 
        Another new approach was using a  checkbox they named \textit{Talked about TD} in every ticket type to remind them of the danger of incurring TD. 
        %This is a similarly new approach created by the practitioners.
        %In the data analysis in~\Cref{fig:alternatives}, we could see that the button immediately led to more thoughts of alternative solutions options in October 2023. 
        Additionally, the team added a text template to all ticket types to record alternatives, benefits, drawbacks, and risks, as presented in the work of Borowa et al.~\cite{borowa_debiasing_2022}.
        %As expected, this also led to an increase in the corresponding prerequisites in our data analysis~\Cref{fig:alternatives} in August 2024.

    \paragraph{TD Prioritization} 
        % TD prioritization was a seemingly new notion of TD to the team. 
        % They valued the new input and added multiple attributes for evaluation and prioritization to the TD ticket type, e.g., interest, interest probability, breaking change, contagiousness, and consequences.
        The most relevant prioritization approaches  were the evolution-based prioritization~\cite{Schmid2013} and a cost-based prioritization~\cite{Tom2013b, McConnell2008a} to identify \textit{low-hanging fruits}.
        These considerations led to a priority based on an \textit{educated guess} rather than pure \textit{gut feeling}.
        
        The new approach of calculating an ROI for priority based on the underlying attributes was met with resistance from the participants.
        They conceded that using an \textit{ROI-based} priority might support the priority estimation if the underlying factors are set carefully and with their effect on the ROI in mind. 
        However, they deemed the ROI insufficient to determine the overall priority, as information on contagiousness, risks (breaking change), and developer availability are not included in the ROI but are essential factors for priority estimation. 
        Finally,  the participants used the \textit{ROI-based} priority to analyze and improve their \textit{educated guesses} in the fifth cycle.
        %During the meetings, the researchers 
        %CUT: We observed that this led the participants to change the attributes in various cases, but they rarely reevaluated their \textit{educated guess}.
        %CUT: We interpret this as an anchoring bias, the preference for the first given value~\cite{borowa_debiasing_2022}. 
        Notably, the participant did not adopt the idea of prioritizing TD against functional requirements~\cite{lenarduzzi_systematic_2020}.
        %CUT: We assume this was because they were under little pressure to process functional requirements. % and were free to decide which tickets to process.
        %The \textit{re-submission date} enabled changing the priority iteratively~\cite{lenarduzzi_systematic_2020}.    
        %Consequently, in the last workshop, the team decided not to include future expectations in the priority and to use the \textit{re-submission date} date to manage changing priorities. 

    \paragraph{TD Repayment} 
        The chosen TD repayment approaches were strongly related to the prioritization approaches.
        The team identified TD items that might impede a functional requirement by examining TD tickets in the same component. 
        This approach equals a mix of ignoring tickets that do not impede functional requirements~\cite{FREIRE2023} and evolution-related repayment before an initiative, feature, or user story's implementation~\cite{Schmid2013}. 
        %They planned to use this method on an initiative level (i.e., the highest hierarchy level in Azure DevOps), but they rarely used initiatives for their daily work.
        An initiative would be their equivalent of a project, but they rarely used initiatives for their daily work. Thus, repaying TD incurred by a project after that project's deadline~\cite{Wiese2022} was not adopted. 
        Additionally, they repaid TD tickets they deemed as \textit{low-hanging fruits}, which equals the cost-based approach~\cite{Tom2013b, McConnell2008a}.
        Finally, they used the \textit{re-submission date} to wait for a planned rewrite or a system's discontinuation  without losing information about the TD item. %, i.e., when TD items are solved because the affected system or component is discontinued or rewritten.
        Remarkably, the team did not adopt the idea of a repayment quota~\cite{McConnell2008a, Wiese2022}. % CUT-Discussion:, which might be because the team was free to decide which tickets to process.

    \paragraph{TD Monitoring/Visualization} 
        % Regarding the TD Visualization \& Monitoring, we did not use the related work on visualizing TD as inspiration as they do not visualize backlog tickets but bugs, self-admitted TD, or static analysis metrics, or they visualize the tickets from a proprietary tool.
        % Instead, we developed a TD visualization based on Microsoft PowerBI that can easily be adapted to individual backlog needs. 
        % Moreover, PowerBI provides pre-im\-ple\-men\-ted interfaces for Azure DevOps and Jira\footnote{\url{https://www.atlassian.com/de/software/jira}}, which are commonly used backlog tools in the industry. 
    
        %First, the participants identified that visualization is only reasonable if the amount of TD items is within an unmanageable range, which was not the case for the algorithm subteam. 
        %For the framework subteam, t
        The most relevant visualizations were the ones that support the chosen repayment methods, i.e., the \textit{low-hanging fruits} (V9, \Cref{fig:lowhanging}) and the TD items influencing a user story (V1, \Cref{fig:influence}).
        %CUT candidate
        The visualization presenting the difference between \textit{educated guess} and \textit{ROI-based} priority (V7) was used to fine-tune the \textit{educated guess}. 
        The \textit{ROI-based} visualization (V3) was deemed the most relevant from an IT management perspective, but it still has to prove its practical usability as, unfortunately, no situations for its use arose during the study.
        
        %For some visualizations, the team found them to be relevant for the IT management, e.g., when negotiating about whether to repay a TD item(\Cref{fig:profitable}). 
        %Still, they were not actively used by the IT management.
        %Some visualizations, such as the one of the \textit{re-submission date}  (\Cref{fig:resubmission}) or the timeline of opened and closed TD items  (\Cref{fig:closedopen}), were not actively used.

    \paragraph{TD Manager} 
        The researchers proposed a new work role for a \textit{TD manager}, which was essential for the study to have a contact person within the team. 
        More importantly, this person's task was to continue the process sustainably after its establishment.
        This is similar to the idea of a ``TD champion''~\cite{jaspan_defining_2023}.  
        %CUT-discussion: A senior developer should perform this role but not in a full position.

    \subsubsection{TD Backlog (quantitative)}
    \label{sec:Results-RQ1-backlog}

        \begin{figure} [t]
            \begin{tabular}{@{}c@{}}\\
                    \subfigure[Open TD items over time ]
                    {	\label{fig:openovertime}
                        \includegraphics[width=0.48\textwidth]{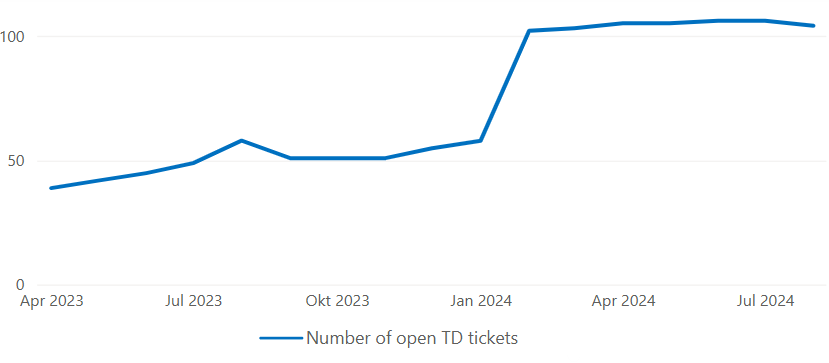}
                    }
                    \\
                    \subfigure[Opened and closed TD items over time]
                    {	\label{fig:closedopen_count}
                        \includegraphics[width=0.48\textwidth]{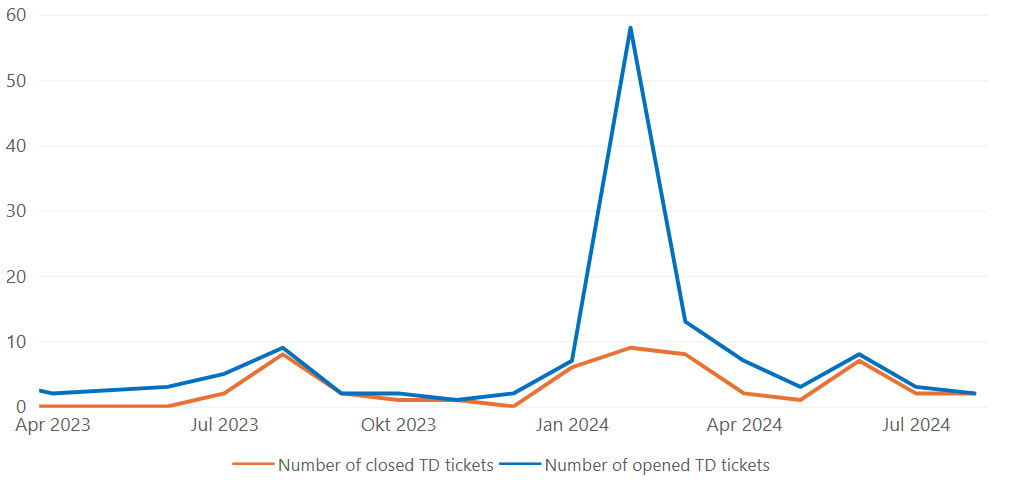}
                    }
            \end{tabular}
            \caption{Backlog Development}
            \label{fig:backlog}
        \end{figure}

        After the consolidation of the backlog, the backlog comprised \BacklogitemsWhenTagged backlog items, of which \TDitemsTagged were identified as TD items. 
        %Five of those \TDitemsTagged TD items have previously been marked as bugs.
        During further work with the backlog, the team deleted (not closed!) \TDitemsDeleted items that were duplicates or related to technology they no longer supported in November 2023.
        This led to a reduced total of \TDitemsTaggedExculdingDeleted instead of \TDitemsTagged TD items at the beginning of our study.
        \Cref{fig:openovertime} shows the amount of TD tickets over time and \Cref{fig:closedopen_count} the number of opened and closed TD tickets.
        The spike in the number of opened tickets in February 2024 was due to the refactoring initiative at that time.
        The team recorded \TDitemsOpenedFeb small refactorings as TD tickets. 
        However, they were only able to repay a few of them in the given time.
        At the end of our study, this led to an increased total of \TDitemsFinal open TD tickets.
        %This shows that the establishment of the TDM process overall led to a higher number of TD items.
        However, apart from the spike from the refactoring initiative, the number of opened and closed TD tickets is similar.
        This means that new TD tickets were created, but nearly the same number of TD tickets were continuously repaid.

    %CUT candidate:
    \begin{absolutelynopagebreak}
        
    \begin{framed}
    \textbf{RQ1.1:}
        Using a backlog for TDM, particularly in combination with a \textit{re-submission date}, is immensely valuable not only for monitoring but also for preventing TD.
        The approach by Schmid~\cite{Schmid2013} to prioritize and repay TD based on the affected system's evolution seems to be the most intuitive for practitioners.
        A TD manager, i.e., a TD champion~\cite{jaspan_defining_2023}, plays an essential role in maintaining the TDM process.
        %Visualizations were particularly beneficial to identify \textit{low-hanging fruits.}
        % Additionally, we identified that participants' biases, such as the ``central tendency bias'' and the ``anchoring bias'' influenced the TDM process at salient points. 
        % We suggest further researching such biases and developing methods to overcome them.
        % A final question arose during the results' analysis that we cannot answer with our research but that seems interesting: 
        % Is the importance that research places on decision-making using quality attributes echoed in practice, or should research focus more on other decision-making factors?
        To answer RQ 1.1.~\Cref{tab:approaches} presents an complete overview.
    \end{framed}
    \begin{framed}
    
        \textbf{RQ1.2:} Using a \textit{Talked about TD} checkbox and a \textit{resubmission date} in the backlog were approaches newly developed and employed by the participants. 
        Using a \textit{low-hanging-fruits} visualization and employing an \textit{educated guess} for prioritization were new additions to their TDM process.
        Using an ROI calculation (\textit{ROI-based} prioritization) was a new approach but only adopted partly, i.e., to optimize the team's process but not as part of the process. 
    \end{framed}

    \begin{framed}
        \textbf{Supplementary finding:} While the amount of TD items in the backlog did increase during the study, the participants were more content with the backlog at the end of the study. 
        Presumably, this shows that the ability to manage TD items is more relevant to practitioners than actually paying them back.
    \end{framed}
    \end{absolutelynopagebreak}

    \subsection{RQ2: How much effort are the practitioners of our case company willing to invest in a TDM process?}
    \label{sec:Results-RQ2}
\noindent

    \begin{figure} [t]
        \begin{tabular}{@{}c@{}}
                \\
                \subfigure[Mean time investment (hrs.) over all participants]
                {	\label{fig:effortAll}
                    \includegraphics[width=0.48\textwidth]{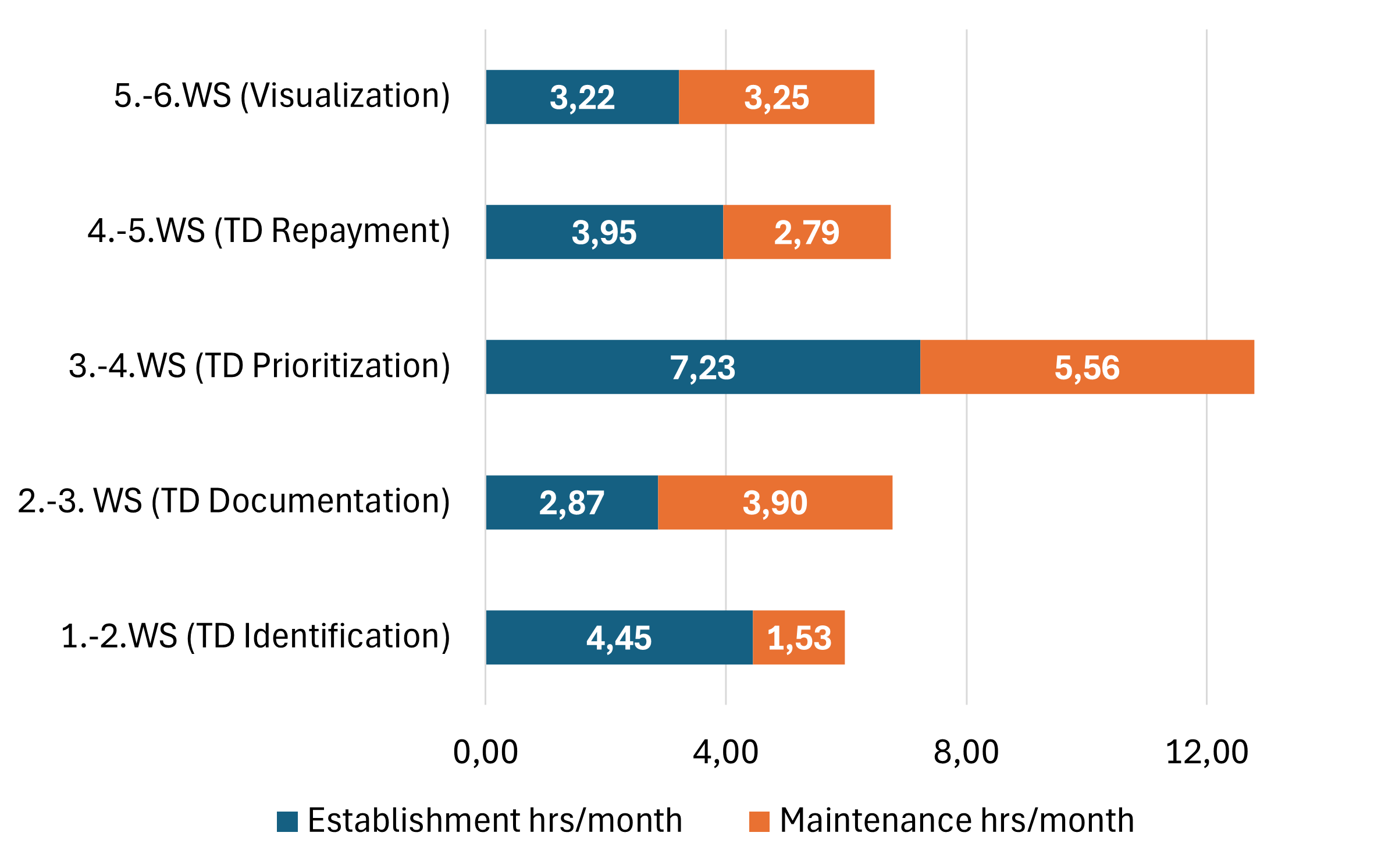}
                }\\
                \subfigure[Time investment (hrs.) of the TD manager]
                {	\label{fig:effortTDM}
                    \includegraphics[width=0.48\textwidth]{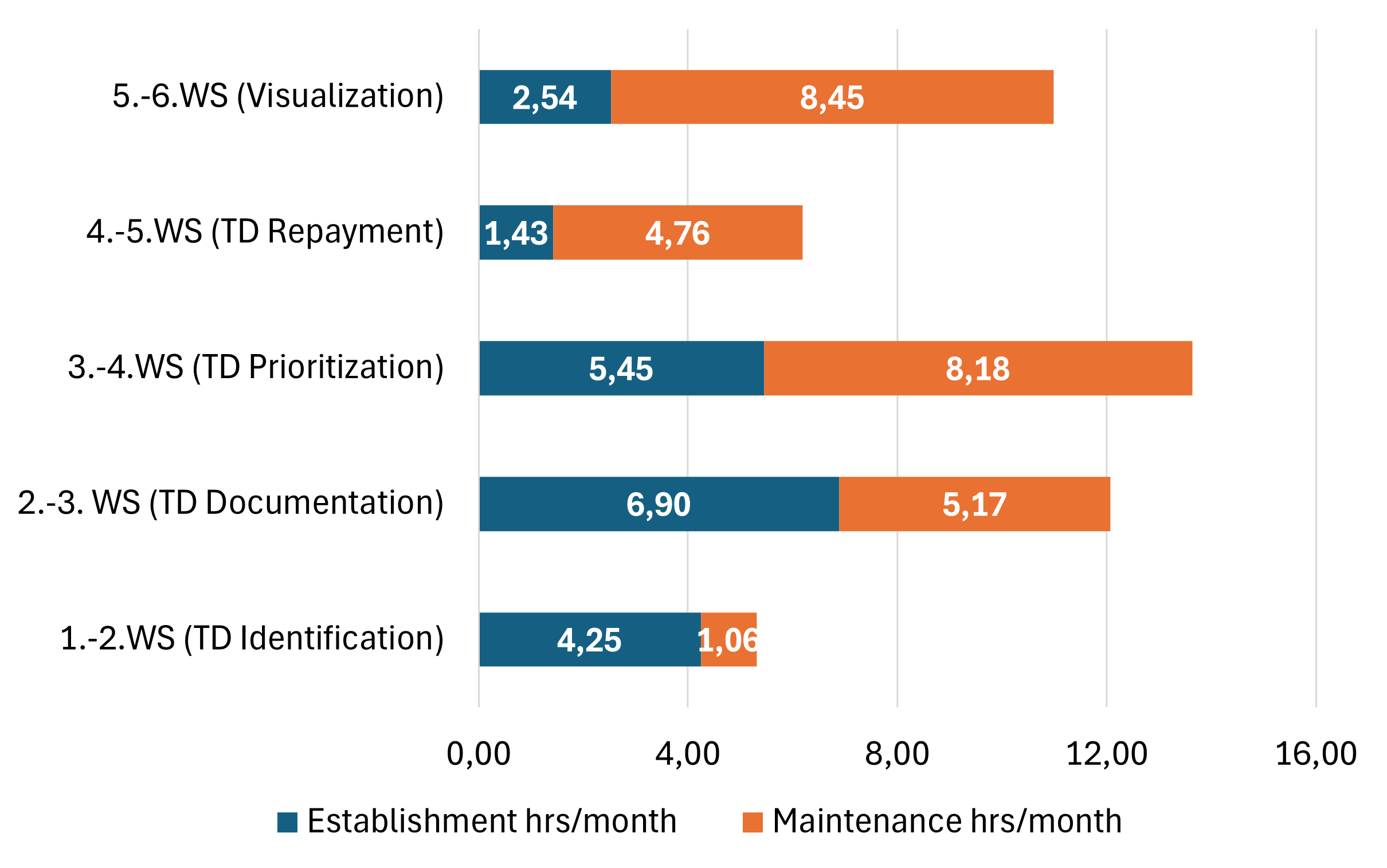}
                }
        \end{tabular}
        \caption{ Participant's effort for establishment and maintenance of the TDM process (excl. TD repayments during this time) (WS -- workshop)}
        \label{fig:effort}
    \end{figure}

    In the questionnaires, we asked the participants to estimate their effort invested in the initial establishment and the continuous maintenance of a TDM process per cycle.
    Notably, the requested effort does not include the effort to repay TD but just the overhead that a process and the establishment of this process creates. 
    We broke the provided estimations down per month for each cycle, and the results are shown in~\Cref{fig:effort}. 
    Across all cycles, the participants invested \MeanEstablishmentEffort hours per month for the establishment and \MeanMaintenanceEffort hours per month to maintain the TDM process.
    This is less than 3\% of the working hours for each, assuming an 8-hour workday and 22 workdays per month.
    
    We found that TD prioritization required the most effort, which is understandable, as all tickets were reviewed to record the cost and prioritization attributes.
    Similarly, the visualization cycle consumed more time because the \textit{educated guess} and the \textit{ROI-based} priority were compared.
    We assume that some participants evaluated the effort for the review as maintenance as it took place during the regular refinement meetings, explaining the higher maintenance share in this cycle.
    %\Cref{fig:effort} shows that the real effort is not as high as some might expect.
    %Assuming an 8-hour workday and 22 workdays per month, the participants invested less than 3\% each in establishing and maintaining the TDM process.

    Notably, the invested effort highly depends on the participant. 
    In our case, the algorithm developers did not invest much time, while, e.g., the \textit{TD manager} invested a high amount of time for obvious reasons.
    In the end, the TD manager needed  less than 5\% of their time, which, nonetheless, should be taken into account when assigning other work packages to the TD manager.

    \begin{framed}
        \textbf{RQ2:} 
        In the mean, the participants invested less than 3\% of their time in establishing and maintaining the process, respectively. 
        The TD manager continuously invested less than 5\% of their time in maintenance.
    \end{framed}

 % \subsection{RESULTS ACROSS ALL CYCLES / EVALUATING}
 %    \label{sec:ResultsAllCycles}

 \subsection{RQ3: Is the team members' awareness of TD in decision-making situations improved sustainably by the TDM process' establishment?}
     \label{sec:Results-RQ3}
\noindent

    This section presents the evaluation results we collected to assess the participants' TD awareness. 
    For this purpose, we analyzed whether the participants considered relevant information (prerequisites) when making decisions (\Cref{sec:Results-RQ3-requirments}), and the unconscious incurrence and correct identification of TD (\Cref{sec:Results-RQ3-incurrence}).

    \begin{figure} [t!]
       %\begin{tabular}{@{}ccc@{}}
        \begin{tabular}{@{}c@{}} 
                \subfigure[Questionnaire: \% of participants stating to having considered prerequisite during the last action cycle]
                {	\label{fig:WSQuestionnaire_Alternatives}
                    \includegraphics[width=0.48\textwidth]{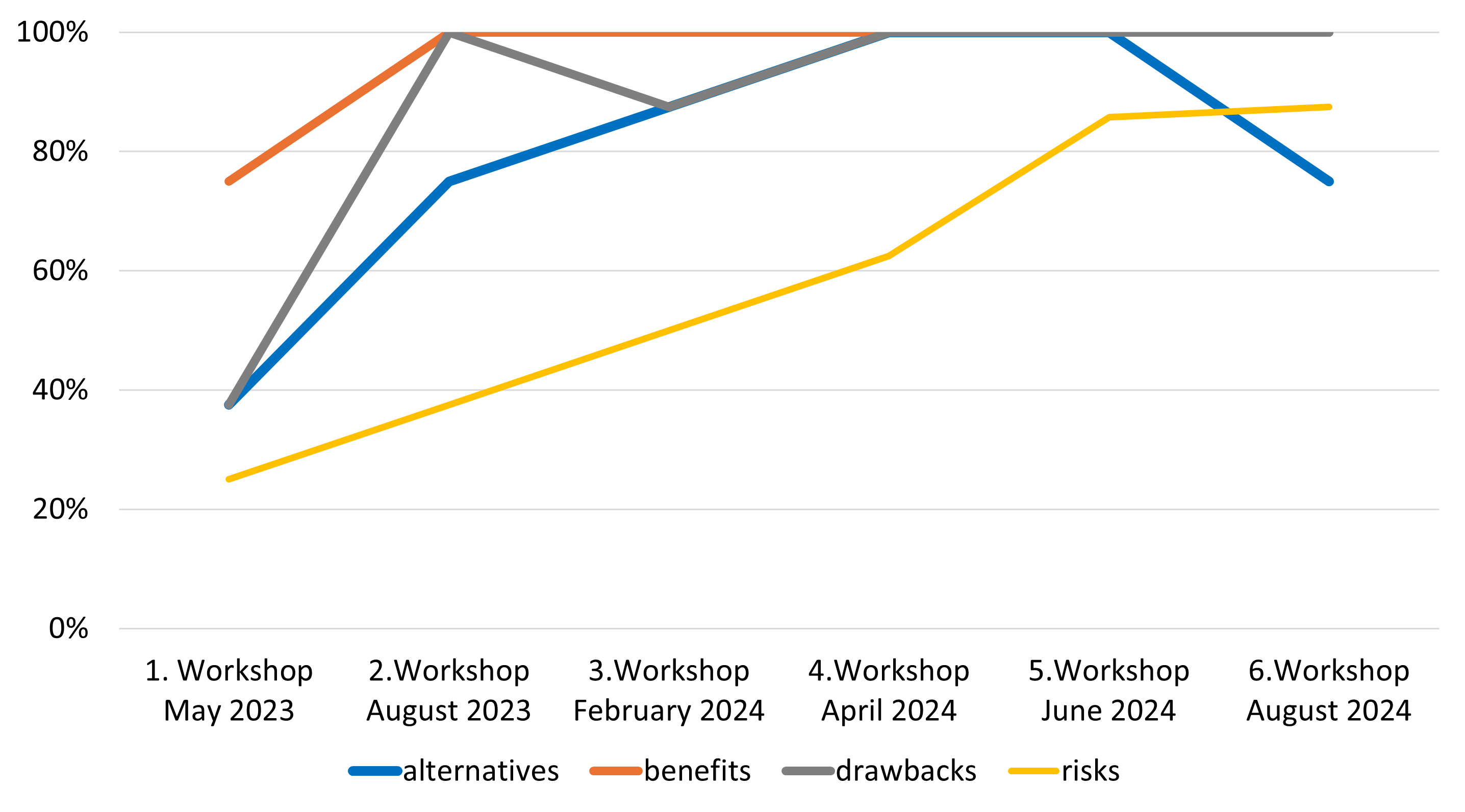}
                }\\
                \subfigure[Observations: \% of ticket where prerequisites were mentioned ]
                {	\label{fig:Observation_Alternatives}
                    \includegraphics[width=0.48\textwidth]{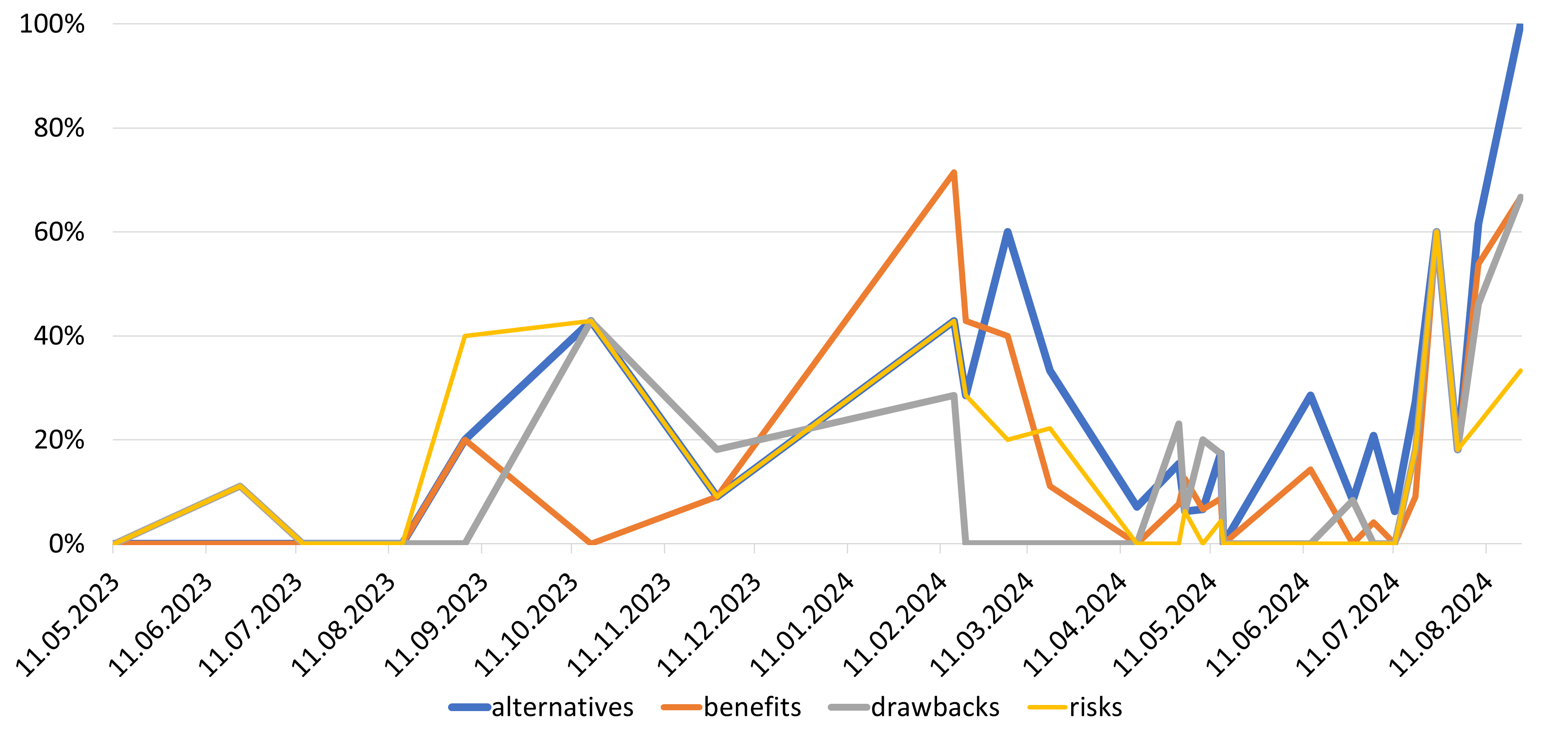}
                }\\
                \subfigure[TD-SAGAT: \% of participants stating to having considered prerequisites during the previous ticket's discussion ]
                {	\label{fig:TDSAGAT_Alternatives}
                    \includegraphics[width=0.48\textwidth]{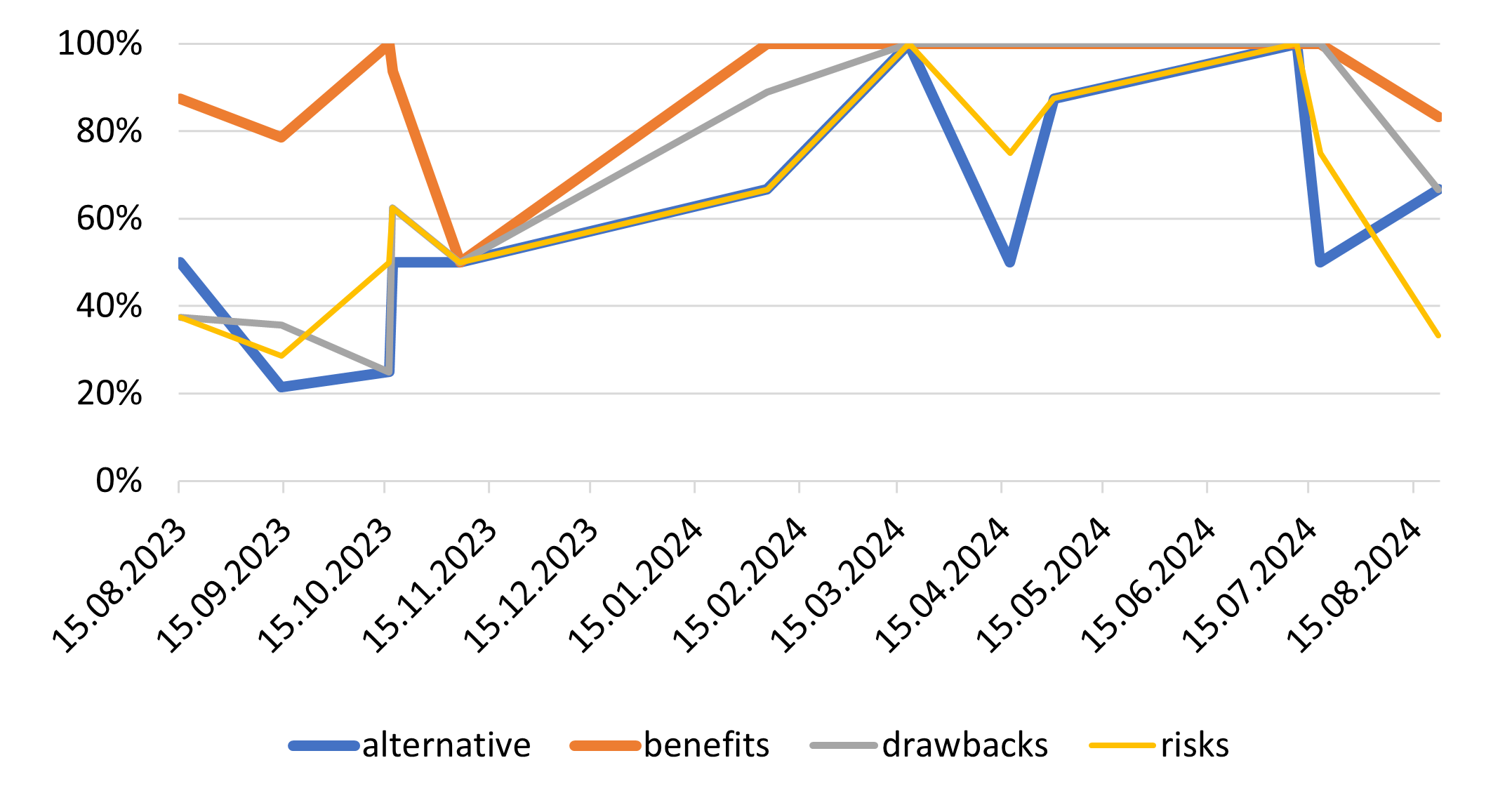}
                }
                
        \end{tabular}
        \caption{Considering prerequisites regarding comparing alternatives}
        \label{fig:alternatives}
    \end{figure}

    \begin{figure} [t!]
        \begin{tabular}{@{}c@{}}
                \\
                \subfigure[Questionnaire: \% of participants stating to having considered prerequisite during the last action cycle]
                {	\label{fig:WSQuestionnaire_Costs}
                    \includegraphics[width=0.48\textwidth]{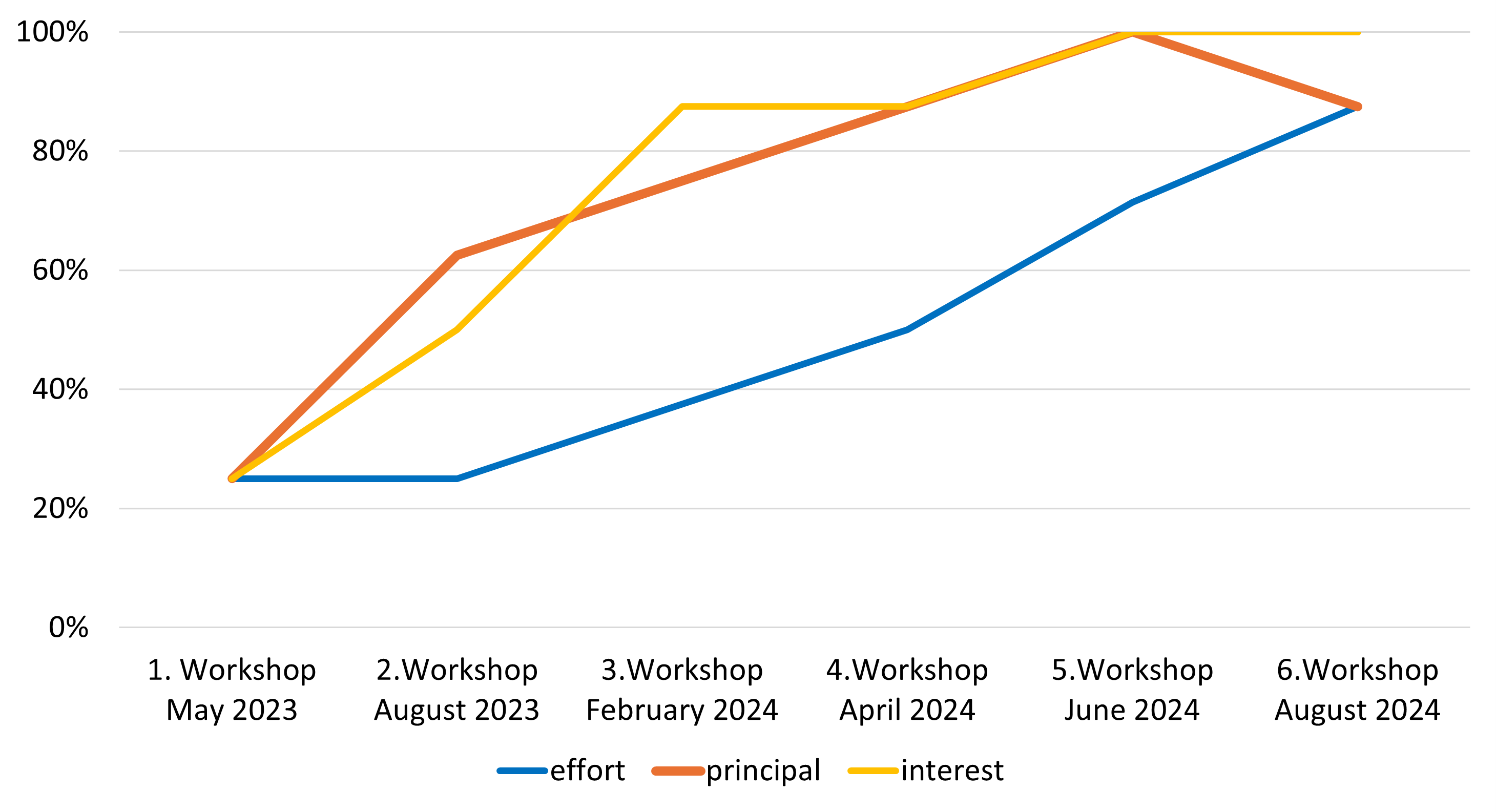}
                }\\
                \subfigure[Observations: \% of ticket where prerequisites were mentioned ]
                {	\label{fig:Observation_Costs}
                    \includegraphics[width=0.48\textwidth]{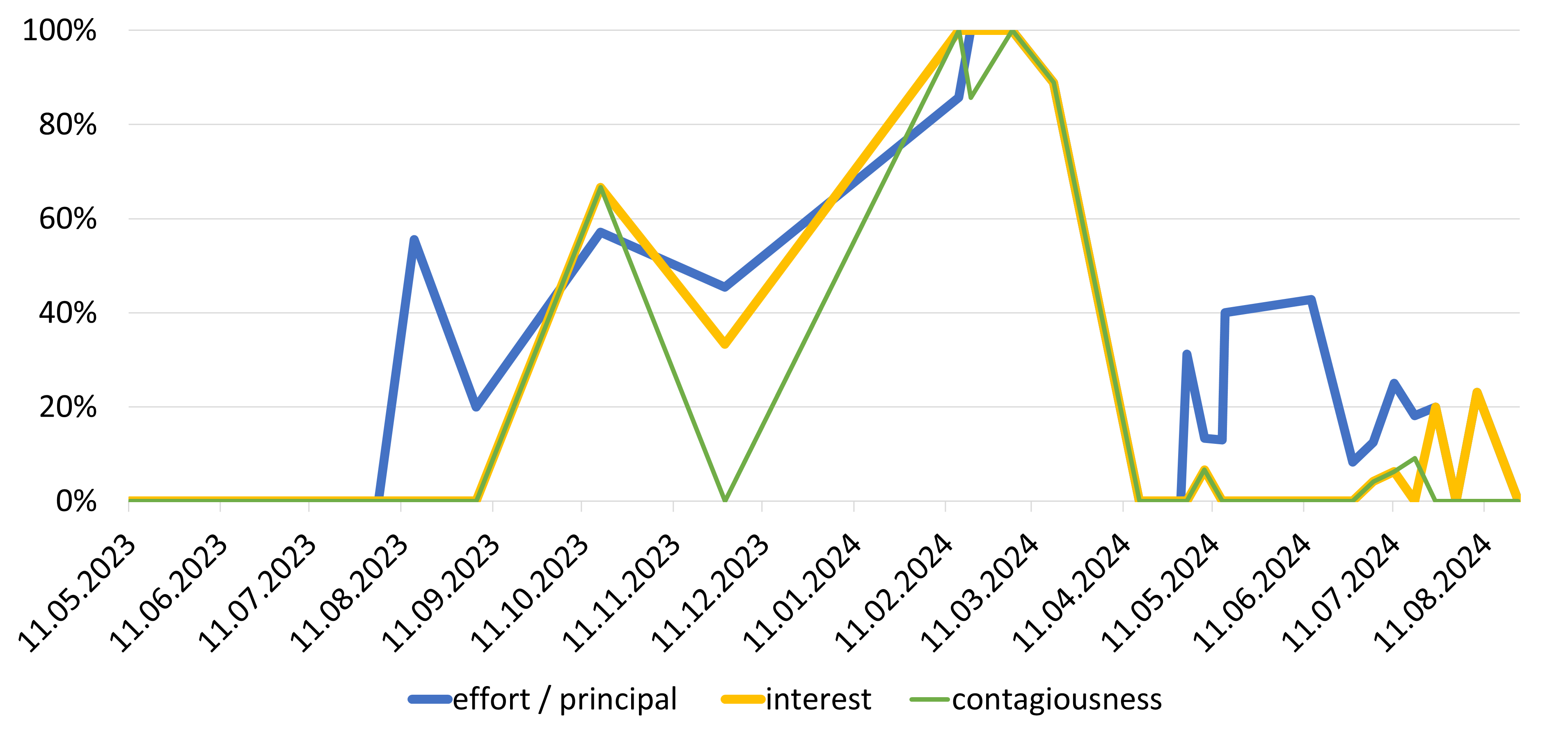}
                }\\
                \subfigure[TD-SAGAT: \% of participants stating to having considered prerequisites during the previous ticket's discussion ]
                {	\label{fig:TDSAGAT_Costs}
                    \includegraphics[width=0.48\textwidth]{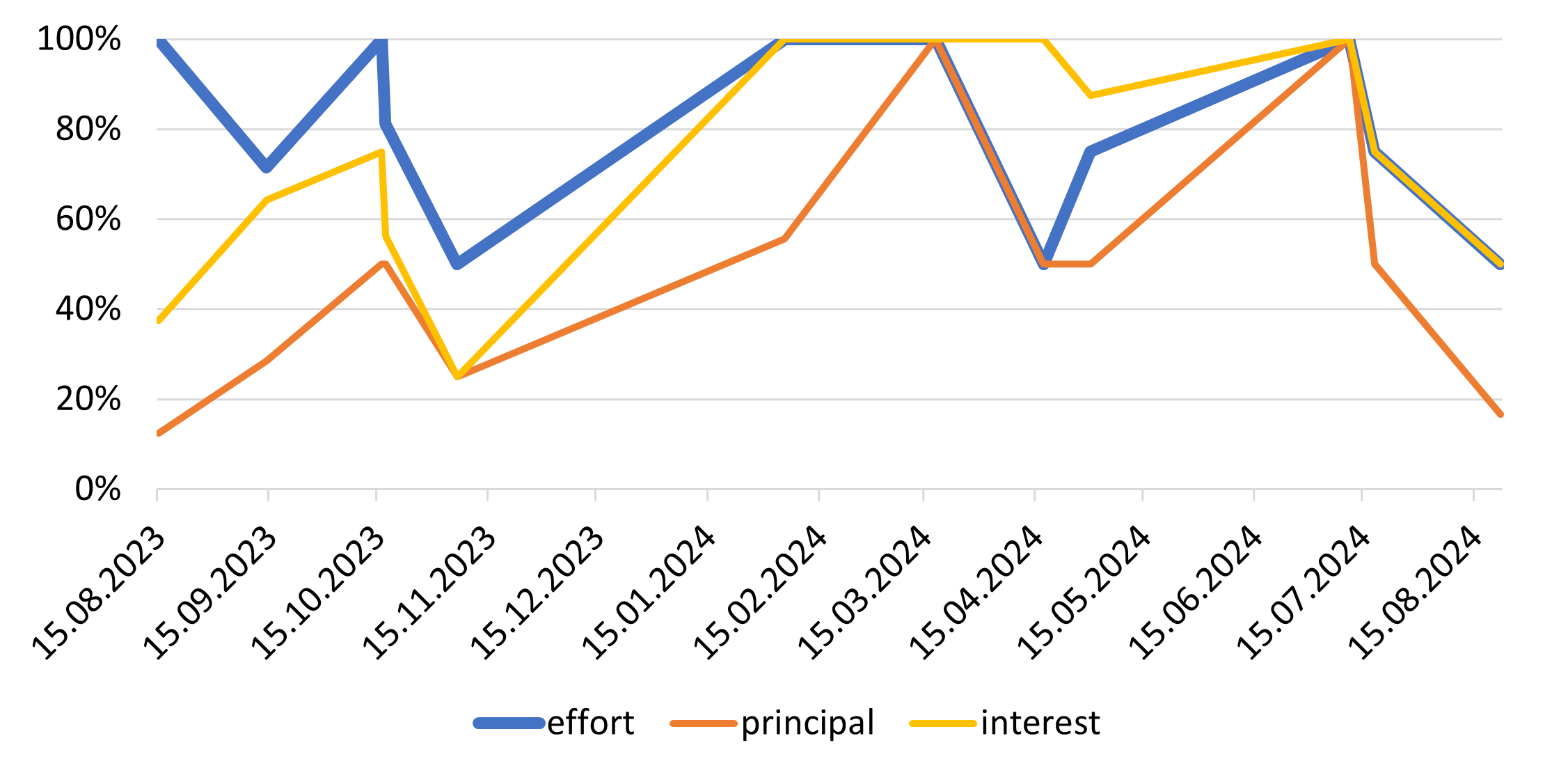}
                }
        \end{tabular}
        \caption{Considering prerequisites regarding  evaluating costs}
        \label{fig:costs}
    \end{figure}
       
    \begin{figure} [t!]
        \begin{tabular}{@{}c@{}}
                \\
                \subfigure[Questionnaire: \% of participants stating to having considered prerequisite during the last action cycle]
                {	\label{fig:WSQuestionnaire_Consequences}
                    \includegraphics[width=0.46\textwidth]{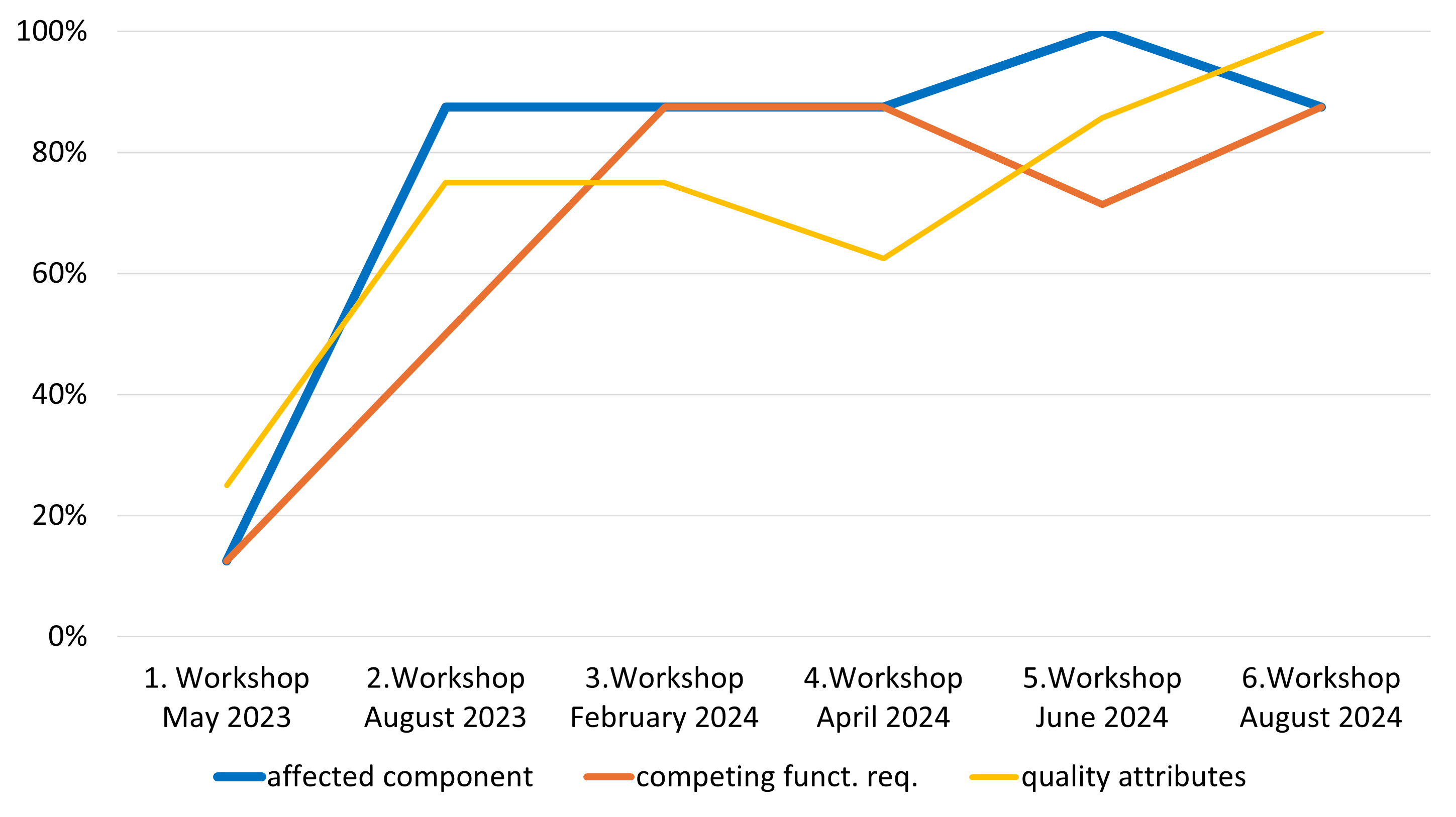}
                }\\
                \subfigure[Observations: \% of ticket where prerequisites were mentioned ]
                {	\label{fig:Observation_Consequences}
                    \includegraphics[width=0.48\textwidth]{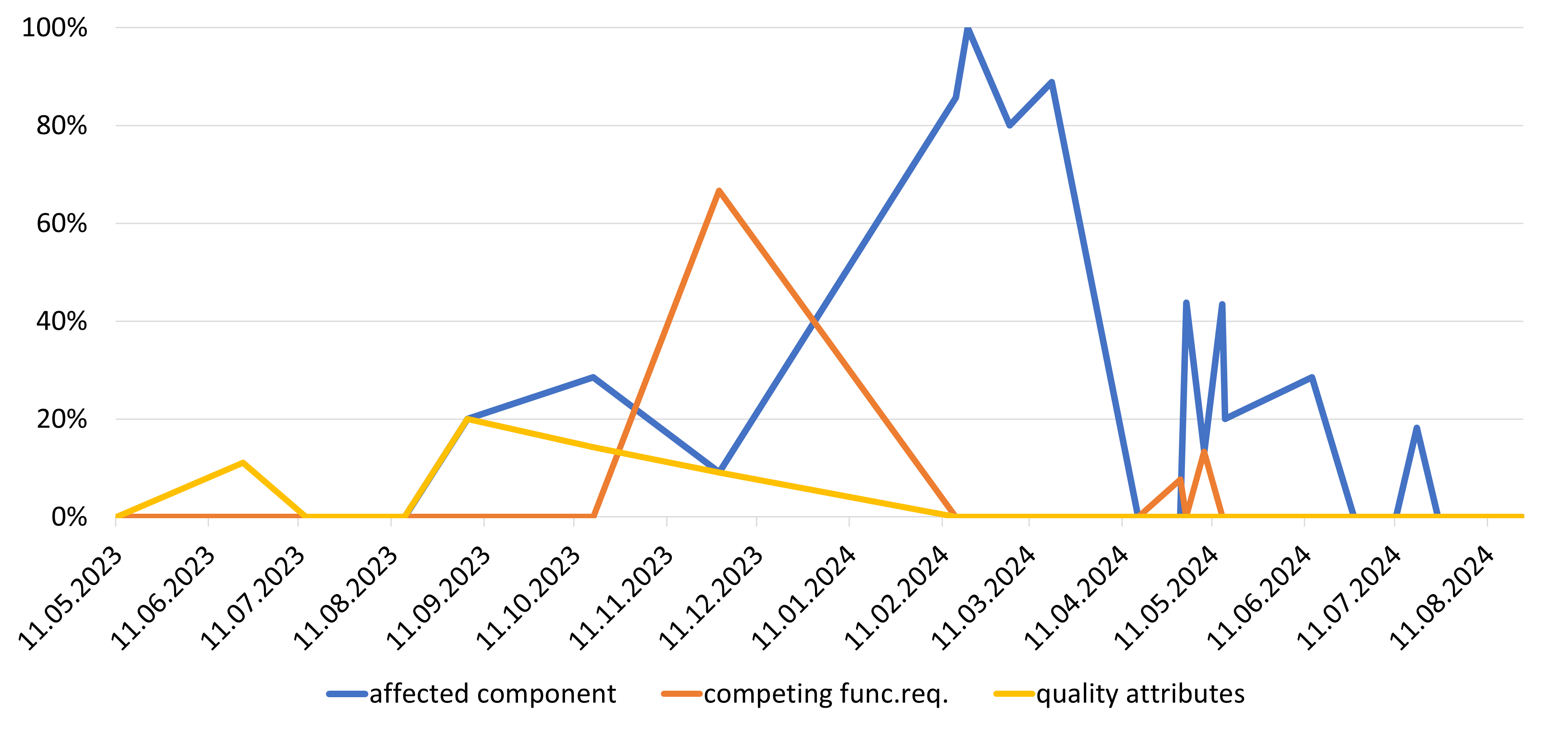}
                }\\
                \subfigure[TD-SAGAT: \% of participants stating to having considered prerequisites during the previous ticket's discussion ]
                {	\label{fig:TDSAGAT_Consequences}
                    \includegraphics[width=0.48\textwidth]{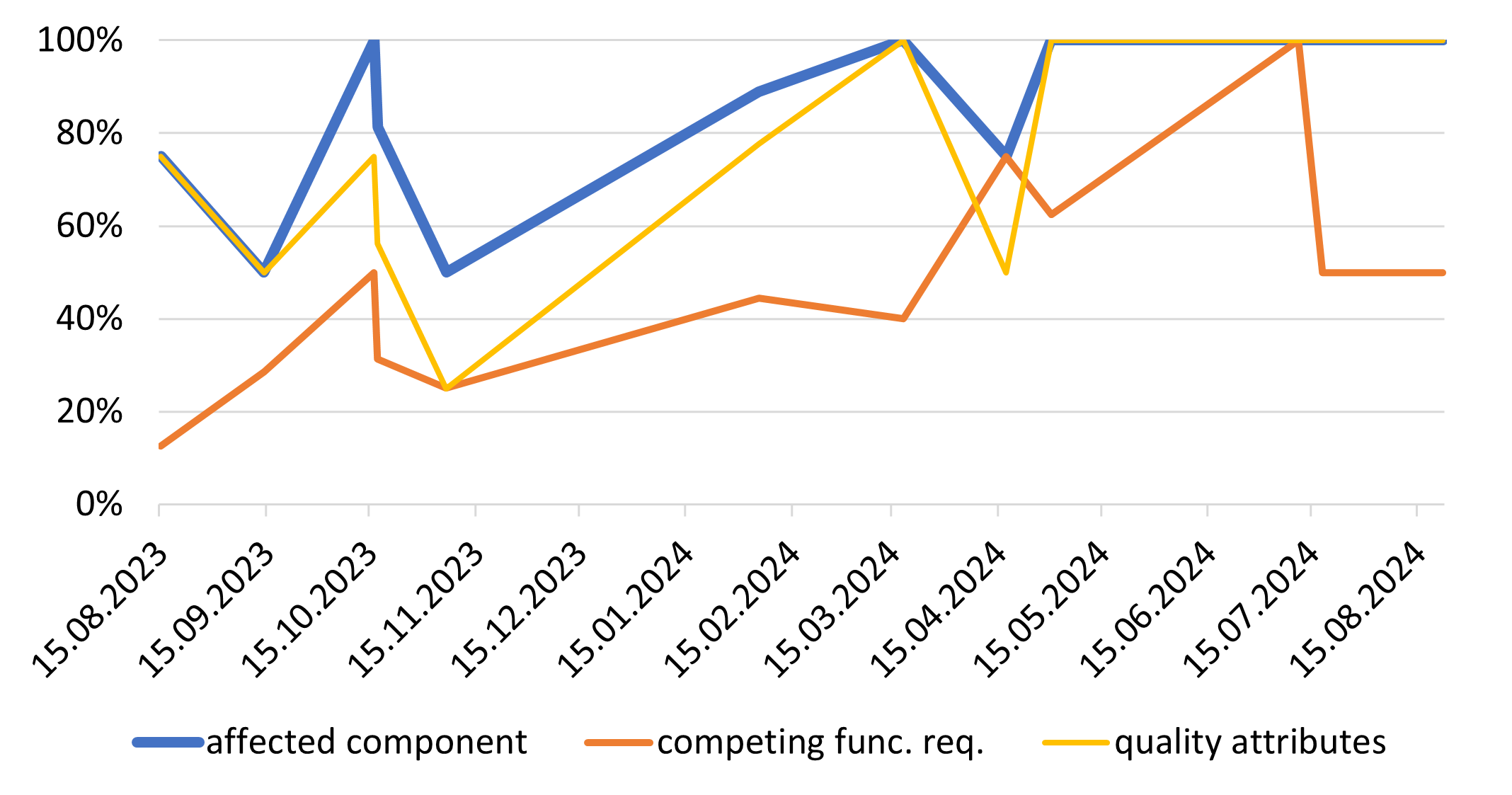}
                }
        \end{tabular}
        \caption{Considering prerequisites regarding  TD's potential consequences }
        \label{fig:consequences}
    \end{figure}

    \subsubsection{Requirements for Decision-Making}
    \label{sec:Results-RQ3-requirments}
    %This section presents the evaluation results we collected to evaluate the participant's TD awareness.
    We evaluated whether the participants assessed the ten prerequisites (identified by the SAGAT approach) when making decisions, utilizing three methods, i.e., the self-assessment surveys (\Cref{sec:method-surveys}, the observations (\Cref{sec:method-observation}, and TD-SAGAT (\Cref{sec:method-tdsagat}.
    The ten prerequisites are organized into three categories (\textit{Comparing alternatives}, \textit{Evaluating costs}, and \textit{Considering potential consequences}) to simplify their analysis as explained in~\Cref{sec:method-tdsagat}. % ``Comparing alternatives'' includes the assessment of alternatives, benefits, drawbacks, and risks; ``Evaluating costs'' includes the assessment of effort, principal, and interest; ``Considering consequences'' includes the assessment of affected components, competing functional requirements, and quality attributes.
    %All three methods used in the study were compared with each other in 
    The results for each category are depicted in \Crefrange{fig:alternatives}{fig:consequences} and separated into charts for each method. % \textit{comparing solutions}, \textit{evaluating costs}, and \textit{evaluating consequences} explained in \Cref{sec:method-tdsagat}.
    %Each figure contains three different charts (a to c). 
    
    % \Cref{fig:alternatives} shows how often solution alternatives, advantages, disadvantages, and risks were discussed or considered during the study. 
    % \Cref{fig:costs}  relates thematically to the costs of TD tickets in terms of effort, principal, and interest. 
    % \Cref{fig:consequences}  visualizes the study group’s awareness of consequences related to affected components, competing functional requirements, and quality attributes.
    
    \paragraph{Comparing Alternatives} $~$
    
    \Cref{fig:alternatives} shows that the awareness for comparing alternatives, including considering benefits, drawbacks, and risks, rose after the second workshop on TD incurrence (including the debiasing techniques) in August 2023.
    This positive development is particularly evident in the results of the self-as\-sess\-ment survey in \Cref{fig:WSQuestionnaire_Alternatives}.
    The team discussed these prerequisites, even though most of the time, they still deemed their initial idea to be the better alternative: \textit{``An alternative to [framework name] would be, maybe you can do it natively, but that's a lot of implementation work.''}
    \Cref{fig:Observation_Alternatives} and \ref{fig:TDSAGAT_Alternatives} show that the awareness rose even further after the introduction of the \textit{Talked about TD} checkbox at the beginning of October 2023.
    The \textit{Talked about TD} checkbox also introduced discussions on the risks of incurring TD:  \textit{``P2: `Do you also have to think about TD here? Because that could entail organizational TD.' P1: `That's actually true, yes. Do you have any ideas?' P3: `Well, if we have to simulate it and there's no one available, then we have a problem. Then, it will take its toll. Then we have to pay interest.'''}
    \Cref{fig:Observation_Alternatives} shows that after a drop in November 2023, the observation values further increased between January and March 2024.
    We assume that the drop was due to a focus on a competing functional requirement (see~\Cref{fig:Observation_Consequences}).
    The increase was presumably  due to the implementation of the TD ticket type and the review of all TD tickets to record all attributes for documentation and prioritization.  
    In parallel, the team worked on a refactoring initiative, increasing the number of new tickets as visible in \Cref{fig:closedopen_count}.
    This figure also shows that new tickets were rarely created from April 2024 onward.   
    Consequently, \Cref{fig:Observation_Alternatives} shows fewer discussions on alternatives for this time as primarily old tickets that were already refined were re-discussed, or only the \textit{re-submission date} was postponed.
    %Where the values in the course of the graphic are then partially at 0, these are probably old tickets that were not discussed but simply postponed to a new resubmission date. 
    %\Cref{fig:closedopen_count} and \ref{fig:closedopen_effort} 
    From July 2024, the values in \Cref{fig:Observation_Alternatives} began to rise again after introducing a reminder to discuss and document alternatives, benefits, drawbacks, and risks to each ticket.
    This led to a re-discussion and documentation of this information for old tickets. % to remind the team members to talk about alternatives, advantages, disadvantages, and risks in meetings. 
    %The values were at 100\% percent when discussing alternatives, over 60\% when discussing advantages and disadvantages, and over 30\% when assessing risks at the end of the study. 
    %Therefore, the reminder helped to stimulate team discussion again. %, which is evident in the observations. 
    This positive development can similarly be witnessed in \Cref{fig:TDSAGAT_Alternatives}, i.e., the TD-SAGAT results for the situation-specific assessments. %, even if the values for the solution alternatives dropped in July 2024.
    
    \paragraph{Evaluating Costs} $~$
    
    A positive trend regarding the assessment of costs, i.e., effort, principal, and interest, is evident in \Cref{fig:WSQuestionnaire_Costs}. 
    The values increased continuously over the course of the study, i.e., more and more participants declared they would consider these prerequisites. % and were over 80\% at the end of the last workshop in August 2024. 
    A similar direction of the values can also be seen in the TD-SAGAT assessment in \Cref{fig:TDSAGAT_Costs}.
    However, there was a notable decrease at the end of the study.
    We regularly observed discussions on the interest: \textit{``P2: `Interest is, you say\' P3: `Disturbing, it's annoying because you have to send everything back and forth to [name of function], and then there are restrictions again and so on.'''}
    Particularly, the discussion on interest probability often led to astounding results: 
    \textit{``P2: Interest probability is kind of the exact opposite for me now because the average case is that people don't look into it [documentation] \ldots Nobody misses it, and probably nobody needs it.''}
    \Cref{fig:Observation_Costs} shows that the trend regarding observations was less linear.
    For~\Cref{fig:Observation_Costs}, we were able to combine effort and principal in one value because the effort for repaying a TD item is the principal, and we had the information on the ticket type for each observation.
    Moreover, we observed and marked discussions on a TD item's contagiousness.
    We included it as a prerequisite of the costs, as it illustrates the increase in principal over time. 
    Unfortunately, this value was not covered in the surveys.
    The progression of the observation graph can be explained in the same way as \Cref{fig:Observation_Alternatives}.
    The significant increase between January and March 2024 probably resulted from reviewing the backlog and recording all attributes in the new ticket type. 
    The drop after that was most likely due to the lack of new tickets, meaning that no discussions about tickets took place. 
    During this time, only the effort was readjusted due to the initiative to estimate the effort on a bigger scale. 
    
    \paragraph{Considering Potential Consequences} $~$
    
    Considering TD consequences included considering the component affected by a TD item, the potentially competing functional requirements when prioritizing a TD repayment, and the TD item's effects on quality attributes. 
    As presented in~\Cref{fig:WSQuestionnaire_Consequences}, the self-assessed awareness for those prerequisites increased mainly at the beginning of the study,  i.e., most participants declared to consider these factors since the second workshop. 
    %, where a strong increase in the values for affected components, competing functional requirements, and quality attributes can be observed for the second workshop. 
    %After completing the last workshop, the participants’ values for self-assessment were all above 80\%, as shown in \Cref{fig:WSQuestionnaire_Costs}. 
    A similar development can be observed in \Cref{fig:TDSAGAT_Consequences} for TD-SAGAT, even if fluctuations occur. % here as in \Cref{fig:TDSAGAT_Alternatives}. 
    Small fluctuations  might be attributed to the TD-SAGAT method, which provides snapshots of the situation instead of a constant long-term development. %as perceived by the participants when discussing TD tickets. %, which might have caused the fluctuation.
    \Cref{fig:Observation_Consequences} shows that quality attributes, despite their relevance, are rarely discussed. 
    The team often discussed this issue and had plans to address it, as mentioned by P2 in February 2024:
    \textit{ ``P1 and I actually wanted to ask all the stakeholders what they expect from our team, because we don't actually know exactly who our stakeholders are and what their expectations are, and the idea was to derive quality attributes from this.''} % [So] that we then see that one person is somehow into maintainability or many are into maintainability, so that's important or things like that.}
    Nevertheless, the team did not manage to alter their behavior regarding quality attributes during this study.
     \Cref{fig:Observation_Consequences} shows a peak for considering the affected component at the beginning of 2024. 
    From this category, the affected component is the only attribute recorded in the TD tickets. 
    Thus, this peak and the slight decrease after it resulted most probably from the reviewing and documentation similar to \Cref{fig:Observation_Alternatives} and \ref{fig:Observation_Costs}. 
    %In contrast, two peaks can be seen in the consequences' observations in  \Cref{fig:Observation_Consequences}.
    
    Notably, in  November 2023, the team focused on a functional ticket, so the value for competing functional requirements increased in \Cref{fig:Observation_Consequences}. 
    At this time, a decrease can be noticed in all other values, including the comparison of alternatives (\Cref{fig:alternatives}), the assessment of costs (\Cref{fig:costs}), and TD's consequences (\Cref{fig:consequences}). 
    Together, this leads us to the assumption that the focus on one functional ticket at this time might have influenced the whole TDM process negatively.
    %, supporting all methods' validity. %Unfortunately, in discussion with the participants, we could not identify reasons for this decrease. 
    %This leads us to assume a development in the change process where such low points are accepted as part of the change.

    \subsubsection{Unconscious TD incurrence and TD identification}
    \label{sec:Results-RQ3-incurrence}
\begin{figure}%[H]
        \centering
        \includegraphics[width=0.5\textwidth]{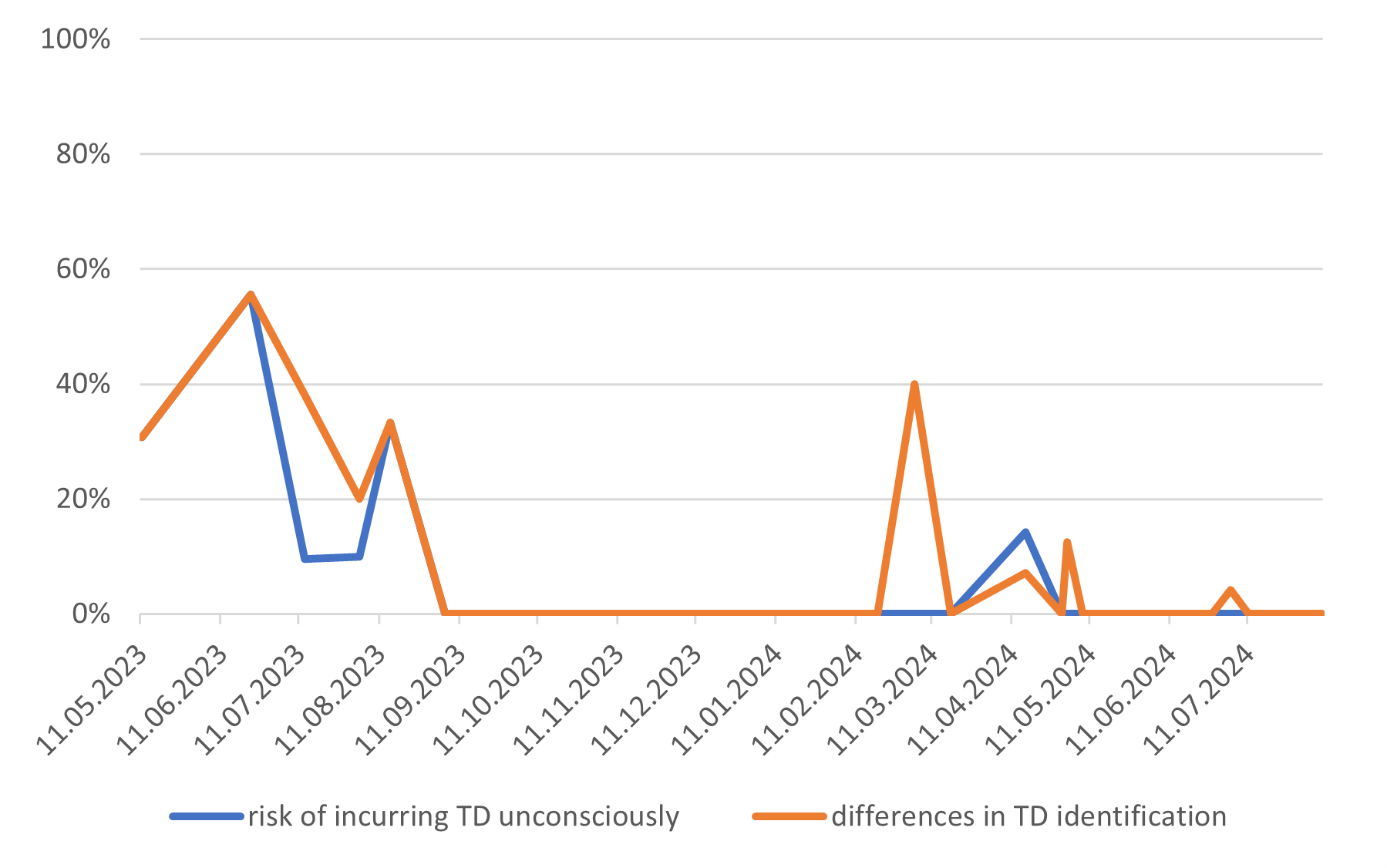}
    \caption{ Percentage of TD items per meeting, where different assessments on TD identification between researchers and team occurred, and where the team runs the risk of incurring TD unconsciously } % Differences in TD identification between researchers and practitioners}
        \label{fig:TDDispute}
        %\vspace{-5mm}
    \end{figure}

    In addition to the requirements, we observed whether there were differences between our and the practitioners' judgments as to whether a ticket could be a TD ticket and whether a functional ticket carried the risk of an unconscious TD incurrence (see \Cref{sec:method-observation-collection}).
    In \Cref{fig:TDDispute}, we can see many assessment differences at the beginning of the study, which we explained in~\Cref{sec:Results-Cylce1} about the action cycles. 
    Again, in the phase of TD prioritization between February and May 2024, there are some assessment controversies.

    Situations with a potential risk of unconscious incurrences are reduced after the second workshop (in August 2023), which focused on TD prevention.
    The knowledge that TD can be incurred consciously or unconsciously led the team to classify their own TD items: 
    \textit{``This is also TD that we didn't consciously incur \ldots We have a timer class, but there are other places where this timer class is not used, but [instead] some other code, which is why it reduces maintainability.''}
     Further, they were able to realize and discuss conscious TD incurrence: 
    \textit{``But that could be discussed. So I think it would be legitimate to say that we're taking on a lot of technical debt now because she [the customer] doesn't have that much time. And then later, when we want to use it again, we'll have to make sure that it's straightened out.''}

    \begin{framed}
        \textbf{RQ3:} 
        The self-assessment shows an increase in awareness of all prerequisites. 
        TD-SAGAT supports these increases overall but shows fluctuations typical for this method.
        The observations show that, in particular, the reminders and documentation of attributes in the backlog template sustainably led to discussions on respective prerequisites.
        We were able to identify how a focus on one important functional requirement was able to affect the whole process negatively.
        During the study, the team learned to identify TD more reliably, and they reduced the risk of unconsciously incurring TD.
    \end{framed}

    %    \begin{figure}%[H]
    %     \centering
    %     \includegraphics[width=0.48\textwidth]{img/closedopen_TD_CountedTickets.png}
    %     \caption{Opened and closed TD items over time}
    %     \label{fig:closedopen_count}
    %     %\vspace{-5mm}
    %  \end{figure}

    %    \begin{figure}%[H]
    %     \centering
    %     \includegraphics[width=0.48\textwidth]{img/openovertime_CountedTickets.png}
    %     \caption{Open TD items over time}
    %     \label{fig:openovertime}
    %     %\vspace{-5mm}
    % \end{figure}
    % shows an even stronger increase in the period from January to March 2024 in relation to affected components. 
    %During this period, there were new tickets that appeared to be increasingly viewed from the aspect of affected components in the context of the consequences of technical debt. 
    
    % As already mentioned, \Cref{fig:closedopen_count} and \ref{fig:closedopen_effort} show that many new tickets were discussed at this time. 
    % The spike in the number of opened tickets in February 2024 was due to the refactoring initiative at that time. 
    % The team recorded many small refactorings as TD tickets. However, they were only able to repay a small amount of them.

\section{DISCUSSION}
\label{sec:Discussion}

In this section, we discuss the TDM establishment, including recommendations for practitioners in~\Cref{sec:Discussion-RQ1}. % notable observations during the TDM establishment and list recommendations for practitioners. % to enable replication of the workshops. % and identify promising research fields.
In~\Cref{sec:Discussion-RQ3}, we discuss the effects the workshops had on TD awareness. 
Finally, we examine the overall study design to enable researchers to replicate and improve the study (\Cref{sec:discussion-studydesign}).

% In this section, we give an overview of which approaches from research were adopted, which new approaches occurred, how much effort the team members were willing to invest, and the effect the establishment of the TDM process had on the participants' awareness of TD.

\subsection{Workshop Approach for TDM Establishment}
\label{sec:Discussion-RQ1}
During the TDM establishment, we made some notable observations, which we discuss below, including ideas on the process's improvement. 
Additionally, based on our observations and analyses, we present a list of actionable recommendations for practitioners who plan to establish a TDM process.

    \subsubsection{Notable Observations during the TDM Establishment} $~$
        
        %Backlog tool
        First, the backlog tool already used by a team seemed to be the natural tool to manage TD. 
        However, two issues occurred during this tool's adoption:
        These backlog tools are commonly administrated company-wide, which makes changes for single teams more unlikely or negotiating them tedious. 
        As TDM is an issue in multiple teams, this can be remedied by explaining the goal and providing added value for multiple teams. 
        Regardless, we advise calculating the additional time this might take when establishing a TDM process.
        As a second issue, adding value calculations or rule-based pre-filling of attributes might be challenging in a proprietary tool.
        We assume these technical issues might be one reason why practitioners prematurely abandon TDM establishment approaches. 
        A solution would be for the backlog tool vendors to integrate TD ticket types by default.

        % %next Workshop
        % Second, holidays and vacations led to a postponement of workshops.
        % Also, sometimes, the team had to postpone their work for the TDM process due to more urgent work topics.
        % We recommend always arranging at least the following two workshop dates while remaining flexible in responding to changes.
        
        % ROI-based
        Second, while the participants used the \textit{ROI-based} priority to analyze and improve their \textit{educated guesses}, we observed that this led them to change the attributes in various cases. 
        Still, they rarely reevaluated their \textit{educated guess}.
        This can be caused by an anchoring bias, i.e., the preference for the first given value~\cite{borowa_debiasing_2022}. 
        Additionally, the participants remarked that they initially did not evaluate the underlying attributes with the ROI in mind.
        For future replication, we suggest determining the ROI from the beginning (i.e., during the second or third cycle) and proposing it as an additional prioritization factor.

        %biases
        Third, apart from the mentioned ``anchoring bias,'' we identified further biases, such as the ``central tendency bias,'' and ``curse of the knowledge bias,'' that influenced the TDM process at salient points. 
        We recommend researching this topic more profoundly and developing methods to mitigate the effects of those biases in software engineering.
        
        % TD manager - knowledge silo
        Fourth, the TD manager in this team was the team architect, which supported this participant's role as a \textit{knowledge silo.}
        While a senior developer should perform this role, we suggest not assigning the most important developer or team architect but distributing such roles within the team. 

        Fifth, workshops three to five were performed online due to the teams' preferences, and we did not detect any drawbacks from this context.
        Regardless, we advise performing the first two or at least the first workshop on-site to get to know each other, build trust between the participants, including workshop supervisors, and sufficiently motivate the participants.

        Finally, we consider two instances of not-adopted approach\-es remarkable:
        %Prioritizing against functional
        This team did not adopt the idea of prioritizing TD against functional requirements, which might be because they were under little pressure to process functional requirements. 
        Additionally, they did not adopt the idea of a repayment quota, which might be because the team was comparably free to decide which tickets to process. 
        However, further research on the reasons for this lack of adoption might be appropriate. 
        It is intriguing to further analyze whether other teams in other contexts might be more willing to adopt those approaches and which contextual factors might lead to adopting which approaches.
        
    %     
    % \subsubsection{Lessons learned regarding the workshop concept}
    % \label{sec:discussion-workshopdesign}

    % \subsubsection{}  $~$
    % First of all, the study was delayed somewhat due to negotiations with the administrators of the backlog tool, . 
    % We would suggest clarifying the access in advance.
    % Similarly, using Microsoft PowerBI or another appropriate tool already used in the company, e.g., Tableau\footnote{\url{https://www.tableau.com/}}, for visualization should be clarified.
    % We assume these technical issues might be why TDM establishment approaches are prematurely abandoned in practice. 
    % Therefore, a project sponsor, i.e., a manager interested in the TDM establishment, should aid this process.

    % Second, holidays and vacations led to a postponement of workshops.
    % Also, sometimes, the team had to postpone their work for the TDM process due to more urgent work topics.
    % We recommend always arranging at least the following two workshop dates while remaining flexible in responding to changes.

    % Finally, we changed the initial plan, which was to introduce the visualizations as the third step.
    % After the second step and the discussion about cost factors, questions regarding prioritization emerged naturally, which made us change this plan to follow the study's flow expressed by the participants. 
    % We suggest retaining this procedure.

    \subsubsection{Recommendations for Practitioners} $~$
    \label{sec:Discussion-RQ1-recommendation}

   % What recommendations would you suggest for a team that does not formally consider technical debt, based on your experience? What challenges might they face? Where and how should they start?
   
    Our proposed workshop approach is flexible and does not provide a strict process to follow. 
    Instead, it offers a way to determine a process that fits the respective team or company.
    Thus, we assume that the workshop approach, as well as the establishment of the TDM processes, might be adopted by multiple other companies. 
    The approach is mainly fitted for teams that are organized in an agile way and are using some form of work items, e.g., user stories or task lists.
    Furthermore, it might be beneficial if some adaptable backlog tool is used.
    However, this process would similarly work for teams that use an Excel sheet as a backlog, e.g., by adding new columns per item.
    Analyzing whether this workshop approach would also work in less agile IT departments provides a fascinating area of future research.
    
    To enable the adoption of this approach, we have collected some practical suggestions for practitioners when employing the workshops or establishing a TDM process in another way based on our experiences during the action research.
    For easy access, we provide the actionable recommendations in the form of a list:
    
    \begin{itemize} [leftmargin=*, itemsep=0pt]
    
        \item Do not try to implement the TDM process with all its intricacies from the beginning, but take the time to improve the process continuously.
        \item Start with identifying and tagging all the TD items in the backlog, then create a separate item type and successively add relevant attributes.

        \item Focus on the TD items identified by the developers (instead of putting too much focus on automatic detection) as those TD items are the effective ones, i.e., they have at least at some point already incurred interest.
        \item Motivate the developers to document TD items in the backlog that are on their minds but not yet documented.
        
        \item To distinguish TD from other ticket types: 
        (1)~Define your customer, e.g., would you define colleagues from other teams who use your system as customers? If they have requests regarding your system, would you determine this to be TD or a functional requirement?
        (2)~Ask yourself whether interest occurs. 
        (3)~Ask yourself who would be willing to pay for the implementation. % of this ticket.
        
        \item When your backlog contains too many items to identify all TD items manually: 
        (1)~Search for specific phrases like ``refactoring,'' ``quality,'' ``test,'' ``document,'' etc. 
        (2)~Start by focusing on the most recent items, e.g., items created in the previous year.
    
        \item Use the tools you are familiar with, e.g., your current backlog tool.
        (1)~Include tool/backlog admins in the project early on and discuss adding a specific TD item template.
        (2)~Analyze whether the backlog tool is flexible enough to allow for calculations (and not only rules).
                
        \item Install a TD manager, i.e., someone who feels responsible for the process. 
        (1)~Assign an experienced and respected developer who is willing to invest time.
        (2)~Do not assign a team member as TD manager who already has many responsibilities.
        \item Plan the additional time to set up and maintain the process in your project/portfolio management, particularly including the time for the TD manager.
        
        \item When discussing new attributes, include discussing the selectable items of combo boxes, e.g., low, medium, and high for interest or interest probability, and their respective meanings and usage.
        \item If you plan to employ an ROI calculation: 
        (1)~Consider this when choosing the selectable items for the effort, interest, and interest probability. 
        (2)~Consider this when assigning those values to a TD item.
        \item Use reminders in the ticket templates to maintain awareness for certain topics (checkboxes are sufficient for this purpose).

        \item Use of a \textit{re-submission date} for TD items. Make sure to define an appropriate process, i.e., define in which meeting a ticket that reached the date will be re-\-eval\-u\-at\-ed.
        \item When using an \textit{educated guess} for priority estimation: 
        (1)~Discuss the various priority factors and decide which factors should be included with the team, e.g., whether to include the effort or not. 
        (2)~ Determine in advance if you use worst-case scenarios or medium-case scenarios when evaluating priorities.
        (3)~Make sure not to include a factor multiple times, e.g., do not include the future expectation in the priority if you have a \textit{re-submission date}.
        \item Determine your system's architectural components. If you plan to use the process throughout the company, define a component hierarchy that fits into the company's hierarchy of products, systems, and components to be able to aggregate the results for the management.
        \item Discuss with the team what the term ``quality'' implies in terms of your system. Employing the task on quality attributes from workshop two might help your team to get to know quality attributes and experience the resulting hardships of deciding on trade-offs. 
        \item Document the TD ticket's purpose and attributes, and explain them to persons outside the team who add (TD) tickets to your backlog.
        \item Structurally refining TD items will presumably lead the developers to acknowledge that not all TD items must be repaid. Hence, discuss suitable repayment methods after having some TD tickets prioritized with the team.
        \item Use visualizations if the number of TD tickets is too big to analyze otherwise.
        
       % \item Always plan the next meeting on this topic to inspect and adapt the current process when one meeting is finished.
        %\item You could embed the TDM process development into the Scrum retrospective meetings. However, we would recommend dedicating every third or fourth retrospective meeting on this topic.
    \end{itemize}

    During our study, we identified that issues in the general work process might incur TD. 
    Thus, we recommend discussing the following questions regarding your meetings and backlog item's states:
    
    \begin{itemize} [leftmargin=*, itemsep=0pt]
        \item Which purpose has each meeting, and how does it differ from other meetings?
        \item In which meeting will you change which attribute?
        \item Is the meeting's purpose transparent to every participant? To answer this, do not rely on the fact that there is a written description, but ask each participant.
        \item Which status is changed in which meeting?
        \item In which (other) situation is each status changed?
        \item What does ``refined'' mean in your context? Which attributes have to be filled out? 
        %\item To remind you of discussing alternatives, benefits, drawbacks, and risks, you could employ a ``definition of ready''\footnote{\url{https://www.atlassian.com/agile/project-management/definition-of-ready}}.
    \end{itemize}

 \subsection{Effects on TD Awareness}
    \label{sec:Discussion-RQ3}

    %We used various methods to estimate the participants' TD awareness: the self-assessment through the workshop questionnaires, the observations of team discussions, and the assessment of the participants' thought processes through the TD-SAGAT method.
    %The results were presented in~\Cref{sec:Results-RQ3}.

    %Overall, we can see that the reminders in the backlog ticket, i.e., TD attributes (e.g., for cost evaluation) and templates for textual descriptions (e.g., benefits, drawbacks, and risks), sustainably led to discussions of decision-relevant aspects.
    
    In the second workshop in August 2023, we introduced the relevance of considering alternatives, benefits, drawbacks, and risks to avoid cognitive biases, which changed the team's observable behavior initially, though not sustainably (\Cref{fig:Observation_Alternatives}).
    Introducing a TD ticket type and respective attributes in February 2024 led to a rise in all recorded attributes like interest, contagiousness, or component (e.g.,\Cref{fig:Observation_Costs}). 
    Further, the \textit{Talked about TD} checkbox implemented led to a rise in discussed alternatives.  % \Cref{fig:Observation_Alternatives} and \ref{fig:Observation_Costs}. 
    %Unfortunately, the observations show low values due to only a few new TD tickets between April and July 2024. %\Cref{fig:Observation_Alternatives} and \ref{fig:Observation_Costs} 
    %However, the TD-SAGAT results show a sustainable rise after introducing the respective attributes. % in the decision-making process.
    The reminder for alternatives, benefits, drawbacks, and risks added in July 2024 led again to a rise in~\Cref{fig:Observation_Alternatives} due to the participants re-discussing the tickets. 
    Information not recorded in attributes, e.g., quality attributes or functional requirements competing with a TD repayment, was less discussed in meetings, even though it seems to have been on the developers' minds, as~\Cref{fig:TDSAGAT_Consequences} indicates. 
    From this, we conclude that the workshop information only had a short-term effect on TD awareness, and long-term awareness can be reached if the learned information is included in the daily process by adding respective attributes to the ticket types.
    %Additionally, this analysis shows the relevance of the TD-SAGAT method to supplement observations.
    
    In the observations, we further analyzed how often the danger of unconscious TD incurrence occurred and how often we identified  a ticket to be TD where practitioners did not.
    While there were various risks of unconscious TD incurrence and controversies on what items were TD at the beginning of the study, we did not identify these instances anymore at the end. 
    This seemingly proves the success of our workshop approach. 
    However, just employing the workshop concept would not include the action research's ``evaluation'' phases. 
    We can not determine whether this success can be reached without a TD expert evaluating the meetings. 
    This implies that we cannot conclude whether this success results from the workshop concept or the action research.
    Thus, evaluating these effects without presenting the evaluation results to the participants might be an intriguing next research step. 
    
%    While the discussions improved over time, the team did not reduce the number of TD tickets.
%    However, they stated they are much more satisfied with the backlog because they can better prioritize TD tickets, have a better overview, and have a better process to repay TD when necessary.
%    We assume that this insight shows an improved TD awareness and is an especially valuable outcome of our study.

    \subsection{Discussion of the Study Design}
    \label{sec:discussion-studydesign}
    
    In this section, we briefly discuss what we learned regarding the study design, which might be helpful for replication. 
    Furthermore, we provide an assessment of the adopted and tested TD-SAGAT approach.

    \subsubsection{Lessons Learned} $~$
    \label{sec:discussion-studydesign-lessonslearned}

    % studies delays
    First, similarly to the first topics in the previous section, the study was delayed due to negotiations with the administrators of the backlog tool and the implementation of the visualization tool. 
    We are aware that sometimes an essential trust must be established before getting access to the company's tools, and thus, delays might be unavoidable.
    %We also mentioned the problems in arranging the workshop. 
    %For both reasons
    Therefore, the study's timetable must be particularly flexible to accommodate these uncertainties. 
    Change management processes~\cite{Fuhring2021a} assume such impasses and have methods to counteract them, which is why they adequately supplement this study design. 

%prerequisites reduction
    Second, we reduced the number of observed prerequisites over time to simplify the evaluation, focus it on the most critical aspects, and make the evaluations of the different methods comparable.
    For example, while we started the initial questionnaire with some pretty particular questions, e.g., whether the evolution of the affected component or its business value was taken into consideration, we ended up only asking whether the affected component was generally considered.

    %tickets mehrfach besprochen
    %Nicht löschen relevant für TD-SAGAT.
    Third, we had problems evaluating the observations, as many tickets were discussed more than once, but usually, not all attributes were discussed. 
    Particularly at the study's end, tickets were often re-discussed due to new insights and the \textit{re-submission date}.
    The normalization of attribute discussions to the number of discussed tickets proved problematic.
    It showed low values only because the tickets had already been discussed, and their attribute's discussions had already taken place in previous meetings.
    We explained this in our analysis of those results, and future adoptions of this evaluation must similarly keep this effect in mind. 
    We found TD-SAGAT to be a valuable supplement to pinpoint this issue, as the values for TD-SAGAT increased while the values for observation seemed to decrease. %and as a basis for an in-depth evaluation and discussion

    % backlog size     
    Fourth, the studied team was newly assembled and had only 120 backlog tickets, which is a small amount that can be manually searched for TD tickets with a small effort. 
    Yet, analyzing more extensive backlogs, particularly with older items in them, might not be easy to do manually.
    We suggest researching how machine learning mechanisms or language processing might help cluster backlog tickets and identify TD tickets.

    Fifth, as mentioned in the previous section, we cannot differentiate whether some of the outcomes result from the workshop concept or the interventions of the action research. 
    Thus, repeating this study without performing the action research's ``evaluating'' phase or at least without presenting those results to the practitioners could, therefore, extend the evaluation of the workshop concept's success.

%Teilnehmer unterliegen starken biases. Sie bevorzugen immer das was aus ihren Reihen kommt (IKEA, re-invent) und aind anderem eher abweisend gegenüber. Wir wirken dem entgegen durch die freie Entscheidung. Wie kann man dem bei der Einführung von TD noch weiter entgegenwirken?

    %backlog data for evaluation
    %\subsubsection{Using backlog data for evaluation}
    Finally, while the amount of TD items did increase over time (\Cref{fig:backlog}), the satisfaction with the backlog improved (\Cref{fig:BacklogSatisfaction}).
    We discussed this seeming contradiction with the practitioners and found out that the improved satisfaction values were not rooted in the amount of TD but in the ability to manage the backlog and distinguish relevant from irrelevant TD. 
    This implies that the reduction of TD items might not be as relevant to a successful TDM process as is currently expected. 
    For future research, we advise that to assess the success of a TDM process, researchers should not focus too much on the number of TD items and should always take the subjective view of team members into account.

    \subsubsection{Assessment of the TD-SAGAT Evaluation} $~$

    % Third, we had problems evaluating the observations as many tickets were discussed more than once, but usually, not all attributes were discussed. 
    % Particularly at the study's end, tickets were often re-discussed due to new insights and the \textit{re-submission date}.
    % The normalization of attribute discussions to the number of discussed tickets proved problematic as it showed low values only because the tickets had already been discussed and discussions have already taken place in previous meetings.
    % We found TD-SAGAT to be a valuable supplement to pinpoint this issue, as the values for TD-SAGAT increased while the values for observation seemed to decrease.
    The TD-SAGAT method is a newly employed method that proved its feasibility for usage in research during this action research.
    The method discloses insights beyond the observable behavior, and it analyzes underlying prerequisites instead of bluntly asking for awareness. 
    Due to the situational interruptions, the information is still in the participants' memory. 
    This reduces bias compared to answering questions in hindsight, where memory loss can lead to more biased assessments.
    However, the method is not intended to be used by practitioners as it would create too much overhead for them due to the analysis and the continuing meeting interruptions.

    The TD-SAGAT results mostly supported the results of the workshop questionnaires and the observations but also showed some deviation (see~\Cref{sec:Results-RQ3}).
    We analyzed deviations and discussed them with the participants to identify whether the method was feasible and if an explanation for the deviation could be found.
    We identified the issues illustrated in the previous subsection regarding our observation's analysis and adjusted our final evaluation based on this.
    Additionally, we discussed why benefits, drawbacks, and risks were on the participants' minds (TD-SAGAT) but were not discussed (observations), which led to two significant changes:
    First, the team added a reminder to discuss these issues to the TD ticket template.
    Second, the participants identified the issue of \textit{knowledge silos} as explained in~\Cref{sec:Results-Cylce5} and changed their behavior to rotating the meeting moderation.

    However, this method also has some shortcomings, particularly the low number of data points in our situation.
    We did not want to disturb and influence the participants too much, so we conducted this method only once a month.
    Additionally, not all participants were able to complete the surveys due to \textit{knowledge silos}.
    Overall, TD-SAGAT provides snapshots of situations resulting in fluctuations in the TD-SAGAT figures.
    For the analysis of TD-SAGAT results, it is consequently essential to identify the long-term direction of the prerequisites, rather than focusing on these fluctuations.
    An issue in the survey creation was that we neglected the prerequisite of a TD item's contagiousness, which we would suggest adding as a prerequisite to future TD-SAGAT usage.
    
    Still, in our opinion, this method is a valuable addition to method triangulation as it researches the thought process in addition to visual behavior (observations) and participants' opinions (surveys).
    %As such, the measurements are much more reliable and unbiased than surveys.
    Further, we were able to identify issues that we would not have determined without this method. 

    Finally, we would like to encourage researchers to adapt and use this method in other situations. 
    For example, developers could be interrupted when implementing code to determine their thought processes in those situations.

    \section{THREATS TO VALIDITY}
    \label{sec:ThreatsToValidity}
        %We base our action research on the specifications in Staron's book on Action Research in Software Engineering~\cite{staron_action_2020}.
        %In his book on action research in software engineering, Staron~\cite{staron_action_2020} presents multiple common potential threats to validity for action research studies. % using the guidelines provided by Wohlin et al. as orientation~\cite{Wohlin2012}.
        As explained in the introduction to this research method (\Cref{sec:method-actionresearch}), \textit{ACM Special Interest Group on Software Engineering}'s empirical standards state that ``example criteria include reflexivity, credibility, resonance, usefulness, and transferability,'' i.e., they are decisive for action research studies~\cite{sigsoft_acm_2025}.
        For the reflexivity, we systematically reflect on our potential biases in the following subsections (\Crefrange{sec:ThreatsToValidity_Construct}{sec:ThreatsToValidity_External}) by analyzing the common potential threats to validity mentioned in Staron's book on action research in software engineering~\cite{staron_action_2020}.
        Regarding credibility, we provided exhaustive additional material~\cite{AdditionalMaterial}.
        The resonance is addressed through the retrospective parts of each workshop (\Cref{sec:actioncycles}).
        To address the usefulness and transferability, we provide details on the workshops in~\Cref {sec:workshops} for replicating the study and recommendations for practitioners in~\Cref{sec:Discussion-RQ1-recommendation}. 
        Generalizability refers to the ability to draw conclusions from quantitative research with a representative sample size, such as when conducting a survey of 1,000 randomly chosen participants to generalize the results to the entire population.
        In contrast, transferability is a key concept in qualitative research, referring to the potential for research results to be applied in other contexts.
        In our case, the transferability is offered by replicating the workshops to establish TDM processes in other companies. 
        In contrast to providing a fixed TDM process, the flexible approach of our workshop concept, where participants choose approaches that best fit their contexts, enables this transferability.
        The empirical standards further note that ``positivist quality criteria such as internal validity, construct validity, generalizability, and reliability typically do not apply''~\cite{sigsoft_acm_2025}.
        Nevertheless, we explicate and discuss these criteria for our study as far as possible in fulfillment of the criterion of reflexivity.
        %\footnote{\url{https://www2.sigsoft.org/EmpiricalStandards/docs/standards?standard=ActionResearch}} 
        We provide information on the threats by the names given to them in the book by Staron~\cite{staron_action_2020}. %) that are relevant in the context of our study.
    	
    	\subsection{Construct Validity}
            \label{sec:ThreatsToValidity_Construct}
            
            First, we fall for the ``mono-operation bias'' because we study only one team of a big company. 
            As with every IT team, this team has some special characteristics, e.g., central development with no real customer contact or the subteam of algorithm developers. 
            % (1)~central development with no real customer contact, (2)~particularly specialized team members, (3)~the team's new composition, and (4)~the subteam of algorithm developers. 
            For future work, we plan to replicate this study in other teams and domains to identify generalizable aspects.
            However, the results of this study have to be considered with this limitation in mind.
            Second, ``interactions of different treatments'' may have occurred. 
            The new composition of this team allowed it to adopt a TDM process while still deciding on the general working processes.
            However, this means that changes in general working processes might have influenced the results as well as the TDM establishment. 
            This threat can be mitigated by replicating this study in other contexts, which we plan for future work. 
            Third, the ``interaction of testing and treatment'' also influenced our study. 
            The recording of the meetings and the researchers' presence in some meetings might have influenced the participants' awareness of the TD topic.
            We mitigated this issue by not attending every meeting and recording meetings instead of attending.           
            Further, for some results, we were not able to pinpoint whether they resulted from the workshop concept or action research (\Cref{sec:Discussion-RQ3}). 
            We were not able to mitigate this threat in this study, but we advise replicating the study with an adapted study design to focus on these issues. 
            %Yet, the influence of the researcher's presence remains a general problem in observation studies.
            %We counteracted the ``restricted generalizability across construct''
            % we measured the effort for TDM init and maintenance
            Similarly, our study might be influenced by ``hypothesis guessing'' and ``evaluation apprehension,'' i.e., the practitioners know the study's goal and that they are the objects of our study, which might result in them adjusting their behavior.
            We separated the action and reference teams and repeated the relevance of the reference team and open criticism at each workshop's beginning to balance some of the effects. 

            %CUT candidate:
            % In addition to the threats mentioned by Staron, it could be noted that the different time lengths of the individual cycles could pose a threat to validity. 
            % However, this is part of action research as it can rarely be avoided when working with practitioners.
            % Staron explicitly mentions that ``a good action research cycle is somewhere between 3 and 6 months long'' and it ``is linked to the schedule of the project or organization (its context) and results in a research paper''(~\cite{staron_action_2020}, p.23).
                
    	\subsection{Internal Validity}
            \label{sec:ThreatsToValidity_Internal}
            %First, the team's ``history'' changed during the long-term study, i.e., the pressure to develop new functionalities increased at the end of the project. 
            %This resulted in less time for the establishment's actions and might have led to sub-optimal research results regarding the visualization part.
            First, the team's ``maturation'' might have had an influence, which we already described for ``interactions of different treatments.'' 
            Particularly, the changing meeting structure might have influenced the observations, which were based on review \& planning meetings for the first three months and on the refinement meetings at the end.
            In the beginning, the team still discussed and refined tickets during the review \& planning meetings.
            Therefore, we assume that the overall impression of the observations is still valid. 
            Nonetheless, we had to be careful when interpreting the results from these meetings. 
            Second, asking about problems beforehand can lead to the perception that there is a problem. 
            Staron called this a ``testing'' threat. 
            Our first questionnaire contained such questions about the current situation. 
            %For example, ticking what the team already does regarding TDM shows that they do almost nothing. 
            However, while this made the problems public, the decision to become active and introduce TDM was already made before the first questionnaire, as the pressure resulting from TD consequences was already high.
            Third, a ``biased selection of subject'' influences our study.
            Commonly, only willing industry teams already suffering from a certain problem will collaborate with researchers. % in industry collaborations.
            This implies that our results might assist other teams willing to invest in a TDM process, but might not assist teams unprepared to invest.
            Finally, we cannot negate the ``John Henry effect,'' i.e., that participants compete with a given baseline. 
            %In our case, this threat is similar to ``hypothesis guessing.''
            However, our baseline was exceptionally low, i.e., a TDM process was nearly non-existent. % (TD was recorded as tickets, but there was no TDM, and the tickets got lost in the backlog). 
            Therefore, we assume that this effect is of little relevance to our results. 
            
           % noch frequency und quality of discussions

    	\subsection{Conclusion Validity}
            \label{sec:ThreatsToValidity_Conclusion}

            First, ``missing the needle in the haystack,'' i.e., missing important observations, and ``seeing things that aren't there,'' i.e., observations unrelated to our intervention, might have influenced our study.
            Both threats were countered by two research\-ers who coordinated their observations with each other. 
            While not all observations were made by the two researchers, they regularly compared their observations to ensure that they were valid and that relevant information was included.
            Regardless, we might have missed some observations, but we assume that the overall tendency is still valid.
            Second, a ``low reliability of measures'' might have affected the TD-SAGAT results.
            We could use TD-SAGAT only about once a month so as not to influence the participants too much. 
            Often, only one or two participants could respond regarding a specific ticket.
            This led to a low number of data points per month.
            This method may be more effective in other teams where team members have a similar background.
            Nevertheless, TD-SAGAT provided valuable information.
            Comparing observations with TD-SAGAT has revealed a problem with the team's know-how transfer, leading to countermeasures like rotating moderation to avoid \textit{knowledge silos}. 
            Third, the ``low reliability of measures'' might also be relevant for our questionnaires.
            The questionnaires were sometimes misleading, leading to questions from the participants. 
            Thus, we improved the questionnaire during our work to simplify the questions.
            % For example, the question about a competing requirement was divided into its existence, business value, and urgency. 
            %In the simplified version, we only ask for its existence, and the previous results are accumulated accordingly. 
            % %CUT candidate:
            % Additionally, we didn't evaluate frequencies in our observations. 
            % We calculated the requirements only once per backlog ticket, even if the requirement was mentioned multiple times. 
            % %Further, we did not evaluate the quality of the discussion.
            % We assumed we could forego the frequency because the observation's effort would be much higher and the benefit low as the teams rarely discussed requirements more than once. 
            % %Regarding the quality, we assume that this would be interesting to observe but would justify a separate paper.
            Finally, regarding the ``random heterogeneity of respondents,'' in rare situations, the questionnaires were initially understood differently. 
            However, research\-ers were present when the questionnaires were completed and were able to clarify uncertainties. %, e.g., regarding the item for which the TD-SAGAT questionnaire should be completed.
            
    	\subsection{External Validity}
            \label{sec:ThreatsToValidity_External}
            First, we examined a team that consciously wanted to tackle the topic of TDM and that had the full support of the team manager, which is a ``selection bias'' as similarly discussed for internal validity.
            This limits the transferability to teams that are similarly willing to establish a TDM process. 
            However, it was not the aim of this study to establish a TDM process in unwilling teams. 
            %mentioned by Staron as ``selection bias'' and ``interaction effects of selection bias and the experimental variable.'' 
            %Furthermore, this is similarly relevant to the ``interaction effects of selection bias and the experimental variable'' and ``biased selection of subject.''
            %Our results are, therefore, only transferable to other teams willing to tackle the topic of TDM and spend time on it.
            Second, if we control the context too much, it becomes difficult to generalize to non-experimental settings, which Staron describes as ``reactive effects of experimental arrangements.''
            We intervened controllingly by presenting certain research results, but we could not present all the research results on TD. 
            Nevertheless, the practitioners could choose for themselves which methods they would adopt, and the whole workshop approach's goal is to enable the participants to choose what fits best into their environment without limiting the scope to one specific process.  %, e.g., the TAP framework developed by the researchers themselves was not adopted.
            Third, we decided to do the workshops with a distance of at least eight weeks to avoid a ``multiple treatment interference,'' i.e., when two or more actions simultaneously make it difficult to control the effects.
            Finally, Staron mentions ``constructs, methods, and confounding factors,'' i.e., that measurements and methods might not allow us to generalize outside the study's context.
            In our study, the method is limited to agile working teams, as we adopt the idea of iterations and team decisions. 
            This approach would work less well in a context where the team is not allowed such decisions, e.g., when a central group like architects or team managers leads such process decisions.
	
\section{CONCLUSION}
\label{sec:Conclusion}

    We conducted an action research study with \ActionCycles action cycles in \StudyMonths months. 
    During this study, we presented TDM approaches from research papers to an IT team in five workshops structured along the TD activities.
    
    This study showed which approaches the IT team adopted and which they dismissed. 
    Further, we showed the effects the establishment of the TDM process had on the participants' TD awareness  in decision-making situations. 
    To evaluate TD awareness, we observed \MeetingCnt team meetings, i.e., \MeetingHrs meeting hours, and evaluated questionnaires from each workshop.
    Additionally, we used TD-SAGAT, a method adopted from the psychology domain to analyze situation awareness. %, based on a method to analyze situation awareness in the psychology domain. 

    We found that two methods for TD prioritization and repayment were preferred, which are (1)~repayment based on the system's evolution~\cite{Schmid2013} and (2)~repayment of \textit{low-hanging fruits}, i.e., based on cost calculations~\cite{McConnell2008a, Tom2013b}.
    For TD prevention, we found that  presenting information, e.g., during the workshops,  is insufficient, but reminders in the backlog's ticket templates, e.g., a \textit{Talked about TD} checkbox, lead to a sustainable change.
    Finally, various new ideas for TDM emerged from this study: (1)~using a \textit{re-submission date}, (2)~using a \textit{Talked about TD} checkbox, (3)~using \textit{educated guesses} for priority instead of calculations, (4)~using a \textit{ROI-based} priority to improve the \textit{educated guess} and (5)~using visualizations for TD prioritization, e.g., to identify the \textit{low-hanging fruits}.

        \textbf{For practitioners,} this study outlined a five-step workshop approach for establishing TDM in practice.
        We demonstrated which research approaches were feasible in an embedded environment and presented five additional (new) approaches. 
        Practitioners can use these as inspiration for establishing their own TDM process.
        We further provided an actionable list of recommendations for establishing a TDM process. 
        %Additionally, we present the five new approaches they might want to adopt.
        
        \textbf{For researchers,} we provided insights into which methods are feasible in practice and which might be less helpful. 
        In particular, the resistance to a calculated, e.g., \textit{ROI-based} priority is striking. 
        Research could address this issue by providing methods to improve the \textit{educated guess} priority instead of researching and providing calculations.
        Further, we identified that attribute fields in a backlog ticket can sustainably support TD awareness, including sound decision-making. %, than just providing information. 
        %It could be interesting to further analyze
        This realization raises the research questions of  which attributes should be part of backlog tickets, including for purposes other than supporting TDM. 
        Including a TD ticket type and the relevant attributes in commercial backlog tools might further support the TDM establishment in practice and could be promoted by researchers.
        Finally, we presented a new method --~TD-SAGAT~-- to analyze TD awareness adopted from psychology and critically discussed its suitability.
        %Finally, we provided a five-step workshop concept for establishing TDM. 
        %While we tested this concept with this one IT team, it is relevant to replicate this study and research how the workshops will work for other teams, e.g., from other domains. 
        %In this way, a guide to best practices for TDM could be developed in the long term.

        \textbf{For future work}, the team wishes to continue the collaboration to distribute the TDM process to other teams. 
        We might approach this topic similarly to our workshops by adopting existing research methods, e.g., TD Guilds~\cite{detofeno_technical_2021}. 
        Further, we plan to evaluate the five-step workshop concept in other teams from other domains. 
        By this, we plan to optimize the workshop concept, compile a guideline comprising approaches that have been demonstrated to be best practices, and supplement these with approaches that work in specific environments.
         This list could be further improved by replicating the study in less agile work environments.
        As this research comprised researchers' interventions, it would be interesting to evaluate if this approach also works with less intervention by researchers. 
        Another intriguing research topic would be to compare the TD awareness evaluation of this study with the results of TDM establishments of pre-defined processes.
        This comparison could determine whether the more open approach of this workshop concept leads to higher TD awareness. e.g., because participants may work more intensively on this topic.
       
\section*{Acknowledgment}
    We thank the case company and its team members for their participation, trust, and the valuable insights they provided.
    In memory of~Prof. Dr.-Ing. André van Hoorn and Prof. Dr.-Ing. Matthias Riebisch.
    
\section*{Funding and Declarations}
    The project on which this report is based was sponsored by the Federal Ministry of Education and Research of Germany under the funding code 01IS24031. 
    Responsibility for the content of this publication lies with the authors.
%We gratefully acknowledge funding from the German Federal Ministry of Research, Technology and Space under the grant 01IS230XX (for the Software Campus project XXX).

    During the preparation of this work, the authors used Grammarly to improve the readability and language of the manu\-script. 
    After using this tool, the authors reviewed and edited the content as needed and take full responsibility for the content of the published article.
    
%    Example see Appendix \ref{sec:sample:appendix}.

%\appendix

%\section{Sample Appendix Section}

% \bibliographystyle{elsarticle-num} 





% \end{thebibliography}
 \bibliography{TDM_ActionResearch}
\end{document}